\documentclass[prd,aps,showpacs]{revtex4}

\usepackage{latexsym, graphicx, color}
\usepackage{amsmath, amssymb}
\usepackage{color}

\begin{document}

\title{Unruh-DeWitt detectors as mirrors: Dynamical reflectivity and Casimir effect}
\author{Shih-Yuin Lin}
\email{sylin@cc.ncue.edu.tw}
\affiliation{Department of Physics, National Changhua University of Education, Changhua 50007, Taiwan}
\date{4 June 2018}

\begin{abstract} 
We demonstrate that the Unruh-DeWitt harmonic-oscillator detectors in (1+1) dimensions derivative-coupled with a massless scalar field can mimic the atom mirrors in free space. Without introducing the Dirichlet boundary condition to the field, the reflectivity of our detector/atom mirror is dynamically determined by the interaction of the detector's internal oscillator and the field. When the oscillator-field coupling is strong, a broad frequency range of the quantum field can be mostly reflected by the detector mirror at late times. Constructing a cavity model with two such detector mirrors, we can see how the quantum field inside the cavity evolves from a continuous to a quasi-discrete spectrum which gives a negative Casimir energy density at late times. In our numerical calculations, the Casimir energy density in the cavity does not converge until the UV cutoff is sufficiently large, with which the two internal oscillators are always separable.
\end{abstract} 


\maketitle

\section{Introduction}

A moving mirror can produce quantum radiation from a vacuum \cite{Fu73, De75, FD76, BD82}; two mirrors at rest can form a cavity with a negative Casimir energy density inside \cite{Ca48, De75, BD82, BMM01, KMM09}, and one or two such cavity mirrors moving in specific ways can create particles in the cavity 
\cite{Mo70, JJ09, JJ10, WJ11}. All these interesting physics can be obtained by simply modeling a perfect mirror as a Dirichlet boundary 
condition for the field at the position of the mirror. Nevertheless,
such simple models may not always be satisfactory either in theoretical or experimental aspects. 
Theoretically, a detector or atom inside a cavity of perfect mirrors would experience endless echoes without 
relaxation if the atom and the field are not started with a steady state of the combined system \cite{LCH16}. 
The equilibrium approach will never apply if one does not introduce an {\it ad hoc} dissipation for the cavity. 
Experimentally, while the incident waves of the fields at all frequencies get total reflection by a perfect mirror in the Dirichlet boundary 
condition, a physical mirror is not perfect anyway: The charges of a realistic mirror responding to the incident electromagnetic waves 
have a finite relaxation time and the reflectivity reaches almost $100\%$ only in a finite working range of frequency. 

To describe more realistic situations and to see how well the results by simply introducing the Dirichlet boundary conditions can do, 
there have been some mirror models which are more sophisticated than the simple, strong boundary conditions for the field. 
For example, Barton and Calogeracos introduced a mass term for the field at the mirror's position which acts 
like a delta-function potential \cite{BC95, CB95}, 
Golestanian and Kardar applied an auxiliary field to constrain the field amplitude around the mirror's position \cite{GK97, GK98},
and Sopova and Ford replaced perfect conducting mirrors by dispersive dielectrics \cite{SF02}. 
Recently Galley, Behunin, and Hu constructed a mirror-oscillator-field (MOF) model with a new internal degree of freedom of the mirror minimally coupled to the field at the mirror's position 
to mimic the microscopic interaction between the field and the surface charges of the mirror \cite{GBH13}. 
Such a microscopic treatment captures the mirror-field interaction in a more physically consistent way. 
The authors of \cite{GBH13} also showed that their MOF model can be connected to the earlier models in Refs. 
\cite{BC95, CB95, GK97, GK98, La95} with different choices of parameters and limits. 
A similar model with the derivative coupling was considered by Wang and Unruh to study the force exerted on the mirror by vacuum 
fluctuations \cite{WU14}. Wang and Unruh further considered a model with the internal oscillator minimally coupled to a massive scalar 
field \cite{WU15} to get rid of the divergent effective mass in \cite{WU14}. 

In Ref. \cite{SLH15} Sinha, Hu and the author of the present paper realized that the mirrors in the MOF models with the minimal and the 
derivative couplings behave like metal and dielectric mirrors, respectively. They introduced a new coupling 
to a harmonic-oscillator bath to describe the interaction between the mirror's internal degree of freedom and the mechanical degrees 
of freedom such as the vibration of the mirror substrate and the environment connected by the suspension of the mirror. 
They also verified that in the strong coupling regime their results are close to those with the Dirichlet boundary conditions
\cite{Ja05, GJ02, GJ03, GJ04}. 

Since the MOF models are nonlinear due to the mirror motion, one usually needs to make some linear approximations in practical calculations.
Among those approximations, restricting the mirror moving along a prescribed worldline may be the simplest one. 
By doing so, the motion of the mirror is not dynamical, and a derivative-coupling MOF model in Ref. \cite{SLH15} reduces to a 
derivative-coupling Unruh-DeWitt (UD$'$) harmonic-oscillator (HO) detector theory \cite{Unr76, DeW79, UZ89, RSG91} with additional HO baths, 
which is the model we are considering in this paper. 

The late-time reflectivity of our ``detector mirror" in the weak oscillator-field (OF) coupling regime is similar to the atom mirrors in the 
cavity and waveguide QED \cite{SF05, ZD08, AZ10, HSHB11, WJ11, CGS11, CJGK12, SMK14}, whose reflectivity are peaked in a narrow band around 
a single frequency of resonance. The cavity of those atom mirrors can only generate few cavity modes inside since the detector and atom mirrors are almost transparent for other harmonics \cite{ZD08}. 
In the field-theoretical derivation for the Casimir effect \cite{Ca48}, however, one needs to sum over all the cavity modes to get the 
Casimir energy density in a perfect cavity. 
Thus in this paper we extend our attention to the detector mirrors in the strong OF coupling regime, where the reflectivity of 
the detector mirror is close to 100\% in a very wide frequency range of the field. 
Later we will see that, while the transient behaviors of the combined system can be significantly different for different coupling strengths,
the late-time renormalized energy density of the field inside a cavity of our detector mirrors is always negative even in the weak
OF coupling regime where the cavity modes are few.

The paper is organized as follows. The classical theory for our single ``detector mirror" is given in Sec. \ref{CT1DM}, where we examine 
the relaxation time and the late-time behavior of the system and then derive the late-time reflectivity determined by the 
interplay between the HO and the field. 
In Sec. \ref{QT1DM} we develop the quantum theory of the detector mirror and show that the energy density of the field outside the 
detector mirror is zero at late times while the equal-time correlations of the field amplitudes at different positions are reduced by the 
mirror. In Sec. \ref{detCav}, we consider a cavity of the detector mirrors, show that there are indeed many cavity modes inside our 
cavity at late times in the strong OF coupling regime, and then calculate the late-time renormalized energy density inside the cavity, which 
turns out to be negative for all nonzero coupling strengths. After addressing the HO-HO entanglement at late times,
a summary of our findings is given in Sec. \ref{Summa}.

\section{Detector mirror: Classical theory}
\label{CT1DM}

A mirror moving along the worldline $z^\mu$ with its internal degree of freedom $Q$ coupled with a quantum field $\Phi$ 
in (1+1)D Minkowski space may be described by the action given in Eq. (1) of Ref. \cite{SLH15}, with $Z$ there replaced by $z^1$ and
with the mechanical damping $\Gamma$ and noise $\xi$ in Eq. (35) there introduced. 
Since the position of the mirror $z^1$ is not considered as a dynamical variable in this paper, we can write down the reduced action as
\begin{eqnarray}
  S &=& -\int dt dx \frac{1}{2}\partial_\mu\Phi_x(t) \partial^\mu\Phi_x(t) 
	    +\frac{1}{2}\int d\tau \left[ \left(\partial_\tau Q(\tau)\right)^2 -\Omega_0^2 Q^2(\tau)\right]
	    \nonumber\\
		&	& -\int d\tau \int dt dx \lambda(\tau) Q(\tau) \partial_\tau\Phi_x(t) \delta(t-z^0(\tau))\delta(x-z^1(\tau)) \nonumber\\
		& & -\int d\tau dy \frac{1}{2}\partial_\nu {\cal Z}_y(\tau) \partial^\nu {\cal Z}_y(\tau)  
		    -\int d\tau dy \tilde{\lambda}(\tau) Q(\tau) \partial_\tau {\cal Z}_y(\tau) \delta(y-\vartheta),
\label{Stot1}
\end{eqnarray}
which is a derivative-coupling UD$'$ detector theory \cite{Unr76, DeW79, UZ89, RSG91}.
Here the natural unit with the speed of light $c=1$ is adopted, $(t,x)$ are the Minkowski coordinates, $\tau$ is the proper time of the 
detector mirror, $Q$ is a HO of mass $m=1$ living in an internal space of the detector, and $\lambda(\tau)$ is the 
switching function of the OF coupling, assumed to be vanishing before the initial time $\tau^{}_I$. The derivative 
coupling is chosen for its well-behaved radiation reaction term, which is the first derivative of the proper time of the detector [e.g. Eq.
(\ref{eomQRR})]. The function $\tilde{\lambda}(\tau)$ corresponds to the coupling between the internal HO and the environmental 
oscillator bath responsible for the mechanical damping and noise. It can be switched on at a different initial moment $\tau'_I\not=\tau^{}_I$.
In the strong OF coupling regime, the absolute value of the OF coupling $|\lambda|$ is much greater than the oscillator-environment(OE) 
coupling $|\tilde{\lambda}|$ so that the former interaction dominates and the detail of the environment would not be important.
Thus for simplicity and consistency, we model the complicated environmental degrees of freedom such as the vibration of the mirror substrate 
and those connected by the suspension of the mirror by a single massless scalar field ${\cal Z}_y(\tau)$ in another internal space $y\in 
{\bf R}^1$, and assume that the internal HO of the mirror also acts as an UD$'$ detector located at $y=\vartheta$ in that internal space 
\footnote{The internal HO $Q$ of a detector in a massless scalar field $\Phi$ in (1+1)D with the $F(Q,\dot{Q})\dot{\Phi}$ coupling
($F(Q,\dot{Q})\Phi$ coupling) behaves like a HO in an Ohmic (sub-Ohmic) bath in the quantum Brownian motion models with a particular set of 
couplings to the bath oscillators \cite{HM94, LH07}. Here $F$ is an arbitrary function.}. In this way the dissipation and fluctuations will be related consistently.
Then the action (\ref{Stot1}) is quadratic and the combined system is linear and solvable.
When considering two or more mirrors, the internal space $y$, the phase parameter $\vartheta$, and the coupling $\tilde{\lambda}$ of 
each mirror will be considered independent of those of the other detector mirrors.

From (\ref{Stot1}) the conjugate momenta of the detector, the field, and the mechanical environment read 
\begin{eqnarray}
  P(\tau) &=& \frac{\delta S}{\delta \partial_\tau Q(\tau)} = \partial_\tau Q(\tau), \\
	\Pi_x(t) &=& \frac{\delta S}{\delta\partial_t\Phi_x(t)} 
		=\partial_t \Phi_x(t) - \lambda(\tau^{}_t) Q(\tau^{}_t) \delta (x - z^1(\tau^{}_t)), \\
	\Upsilon_y(\tau) &=& \frac{\delta S}{\delta\partial_\tau {\cal Z}_y(\tau)} = \partial_\tau {\cal Z}_y(\tau) - 
	   \tilde{\lambda}(\tau) Q(\tau) \delta(y-\vartheta),
\end{eqnarray}
respectively, with which the Hamiltonian on a $t$-slice is given by 
\begin{eqnarray}
  H(t) &=& \frac{1}{2 v^0(\tau^{}_t)}\left[ P^2(\tau^{}_t) + \Omega_0^2 Q^2(\tau^{}_t) \right] \nonumber\\
	& &+\,\frac{1}{2}\int dx \left\{ \left[ \Pi_x(t)+\lambda(\tau^{}_t) Q(\tau^{}_t) \delta(x-z^1(\tau^{}_t))\right]^2 + 
	\left[ \partial_x \Phi_x(t)\right]^2 \right\} \nonumber\\
	& &+\,\frac{1}{2v^0(\tau^{}_t)}\int dy \left\{ \left[ \Upsilon_y(\tau^{}_t)+ \tilde{\lambda}(\tau^{}_t) Q(\tau^{}_t) \delta(y-\vartheta) 
	  \right]^2 + \left[ \partial_y {\cal Z}_y(\tau^{}_t)\right]^2 \right\}\nonumber\\
	& &+\,\lambda(\tau^{}_t) Q(\tau^{}_t) \frac{v^1(\tau^{}_t)}{v^0(\tau^{}_t)}\partial_x\Phi_{z^1(\tau^{}_t)}(t), 
	\label{Hamil}
\end{eqnarray}
where $\tau^{}_t$ is obtained by solving $t=z^0(\tau^{}_t)$ and $v^\mu(\tau)\equiv \partial_\tau z^\mu(\tau)$ is the 
two-velocity \footnote{The last term of the Hamiltonian (\ref{Hamil}) is overlooked in Eq.(2.6) of Ref.\cite{LCH16}. 
This does not affect the results in \cite{LCH16} since 
$(v^0(\tau),v^1(\tau))=(1,0)$ in the cases considered there.}.

Suppose the detector is at rest at $x=0$ in the external Minkowski space, so that $z^1(\tau)=0$, $z^0(\tau)=\tau=t$, $v^1(\tau)=0$ and 
$v^0(\tau)=1$. Then the value of the Hamiltonian (\ref{Hamil}) equals
\begin{eqnarray}
  E(t) &=& \frac{1}{2}\left[ \left(\partial_t Q(t)\right)^2 + \Omega_0^2 Q^2(t) \right] + 
	\frac{1}{2}\int dx \left[ \left(\partial_t\Phi_x(t)\right)^2 + \left(\partial_x \Phi_x(t)\right)^2 \right]\nonumber\\
	& &+\,\frac{1}{2}\int dy \left[ \left(\partial_\tau {\cal Z}_y(t)\right)^2 + \left(\partial_y {\cal Z}_y(t)\right)^2\right] , \label{Eden}
\end{eqnarray}
which appears no cross term between different kinds of the degrees of freedom. 
Anyway, the Euler-Lagrange equations in this case, 
\begin{eqnarray}
	\left( \partial_t^2 - \partial_x^2\right)\Phi_x(t) &=& 
	  \partial_t \left(\lambda(t) Q(t)\right)\delta(x), \label{eomPhi} \\
	\left( \partial_t^2 - \partial_y^2\right){\cal Z}_y(t) &=& \partial_t \left(\tilde{\lambda}(t) Q(t)\right)\delta(y-\vartheta),\label{eomX}\\
  \left(\partial_t^2+ \Omega_0^2\right) Q(t) &=& -\lambda(t) \partial_t \Phi_0(t) -\tilde{\lambda}(t) \partial_t {\cal Z}_\vartheta(t),\label{eomQ}
\end{eqnarray}
from (\ref{Stot1}), are still coupled. 
The general solution of the field $\Phi$ in (\ref{eomPhi}) can be expressed as
\begin{equation}
  \Phi_x(t) = \Phi_x^{^{[0]}}(t) + \Phi_x^{^{[1]}}(t) ,
\end{equation} 
where $\Phi_x^{^{[0]}}(t)$ is the homogeneous solution satisfying $\Box\Phi^{^{[0]}}=0$ and $\Phi_x^{^{[1]}}(t)$ is the inhomogeneous 
solution given by
\begin{eqnarray}
  \Phi_x^{^{[1]}}(t) &=& \int_{-\infty}^\infty d\tau\partial_{\tau}\left( \lambda(\tau)Q(\tau)\right) G_{\rm ret}(t,x;z^0(\tau),z^1(\tau)) 
	\nonumber\\	&=& \frac{1}{2} \lambda(t-|x|)Q(t-|x|)  , \label{Phisol}
\end{eqnarray}
after an integration by part. Here the retarded Green's function for a massless scalar field in (1+1)-dimensional Minkowski space 
${\bf R}_1^1$ reads $G_{\rm ret}(t,x;t',x') = \theta[(t+x)-(t'+x')]\theta[(t-x)-(t'-x')]/2$, where $\theta(x)$ is the Heaviside step 
function with the convention $\theta(0)=1/2$. The surface terms in (\ref{Phisol}) have been dropped since $\lim_{\tau\to\infty}$ $G_{\rm ret}
(t,x;$ $z^0(\tau),z^1(\tau))=0$ for all finite $t$ and $x$, and we assume $\lim_{\tau\to-\infty} \lambda(\tau)=0$ long before the coupling 
is switched on. Similarly, the general solution of ${\cal Z}$ in (\ref{eomX}) is
\begin{equation}
  {\cal Z}_y(t) ={\cal Z}_y^{^{[0]}}(t)
			+\frac{1}{2}\tilde{\lambda}(t-|y-\vartheta|)Q(t-|y-\vartheta|) \label{Xsol}
\end{equation}
where ${\cal Z}_y^{^{[0]}}(t)$ is the homogeneous solution.
Inserting the solutions of the field $\Phi$ and the mechanical environment ${\cal Z}$ into (\ref{eomQ}), one obtains
\begin{eqnarray}
  &&\ddot{Q}(t) + \left(\frac{\lambda^2(t)}{2} +\frac{\tilde{\lambda}^2(t)}{2}\right)\dot{Q}(t) + 
	\left(\Omega_0^2 + \frac{\lambda(t)\dot{\lambda}(t)}{2}+\frac{\tilde{\lambda}(t)\dot{\tilde{\lambda}}(t)}{2}\right) Q(t)\nonumber\\ 
	&=& -\lambda(t) \dot{\Phi}_0^{^{[0]}}(t) - \tilde{\lambda}(t)\dot{\cal Z}_\vartheta^{^{[0]}}(t),
\label{eomQRR}
\end{eqnarray}
which shows that $Q$ behaves like a driven, damped HO with a time-dependent frequency.
 
\subsection{Relaxation}
\label{relax}

Suppose the OF coupling is switched on at $t=t_0$, namely, $\lambda(t)=\lambda \theta(t-t_0)$ with $\theta(0)=1/2$, while $\tilde{\lambda}$ has become a positive constant long before $t_0$, and initially $Q(t_0)=\dot{Q}(t_0)=0$. Integrating (\ref{eomQRR}) from 
$t=t_0-\epsilon$ to $t_0+\epsilon$ for $\epsilon \to 0+$, one has $0 = \dot{Q}(t_0+\epsilon)-\dot{Q}(t_0-\epsilon)+ (\lambda^2/4) Q(t_0)$
for continuous $Q(t)$. This implies that $\dot{Q}$ is continuous around $t=t_0$ since $Q(t_0)=0$, and
so the solution for (\ref{eomQRR}) reads 
\begin{equation}
  Q(t) = \int_{t_0}^t d\tilde{\tau} K(t-\tilde{\tau}) \left[ -\lambda \dot{\Phi}_0^{^{[0]}}(\tilde{\tau}) - 
	\tilde{\lambda}\dot{\cal Z}_\vartheta^{^{[0]}}(\tilde{\tau})\right], \label{Qsol}
\end{equation}
for $t \ge t_0$, where the propagator $K$ is defined by
\begin{equation}
  K(s) \equiv  \frac{1}{2\Gamma} e^{-(\gamma+\tilde{\gamma})s} \left(e^{\Gamma s}-e^{-\Gamma s}\right)
	 = e^{-(\gamma+\tilde{\gamma})s}  \Gamma^{-1} \sinh \Gamma s ,	\label{Kpropa}
\end{equation}
with the coupling strengths $\gamma \equiv\lambda^2/4 >0$ and $\tilde{\gamma}\equiv\tilde{\lambda}^2/4 >0$, the parameter
$\Gamma\equiv\sqrt{(\gamma+\tilde{\gamma})^2-\Omega_0^2}$ in the over-damping cases, and $\Gamma = i\Omega \equiv i \sqrt{\Omega_0^2 - 
(\gamma+\tilde{\gamma})^2}$ in the under-damping cases. 
In the cases of critical damping, $\Gamma^{-1}\sinh\Gamma s$ in (\ref{Kpropa}) reduces to $s$ as $\Gamma \to 0$ . 

In the integrand of (\ref{Qsol}) one can see that there are two channels of relaxation proportional to $e^{-(\gamma+\tilde{\gamma}-\Gamma)
(t-\tau)}$ and $e^{-(\gamma+\tilde{\gamma}+\Gamma)(t-\tau)}$ after (\ref{Kpropa}) is inserted. 
In the cases of under- and critical damping, one has the relaxation time-scale $1/(\gamma+\tilde{\gamma})$, which gets shorter for larger 
$\gamma$ and $\tilde{\gamma}$. In the over-damping cases, however, $\Gamma$ is a positive real number and so $e^{-(\gamma+\tilde{\gamma}-
\Gamma)(t-\tau)}$ sets a time-scale of relaxation, 
\begin{equation}
    t_{\rm rlx}=(\gamma+\tilde{\gamma}-\Gamma)^{-1}, \label{rlxtime} 
\end{equation}
which will be longer for a stronger coupling strength $\gamma$ and/or $\tilde{\gamma}$ if $\Omega_0$ is fixed. For $\gamma\gg \Omega_0$, 
one has $t_{\rm rlx} \approx 2(\gamma+\tilde{\gamma})/\Omega_0^2$.

\subsection{Late-time solutions}

Introducing a right-moving wave $\Phi_x^{^{[0]}}(t)= e^{-i \omega t + i k x}$, $\omega = k >0$, as the driving force in (\ref{Qsol}) and 
assuming ${\cal Z}_y^{^{[0]}}(t) = 0$ for simplicity,
once the OF coupling $\lambda$ has become a positive constant for a sufficiently long time for relaxation ($t-t_0 \gg t_{\rm rlx}$), one has 
$Q(t)\propto e^{-i\omega t}$ at late times according to (\ref{Qsol}). Suppose the time scale of switching-on the OF coupling is much shorter 
than $t_{\rm rlx}$. Inserting the late-time ansatz $Q(t)\to\tilde{Q}_\omega e^{-i\omega t}$ into (\ref{eomQRR}), one can solve 
$\tilde{Q}_\omega$ and find the late-time solution
\begin{equation}
  Q(t) \to  \chi^{}_\omega \left[ -\lambda \Phi_{0}^{^{[0]}}(t)-\tilde{\lambda} {\cal Z}_{\vartheta}^{^{[0]}}(t)\right]
	= - \lambda e^{-i\omega t}\chi^{}_\omega ,
\end{equation}
with the susceptibility function
\begin{equation}
   \chi^{}_\omega \equiv  \frac{-i\omega}{\Omega_0^2-\omega^2 -2i\omega(\gamma + \tilde{\gamma})}, \label{chiw1}
\end{equation}
which implies that
\begin{eqnarray}
  \Phi_x(t) &\to& 
	e^{-i \omega (t-x)} - 2\gamma e^{-i\omega (t-|x|)}\chi^{}_\omega \label{Philate}\\
	&\equiv& \theta(-x)\left[\Phi_x^{^{[0]}}(t) + \Phi_x^{^{[R]}}(t)\right] + \theta(x) \Phi_x^{^{[T]}}(t) \label{PhiIRT}
\end{eqnarray}
from (\ref{Phisol}).

\subsection{Reflectivity}
\label{secReflect}

In (\ref{Philate}), the first term and the second term in the $x<0$ region can be interpreted as the incident and 
reflected waves $\Phi_x^{^{[0]}}$ and $\Phi_x^{^{[R]}}$, respectively, while the superposition of $\Phi_x^{^{[0]}}$ and $\Phi_x^{^{[1]}}$ 
in the $x>0$ region can be interpreted as the transmitted wave $\Phi_x^{^{[T]}}$, as in (\ref{PhiIRT}). 
Thus at late times we can define the reflectivity as 
\begin{equation}
   |{\cal R}(k)|^2 \equiv \frac{|\Phi_{x}^{^{[R]}}(t)|^2}{|\Phi_{x}^{^{[0]}}(t)|^2} \to
   \left| 2\gamma \chi^{}_\omega \right|^2\label{RR}
\end{equation}
and the transmittivity as
\begin{equation}
   |{\cal T}(k)|^2 \equiv \frac{|\Phi_{x}^{^{[T]}}(t)|^2}{|\Phi_{x}^{^{[0]}}(t)|^2} \to 
   \left| 1- 2\gamma\chi^{}_\omega\right|^2. \label{TT}
\end{equation}

\begin{figure}
\includegraphics[width=8cm]{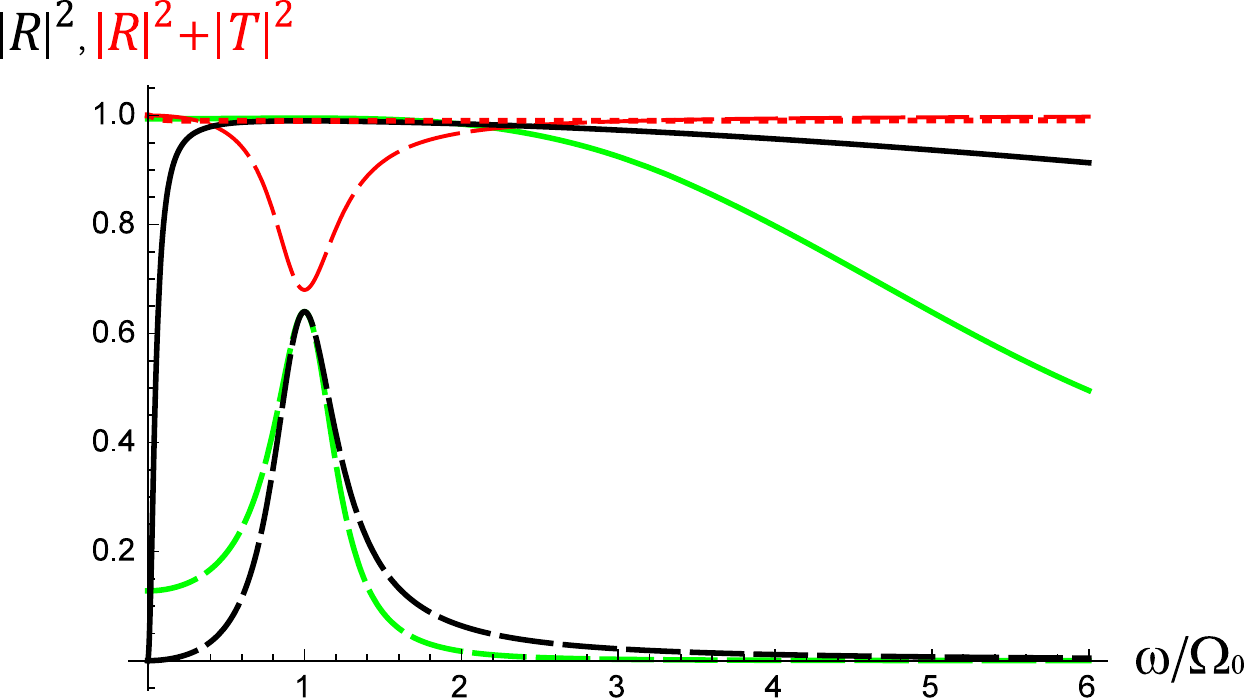}
\caption{The late-time reflectivity $|{\cal R}|^2$ (black lines) and the sum of the reflectivity and transmittivity $|{\cal R}|^2+|{\cal T}|^2$ (red lines) against $\omega=|k|$, given in (\ref{RR}) and (\ref{TT}). Here $\Omega_0=1$, $\tilde{\gamma}=0.05$, and $\gamma = 0.2$ (dashed lines) and $10$ (solid or dotted lines). We plot the reflectivity of the minimal-coupling model [Eq.(17) in Ref.\cite{SLH15}] with the same parameters (green lines) for 
comparison. One can see that the derivative-coupled and the minimal-coupled detectors act like a dielectric and a metal mirror, respectively, in the regime of $\omega\to 0$.}
\label{reflec}
\end{figure}

An example of the above late-time reflectivity and transmittivity is shown in Figure \ref{reflec}. One can see that the reflectivity is 
peaked around $\omega =\Omega_0$ where the internal HO of the detector mirror and the incident wave of the field are resonant.
Observing that $|{\cal R}(k)|^2 = [\gamma/(\gamma+\tilde{\gamma})]^2$ and $|{\cal T}(k)|^2 = [\tilde{\gamma}/(\gamma+\tilde{\gamma})]^{2}$ 
at $\omega=\Omega_0$, the UD$'$ detector will be a perfect mirror ($|{\cal R}(k)|^2 =1$) for the incident monochrome wave 
if the internal HO is decoupled from the mechanical environment ($\tilde{\gamma}=0$). 
When the OE coupling $\tilde{\gamma}$ is not negligible, however, the energy of the field $\Phi$ around the resonant frequency 
will be significantly absorbed by the environment ${\cal Z}$, so that $|{\cal R}(k)|^2 +|{\cal T}(k)|^2$ becomes lower than $1$ around 
$\omega\approx \Omega_0$.

In Figure \ref{reflec} one can also see that when the frequency of the incident wave is far off resonance, the reflectivity is small and so 
the mirror is almost transparent for that incident wave. With weak couplings ($\gamma, \tilde{\gamma} < \Omega_0$), large reflectivity 
occurs only in a narrow frequency range of the width $O(\gamma+\tilde{\gamma})$ around the resonance (dashed curves). This feature is 
similar to the usual dielectric mirrors and atom mirrors \cite{SF05, ZD08, AZ10, HSHB11, WJ11}. 
The cavities of these kind of mirrors can produce only one or a few pairs ($k=\pm \omega$) of resonant modes inside \cite{ZD08}, since the 
detector mirrors are nearly transparent for other harmonics.
In constructing a cavity model for comparing with the conventional approach to the Casimir effect, one may need a detector mirror with a 
very wide working range of frequency to form an effective Dirichlet boundary condition $\Phi_{x_0}(t)\approx 0$ at the mirror's position 
$x=x_0$. This could be done by carefully arranging a collection of detectors or atoms to form the mirror \cite{CGS11, CJGK12}.
Alternatively, one can simply raise the OF coupling of a single detector all the way to the over-damping regime for the internal HO
($\gamma \gg \Omega_0, \tilde{\gamma}$) to achieve it.
As shown by the solid curve in Figure \ref{reflec}, the reflectivity of a detector mirror in this 
over-damping regime will go to approximately $1$ in a wide frequency range at late times,
though it may take a very long relaxation time to reach this stage, as discussed in Sec. \ref{relax}. 
Later we will see explicitly in the quantum theory that a cavity of the detector mirrors in the over-damping regime can indeed generate many 
cavity modes and the field spectrum inside the cavity is quasi-discrete at late times. 

Note that the definition of reflectivity (\ref{RR}) makes sense only at late times. $|\Phi_{x}^{^{[R]}}(t)|^2/$ $|\Phi_{x}^{^{[0]}}(t)|^2$ 
can be greater than 1 in transient when the initial zero-point fluctuations of the detector burst out right after the OF coupling is 
switched on (see, e.g., the left plots of Figure \ref{FxkEvo}).  

\section{Detector mirror: Quantum theory}
\label{QT1DM}

The Heisenberg equations of motion in the quantum theory of our model (\ref{Stot1}), which is a linear system, have the same form as the 
Euler-Lagrange equations (\ref{eomPhi})-(\ref{eomQ}):
\begin{eqnarray}
	\left(\partial_t^2 -\partial_x^2\right)\hat{\Phi}_x(t) &=& \partial_t \left(\lambda(t) \hat{Q}(t)\right)\delta(x), \label{HeomPhi} \\
	\left(\partial_t^2 -\partial_y^2\right)\hat{\cal Z}_y(t) &=& \partial_t \left(\tilde{\lambda}(t) \hat{Q}(t)\right)\delta(y-\vartheta),
	\label{HeomX}\\
  \left(\partial_t^2+\Omega_0^2\right)\hat{Q}(t) &=& -\lambda(t) \partial_t \hat{\Phi}_0(t) 
	-\tilde{\lambda}(t)\partial_t \hat{\cal Z}_\vartheta(t). \label{HeomQ}
\end{eqnarray}
One can see that each operator will gradually evolve to other operators whenever the couplings are on.
To deal with, we write the operators of the dynamical variables at finite $t$ in terms of the linear combinations of the free operators 
defined before the couplings are switched on, each multiplied by a time-dependent c-number coefficient called the ``mode function," namely,
\begin{eqnarray}
  \hat{Q}^{}_A (t) &=& \sqrt{\frac{\hbar}{2\Omega^{}_{0}}}
	      \left[ q_{A}^{A}(t) \hat{a}^{}_A + q_{A}^{{A}*}(t) \hat{a}^\dagger_{A}\right] +
			\int \frac{d{\rm k}}{2\pi}\sqrt{\frac{\hbar}{2{\rm w}}} 
			  \left[ q_{A}^{\rm k}(t) \hat{a}^{}_{p} + q_{A}^{{\rm k}*}(t) \hat{a}^\dagger_{\rm k}\right]\nonumber\\
			& &+ \int \frac{d\tilde{\rm k}}{2\pi}\sqrt{\frac{\hbar}{2\tilde{\rm w}}} \left[ q_{A}^{\tilde{\rm k}}(t) 
			  \hat{a}^{}_{\tilde{\rm k}} + q_{A}^{\tilde{\rm k}*}(t) \hat{a}^\dagger_{\tilde{\rm k}}\right] 
			\equiv \sum_{\kappa}\sqrt{\frac{\hbar}{2\Omega_{\kappa}}}
	      \left[ q_{A}^{\kappa}(t) \hat{a}^{}_{\kappa} + q_{A}^{\kappa *}(t) \hat{a}^\dagger_{\kappa}\right],\label{QAexpan}\\
  \hat{\Phi}^{}_x(t) &=& \sum_{\kappa}\sqrt{\frac{\hbar}{2\Omega_{\kappa}}}
	      \left[ \varphi_{x}^{\kappa}(t) \hat{a}^{}_{\kappa} + \varphi_{x}^{\kappa *}(t) \hat{a}^\dagger_{\kappa}\right],\label{Phiexpan}\\
  \hat{\cal Z}^{}_{y}(t) &=& \sum_{\kappa}\sqrt{\frac{\hbar}{2\Omega_{\kappa}}}
	      \left[ \zeta_{y}^{\kappa}(t) \hat{a}^{}_{\kappa} + \zeta_{y}^{\kappa *}(t) \hat{a}^\dagger_{\kappa}\right],\label{Xexpan}
\end{eqnarray}
where $\kappa$ runs over $A$, $\{{\rm k}\}$, and $\{\tilde{\rm k}\}$, which are the indices for the free HO labeled $A$, the free field mode 
of wave-number ${\rm k}$, and the free mechanical environment mode of wave-number $\tilde{\rm k}$, respectively. 
Here we have renamed $\hat{Q}$ to $\hat{Q}^{}_A$ to be consistent with the multi-detector cases later in this paper, and we denote
$\sum_{\rm k} \equiv \int d{\rm k}/(2\pi)$, $\sum_{\tilde{\rm k}}\equiv \int d\tilde{\rm k}/(2\pi)$, $\Omega^{}_{A}\equiv \Omega_0$, 
$\Omega_{\rm k}\equiv {\rm w} \equiv |{\rm k}|$, and $\Omega_{\tilde{\rm k}}\equiv \tilde{\rm w}\equiv |\tilde{\rm k}|$. 
The raising and lowering operators of the free internal HO have the commutation relation $[\hat{a}^{}_{A}, \hat{a}^\dagger_{A}]= 1$, 
while the creation and annihilation operators for the free massless scalar field and the free environment satisfy $[\hat{a}^{}_{\rm k}, 
\hat{a}^\dagger_{\rm k'}]=2\pi\delta ({\rm k}-{\rm k'})$ and $[\hat{a}^{}_{\tilde{\rm k}}, \hat{a}^\dagger_{\tilde{\rm k}'}]=
2\pi\delta(\tilde{\rm k}-\tilde{\rm k}')$, respectively.

Applying these commutation relations of $\hat{a}$ and $\hat{a}^\dagger$ to the Heisenberg equations (\ref{HeomPhi})-(\ref{HeomQ}), 
one obtains the equations of motion for the mode functions,
\begin{eqnarray}
	\left(\partial_t^2-\partial_x^2\right)\varphi^\kappa_x(t)&=&\partial_t\left(\lambda(t) q_A^\kappa(t)\right)\delta(x), \label{eomvarphi} \\
	\left(\partial_t^2-\partial_y^2\right)\zeta^\kappa_y(t) &=& \partial_t \left(\tilde{\lambda}(t)q_A^\kappa(t)\right)\delta(y-\vartheta), 
	  \label{eomchi}\\
  \left(\partial_t^2+\Omega_0^2\right)q_A^\kappa(t)&=&-\lambda(t)\partial_t \varphi_0^\kappa(t)-
	\tilde{\lambda}(t)\partial_t\zeta^\kappa_\vartheta(t).  \label{eomq}
\end{eqnarray}
Again they have the same form as the Euler-Lagrange equations,
while the initial conditions will be different from those in the classical theory.
The solutions for $\varphi$ and $\zeta$ are similar to (\ref{Phisol}) and (\ref{Xsol}):
\begin{eqnarray}
  \varphi_x^\kappa(t) &=& \varphi_x^{\kappa^{[0]}}(t) +\frac{1}{2} \lambda(t-|x|)q_A^\kappa (t-|x|), \label{varphisol}\\
  \zeta_y^\kappa(t) &=& \zeta_y^{\kappa^{[0]}}(t) +\frac{1}{2} \tilde{\lambda}(t-|y-\vartheta|)q_A^\kappa (t-|y-\vartheta|). \label{chisol}
\end{eqnarray}
where $\varphi_x^{{\rm k}^{[0]}}(t) = e^{-i{\rm w} t + i{\rm k}x}$, $\zeta_y^{\tilde{\rm k}^{[0]}}(t) = e^{-i \tilde{\rm w} t +
i \tilde{\rm k} y}$, and $\varphi_x^{A^{[0]}}(t) = \varphi_x^{\tilde{\rm k}^{[0]}}(t) = \zeta_y^{A^{[0]}}(t) = \zeta_y^{{\rm k}^{[0]}}(t) =0$.
Thus, similar to (\ref{eomQRR}), Eq.(\ref{eomq}) becomes
\begin{eqnarray}
    &&\left[ \partial_t^2+ \left( \frac{\tilde{\lambda}^2(t)}{2} + \frac{\lambda^2(t)}{2} \right)\partial_t +
		\left( \Omega_0^2 + \frac{\tilde{\lambda}(t)\partial_t \tilde{\lambda}(t)}{2} + 
		\frac{\lambda(t)\partial_t \lambda(t)}{2}\right)\right]q_A^\kappa(t)\nonumber\\ &=& 
		-\lambda(t) \partial_t \varphi_0^{\kappa^{[0]}}(t) -\tilde{\lambda}(t)\partial_t \zeta^{\kappa^{[0]}}_\vartheta(t), \label{eomq2}
\end{eqnarray}
after including the back-reactions of the field and the environment.

\subsection{Mode functions for internal HO}
\label{QTUD}

Since the environmental effect on the system is inevitable even at the stage of experiment preparation
while the details of the environment are uncontrollable in laboratories, we
assume the OE coupling $\tilde{\lambda}(t)$ was switched on in the far past $t = \tilde{t}_0 \ll -\tilde{\gamma}^{-1} < 0$ and
then settled to a constant $\tilde{\lambda}$, and the OF coupling $\lambda(t)$ is not switched on until $t=0$ 
\footnote{For theoretical interests, one may first turn on the OF coupling $\lambda(t)$ in the far past then switch on the OE coupling $\tilde{\lambda}(t)$ at $t=0$, though this is hard to be realized in laboratories. Since the model is linear, the switching functions are regular, and the cutoffs for the OE and the OF couplings are the same, the late-time results in this alternative scenario with the same initial state (\ref{IS1mir}) are expected to be the same as those we obtained in this paper. In transients, anyway, the evolutions of the system will be different in different scenarios.}.
Suppose the combined system started with a factorized state:
\begin{equation}
  |\psi(t\le \tilde{t}_0)\rangle = |0\rangle^{}_Q \otimes |0\rangle^{}_{\cal Z} \otimes |0\rangle^{}_\Phi, \label{IS1mir}
\end{equation}
which is a product of the ground state of the free internal HO $|0\rangle^{}_Q$, the vacuum state of the free environment 
$|0\rangle^{}_{\cal Z}$, and the vacuum state of the free field $|0\rangle^{}_\Phi$. 
Then, right before $t=0$, the quantum state of the combined system has become
\begin{equation}
  \rho(t\to 0-) = \rho^{}_{Q{\cal Z}} \otimes \rho^{}_{\Phi}, \label{rho0m}
\end{equation}
where $\rho^{}_{Q{\cal Z}}$ is the late-time state of the HO-environment subsystem and $\rho^{}_\Phi = |0\rangle^{}_\Phi\langle 0|$ is still
the vacuum state of the field. Between $t=\tilde{t}_0$ and $t=0$, $q_A^A(t)$ follows the equation of motion
\begin{equation}
    \left[ \partial_t^2+ 2\tilde{\gamma}\partial_t + \Omega_0^2 \right]q_A^A(t) = 0 \label{EOMqAA0m}
\end{equation}
and behaves like a damped harmonic oscillator, while $q_A^{\tilde{\rm k}}(t)$ follows the equation
\begin{equation}
    \left[ \partial_t^2+ 2\tilde{\gamma}\partial_t + \Omega_0^2\right]q_A^{\tilde{\rm k}}(t) = 
	 -\tilde{\lambda}\partial_t \zeta^{\tilde{\rm k}^{[0]}}_\vartheta(t). \label{EOMqAp0m}
\end{equation}
Thus, as $\tilde{t}_0\to -\infty$, $\rho^{}_{Q{\cal Z}}$ in (\ref{rho0m}) is characterized by the two-point correlators with 
the late-time solutions, $q_A^A(0)= \partial_t q_A^A(0)=0$ for (\ref{EOMqAA0m}), and
\begin{equation}
  \left. q_A^{\tilde{\rm k}}(t)\right|_{t \to 0-}=\left.\frac{ \tilde{\lambda} i\tilde{\rm w} e^{-i\tilde{\rm w}t 
	+i\tilde{\rm k}\vartheta}}{\Omega_0^2 - \tilde{\rm w}^2 -2 i \tilde{\rm w}\tilde{\gamma}}\right|_{t \to 0-} \label{qaptm}
\end{equation}
for (\ref{EOMqAp0m}), which implies $\partial_t q_A^{\tilde{\rm k}}(0) = -i\tilde{\rm w} q_A^{\tilde{\rm k}}(0)$.

Suppose the OF coupling is suddenly switched on at $t=0$ like $\lambda(t) = \lambda \theta(t)$. 
Integrating (\ref{eomq2}) from $t=-\epsilon$ to $\epsilon$ for $\epsilon \to 0+$, one has $\dot{q}^\kappa_A(\epsilon)-
\dot{q}^\kappa_A(-\epsilon)+(\lambda^2/4)q^\kappa_A(0)= 0$ provided that $q^\kappa_A(t)$ and $\varphi^{^{[0]}\kappa}_0(t)$ are continuous.
Then introducing the conditions $q_A^A(0)= \partial_t q_A^A(0)=0$, $q_A^{\rm k}(0)= \partial_t q_A^{\rm k}(0)=0$, and those from 
(\ref{qaptm}) for $q_A^{\tilde{\rm k}}$ and $\partial_t q_A^{\tilde{\rm k}}$ around $t=0$, 
the solutions of (\ref{eomq2}) for $t>0$ are found to be 
\begin{eqnarray}
  q_A^{\rm k}(t) &=& -\lambda\int_{t^{}_0}^t d\tilde{\tau} K(t-\tilde{\tau}) \dot{\varphi}_0^{{\rm k}^{[0]}}(\tilde{\tau}) \nonumber\\
	  &=& \frac{\lambda i{\rm w}}{2\Gamma}\left[ 
	  \frac{e^{-i{\rm w} t}-e^{-(\gamma+\tilde{\gamma}-\Gamma)\eta-i{\rm w}t^{}_0}}{\gamma+\tilde{\gamma}-\Gamma-i{\rm w}} 
	 -\frac{e^{-i{\rm w} t}-e^{-(\gamma+\tilde{\gamma}+\Gamma)\eta-i{\rm w}t^{}_0}}{\gamma+\tilde{\gamma}+\Gamma-i{\rm w}}
		\right], \label{qAksol}\\
	q_A^{\tilde{\rm k}}(t) &=& -\tilde{\lambda}\int_{t^{}_0}^t d\tilde{\tau} K(t-\tilde{\tau})\dot{\zeta}_\vartheta^{\tilde{\rm k}^{[0]}}
	(\tilde{\tau})\nonumber\\& & +\,\frac{\tilde{\lambda} i\tilde{\rm w} e^{-(\gamma+\tilde{\gamma})\eta+i\tilde{\rm k}\vartheta}}
		   {2\Gamma(\Omega_0^2-\tilde{\rm w}^2-2i\tilde{\gamma}\tilde{\rm w})}\left[ 
		\left(\tilde{\gamma}-i\tilde{\rm w}+\Gamma\right)e^{\Gamma \eta}-
		\left(\tilde{\gamma}-i\tilde{\rm w}-\Gamma\right)e^{-\Gamma \eta}\right], \label{qApsol}
\end{eqnarray}
and $q_A^A(t)=0$. Here $\eta\equiv t-t^{}_0 \ge 0$ with $t_0=0$, and the propagator $K(s)$ has been given in (\ref{Kpropa}).
The integral in the first line of (\ref{qApsol}) can be worked out to get an expression similar to the second line of (\ref{qAksol}).

In our numerical calculation, we replace $\theta(t)$ in $\lambda(t)$ by a $C^1$ function
\begin{equation}
  \theta^{}_T(t) = \left\{  \begin{array}{lcc}
	0 & & t \le 0 \\ 
	\left[ 1 -\cos (\pi t/T)\right]/2 & \;{\rm for}\; & 0 < t < T\\
	1 & & t \ge T
\end{array}\right. \label{thetaT}
\end{equation}
to regularize the delta function $\delta(t) = \partial_t \theta(t)$. Then we find $q_A^\kappa(t)$ are always continuous, and our numerical results do approach to (\ref{qAksol}) and (\ref{qApsol}) in the small $T$ limit.
Note that our $\theta^{}_T(t)$ is not smooth or normalizable ($\int_{-\infty}^\infty\theta^{}_T(t)$ diverge), and thus our results are not 
restricted by the quantum inequalities for smooth and normalizable switching functions \cite{Fo91, FR95, Fl97}.

\subsection{Detector energy and HO-field entanglement}
\label{OFEnt}

With the operator expansion (\ref{QAexpan}) and the initial state (\ref{IS1mir}), the symmetric two-point correlators of the internal 
oscillator of the detector read
\begin{eqnarray}
  \langle \hat{Q}_A^2(t) \rangle &=&\lim_{(t',t'_0)\to (t,t_0)}{\rm Re}\left[ \frac{\hbar}{2\Omega_0}q_A^A(t)q_A^{A*}(t') \right.\nonumber\\ 
	&&\left. + \int \frac{d{\rm k}}{2\pi} \frac{\hbar}{2{\rm w}} q_A^{\rm k}(t)q_A^{\rm k*}(t')+ 
	\int \frac{d\tilde{\rm k}}{2\pi} \frac{\hbar}{2\tilde{\rm w}} q_A^{\tilde{\rm k}}(t) q_A^{\tilde{\rm k}*}(t') \right],\label{QA2formal}\\
	\langle \hat{P}_A^2(t) \rangle &=&\lim_{(t',t'_0)\to (t,t_0)}{\rm Re}\left[\frac{\hbar}{2\Omega_0} \dot{q}_A^A(t)\dot{q}_A^{A*}(t')\right.
	\nonumber\\ && \left.+ \int \frac{d{\rm k}}{2\pi} \frac{\hbar}{2{\rm w}}\dot{q}_A^{\rm k}(t)\dot{q}_A^{\rm k*}(t')+ 
	\int \frac{d\tilde{\rm k}}{2\pi} \frac{\hbar}{2\tilde{\rm w}} \dot{q}_A^{\tilde{\rm k}}(t) \dot{q}_A^{\tilde{\rm k}*}(t') \right],
	\label{PA2formal}
\end{eqnarray}
and $\langle \hat{Q}_A (t), \hat{P}_A(t)\rangle \equiv \langle (\hat{Q}_A(t)\hat{P}_A(t)+\hat{P}_A(t)\hat{Q}_A(t))\rangle/2 = 
\partial_t \langle \hat{Q}(t)^2\rangle/2$.
For $t>t_0$, $q_A^A = 0$ and so only the integrals in the above expressions contribute. The closed form of these integrals can be obtained 
straightforwardly after the mode functions are inserted.
For example, by inserting (\ref{qAksol}) we get
\begin{eqnarray}
  && \lim_{(t',t'_0)\to (t,t_0)} \int \frac{d{\rm k}}{2\pi} \frac{\hbar}{2\rm w} q_A^{\rm k}(t)q_A^{\rm k*}(t') 
	= \nonumber\\ 
	&& \frac{\hbar\gamma}{2\pi\Gamma^2} \left\{ 
	  \frac{\Gamma}{\gamma^{}_2}\left[\left(1+e^{-2\gamma^{}_2 \eta}\right) \ln \frac{\gamma^{}_2+\Gamma}{\gamma^{}_2-\Gamma} 
	  +{\rm Ei}[-(\gamma^{}_2-\Gamma)\eta] - {\rm Ei}[-(\gamma^{}_2+\Gamma)\eta]\right] \right. \nonumber\\
  &&\hspace{.5cm} + e^{-2 \gamma^{}_2 \eta}\left[ 
	  \left(e^{2\Gamma \eta} -1+\frac{\Gamma}{\gamma^{}_2} \right){\rm Ei} [(\gamma^{}_2-\Gamma)\eta]
		+\left(e^{-2\Gamma \eta} -1-\frac{\Gamma}{\gamma^{}_2} \right){\rm Ei} [(\gamma^{}_2+\Gamma)\eta]\right.\nonumber\\
	&& \hspace{1.5cm}\left.\left.  + 4 \Lambda_0 \sinh^2\Gamma \eta -e^{2\Gamma \eta}\ln\frac{\gamma^{}_2-\Gamma}{\Omega_0}  
	  -e^{-2\Gamma \eta}\ln\frac{\gamma^{}_2+\Gamma}{\Omega_0}  \right] \right\} \label{IntqAk2}
\end{eqnarray}
for real $\Gamma$ in the over-damping cases. Here  Ei$(s)$ is the exponential integral function, 
$\gamma^{}_2\equiv\gamma +\tilde{\gamma}$, and $\Lambda_0\equiv -\gamma^{}_e
-\ln\Omega_0|t'_0-t_0|$ with the Euler's constant $\gamma^{}_e$. At late times ($\eta=t-t_0 \gg 1/(\gamma+\tilde{\gamma}-\Gamma)$), 
(\ref{IntqAk2}) becomes
\begin{equation}
  \lim_{(t',t'_0)\to (t,t_0)} \int \frac{d\rm k}{2\pi} \frac{\hbar}{2\rm w} q_A^{\rm k}(t)q_A^{\rm k*}(t') 
	\to \frac{\hbar\gamma}{2\pi\Gamma\gamma^{}_2} \ln \frac{\gamma^{}_2+\Gamma}{\gamma^{}_2-\Gamma} . \label{IntqAk2LT}
\end{equation}

If the environment is excluded in our consideration, 
(\ref{IntqAk2}) will be identical to the v-part of the detector correlator 
$\langle \hat{Q}^2(t)\rangle_{\rm v}$ defined in refs. \cite{LH07, LCH16}, where their closed forms in the under-damping regime have been 
given. Indeed, (\ref{IntqAk2}) with $\tilde{\gamma}=0$ can be obtained from Eq.(A9) in Ref. \cite{LH07} with Re $f$ there written 
as Re $[f+f^*]/2$, then replacing the renormalized natural frequency $\Omega_r$ there for the minimal-coupling Unruh-DeWitt HO detector theory in (3+1)D Minkowski space by $\Omega_0$ here for the derivative-coupling detector model in (1+1)D
(also see the Appendix of Ref. \cite{LCH16}), 
and finally replacing every $i\Omega$ there by $\Gamma$ here while noticing that ${\rm Re}\{\Gamma(0,s)\} = -{\rm Re}\{{\rm Ei}(-s)\}$ with the incomplete gamma function $\Gamma(0,s)$.
Note that in this paper we have changed the definitions of $\Lambda_0$ and $\Lambda_1$, corresponding to the UV cutoffs, from 
$-\gamma^{}_e-\ln\Omega |\Delta\tau|$ with $\Delta\tau\to 0$ in our earlier works to $-\gamma^{}_e-\ln\Omega_0|\Delta\tau|$ 
here since the latter is more convenient in the over-damping regime (one cannot simply replace $\Omega$ in the former by $-i\Gamma$, 
which leads to complex values of $\Lambda_0$ and $\Lambda_1$). From now we will use these new definitions for $\Lambda_0$ 
and $\Lambda_1$ even in the under- and critical-damping cases. Associated with this change, the $\ln [(\gamma/\Omega)\pm i] = 
\ln[(\gamma\pm i\Omega)/\Omega]$ terms in (A9)-(A12) of Ref. \cite{LH07} should be replaced by $\ln[(\gamma\pm i\Omega)/\Omega_0]$ here. 

The closed form of the integral $\int \frac{d\rm k}{2\pi}\frac{\hbar}{2\tilde{\rm w}} q_A^{\tilde{\rm k}}(t)q_A^{\tilde{\rm k}*}(t')$ is 
much more lengthy than (\ref{IntqAk2}) due to the second line of (\ref{qApsol}). Fortunately all these extra terms decay out at late times, 
and the late-time result of the integral with $q_A^{\tilde{\rm k}}$ in the over-damping regime is simply (\ref{IntqAk2LT}) with the overall 
factor $\hbar\gamma$ replaced by $\hbar\tilde{\gamma}$. Summing these two integrals together we find
\begin{equation}
  \langle \hat{Q}_A^2 \rangle \to {\rm Re}\,\frac{\hbar}{2\pi\Gamma} \ln \frac{\gamma^{}_2+\Gamma}{\gamma^{}_2-\Gamma} \label{Q2LT}
\end{equation}
at late times, which also applies to the under- and critical-damping cases for $\Gamma = i\Omega$ and $\Gamma\to 0$, respectively.
In the latter case, $\langle \hat{Q}_A^2 \rangle \to \hbar/(\pi\gamma_2)$ at late times.

The late-time result for the correlators can also be obtained by inserting the late-time mode functions  
\begin{equation}
  q_A^{\rm k}(t) \to - \lambda \chi^{}_{\rm w} e^{-i{\rm w}t}, \hspace{1cm}
	q_A^{\tilde{\rm k}}(t) \to - \tilde{\lambda} \chi^{}_{\tilde{\rm w}} e^{-i\tilde{\rm w}t+i \tilde{\rm k} \vartheta}, \label{qAkpLT}
\end{equation}
with the susceptibility function $\chi^{}_\omega$ given in (\ref{chiw1}), into the integrals in (\ref{QA2formal}) and (\ref{PA2formal}). 
Then we get $\langle \hat{Q}_A^2 \rangle$ in (\ref{Q2LT}), $\langle\hat{Q}_A, \hat{P}_A\rangle =\partial_t \langle \hat{Q}^2_A\rangle/2 
\to 0$, and
\begin{equation}
	\langle \hat{P}_A^2 \rangle \to \frac{\hbar}{2\pi} {\rm Re} \left[ 4 \gamma^{}_2 \Lambda_1 -
	\left(2\Gamma+\frac{\Omega_0^2}{\Gamma}\right)\ln \frac{\gamma^{}_2+\Gamma}{\gamma^{}_2-\Gamma}\right] \label{P2LT}
\end{equation}
at late times. 

Note that $\Lambda_1$ has to be large enough to make $\langle \hat{P}_A^2 \rangle$ positive and the uncertainty relation ${\cal U} \ge \hbar/2$ valid, where ${\cal U}\equiv [\langle \hat{Q}_A^2 \rangle\langle \hat{P}_A^2 \rangle -\langle \hat{Q}_A,\hat{P}_A\rangle^2]^{1/2}$ is the uncertainty function \cite{LH07}. This is not pathological, anyway.
Recall that $\Lambda_1 \equiv -\ln\Omega_0 |\Delta\tau| - \gamma_e$ is defined in the coincidence limit $\Delta\tau \to 0$. For a lower UV cutoff $\Lambda_1$, the time resolution for the internal oscillator of the detector is poorer. If $\Lambda_1$ or $\omega_M\equiv \Omega_0 e^{\Lambda_1}$ is too small, the correlators of the oscillators will actually represent the nonlocal correlations of dynamical variables at different proper times (e.g. $\langle \hat{Q}_A(\tau),\hat{Q}_A(\tau + \Delta \tau)\rangle$) with a large time-difference $\Delta\tau \sim 2\pi/\omega_M$. In this case quantum anti-correlation of vacuum fluctuations will enter and reduce the values of the correlators and the uncertainty function. This leads to violation of the uncertainty relation while the uncertainty function ${\cal U}$ has lost its equal-time sense.

From the detector sector of (\ref{Eden}), the expectation value of the energy of the internal oscillator of the UD$'$ detector is 
\begin{equation}
  E^{}_A =\frac{1}{2} \left(\langle \hat{P}_A^2 \rangle + \Omega_0^2 \langle \hat{Q}_A^2 \rangle \right) 
	\to \frac{\hbar}{2\pi} \left[ 4 \gamma^{}_2 \Lambda_1 -	2\Gamma\ln \frac{\gamma^{}_2+\Gamma}{\gamma^{}_2-\Gamma}\right]
\end{equation}
at late times from (\ref{Q2LT}) and (\ref{P2LT}).
It also depends on $\Lambda_1$ and will be positive if $\Lambda_1$ is sufficiently large.

The HO-field entanglement will be strong if the direct coupling $\gamma$ between them is strong. In this case the linear entropy $S_L=1/
(2{\cal U})$, where ${\cal U}\equiv\sqrt{\langle\hat{Q}_A^2\rangle\langle\hat{P}_A^2\rangle -\langle\hat{Q}_A,\hat{P}_A\rangle^2}$, would be 
very close to 1 since $\langle\hat{P}_A^2\rangle$ can be very large in the strong OF coupling limit with a sufficiently large $\Lambda_1$.

\subsection{Reduction of late-time field correlations}

A perfect mirror placed at $x=0$ forces a Dirichlet boundary condition $\Phi^{}_{x=0}(t)=0$ at its position. This would cut the equal-time correlations of the field amplitudes on different sides of the mirror, namely, $\langle \hat{\Phi}^{}_{x}(t)\hat{\Phi}^{}_{x'}(t)\rangle =0$ 
for $x x' <0$. Our detector mirror is not perfect, but it still can reduce the correlations of the field on different sides.

From (\ref{Phiexpan}) and (\ref{IS1mir}), the two-point correlators of the field are given by 
\begin{equation}
  \langle \hat{\Phi}_x(t)\hat{\Phi}_{x'}(t')\rangle = 
	    \frac{\hbar}{2\Omega_0} \varphi_x^A(t)\varphi_{x'}^{A*}(t') + \int \frac{d\rm k}{2\pi} \frac{\hbar}{2\rm w} 
	    \varphi_x^{\rm k}(t)\varphi_{x'}^{k*}(t') + \int \frac{d\tilde{\rm k}}{2\pi} \frac{\hbar}{2\tilde{\rm w}}
			\varphi_x^{\tilde{\rm k}}(t)\varphi_{x'}^{\tilde{\rm k}*}(t') .
\label{FFformal}
\end{equation}
At late times, in the presence of the detector mirror at $x=0$, one has $\varphi_x^A = 0$ and
\begin{eqnarray}
  \varphi_x^{\rm k} (t) &\to& e^{-i{\rm w} t +i {\rm k} x}-
	  2\gamma e^{-i {\rm w} (t-|x|)} \chi^{}_{\rm w}, \\
	\varphi_x^{\tilde{\rm k}}(t) &\to& -2\sqrt{\gamma\tilde{\gamma}}e^{i\tilde{\rm k}\vartheta -i\tilde{\rm w}(t-|x|)} 
	\chi^{}_{\tilde{\rm w}},
\end{eqnarray}
from (\ref{varphisol}), (\ref{chisol}), and (\ref{qAkpLT}).
Inserting these mode functions into (\ref{FFformal}), one obtains a sum of two integrals of dummy variables ${\rm k}$ and $\tilde{\rm k}$.
One can rename both ${\rm k}$ and $\tilde{\rm k}$ to $k$ to get
\begin{eqnarray}
  & &\langle\hat{\Phi}_x(t) \hat{\Phi}_{x'}(t')\rangle \to \int_{-\infty}^\infty \frac{dk}{2\pi}\frac{\hbar}{2\omega}
	\left\{e^{ik(x-x')-i\omega(t-t')} - \gamma e^{-i\omega(t-t')} \times \right.\nonumber\\
 	& &\left. \left[ 
	    \left(2 e^{i(\omega|x|-kx')}-e^{i\omega(|x|-|x'|)}\right)\chi^{}_{\omega} + 
	    \left(2 e^{-i(\omega|x'|-kx)}-e^{i\omega(|x|-|x'|)}\right)\chi^*_{\omega} \right] \right\} \label{FFkint}
\end{eqnarray}
with $\omega = |k| >0$, by applying the identity straightforwardly from (\ref{chiw1}),
\begin{equation}
   \chi^{}_\omega + \chi^*_\omega = 4(\tilde{\gamma}+\gamma)|\chi^{}_{\omega}|^2 , \label{FDR}
\end{equation}
which has the form of the fluctuation-dissipation relation. 
The first term in the integrand of (\ref{FFkint}) gives the correlator of 
the free field; thus, the late-time renormalized two-point correlator of the field reads
\begin{eqnarray}
 & & \langle\hat{\Phi}_x(t) \hat{\Phi}_{x'}(t')\rangle_{\rm ren}\equiv \langle\hat{\Phi}_x(t) \hat{\Phi}_{x'}(t')\rangle -
    \langle\hat{\Phi}^{[0]}_x(t) \hat{\Phi}^{[0]}_{x'}(t')\rangle \nonumber\\
 &\to& -\frac{\hbar \gamma}{2\pi} \int_0^\infty \frac{d\omega}{\omega} e^{-i\omega(t-t')} \left[ e^{i\omega(|x|+|x'|)}\chi^{}_{\omega} + 
	      e^{-i\omega(|x|+|x'|)}\chi^*_{\omega} \right] \label{FFrenInt}
\end{eqnarray}
after we split $\int_{-\infty}^\infty dk(\cdots)$ into $\int_{-\infty}^0 dk (\cdots) + \int_0^\infty dk(\cdots)$ and then 
express both terms in $\int_0^\infty d\omega (\cdots)$. The above integral can be done analytically, which yields
\begin{eqnarray} 
& &\langle\hat{\Phi}_x(t) \hat{\Phi}_{x'}(t')\rangle_{\rm ren} \nonumber\\
 &\to & \frac{\hbar\gamma}{4\pi\Gamma} \left\{ 
    e^{\gamma^{}_-\Delta^{}_-}{\rm Ei}\left(-\gamma^{}_-\Delta^{}_-\right) -
		e^{\gamma^{}_+\Delta^{}_-}{\rm Ei}\left(-\gamma^{}_+\Delta^{}_-\right) \right. \nonumber\\
& & \hspace{.7cm} +\, e^{\gamma^{}_-\Delta^{}_+}{\rm Ei}\left(-\gamma^{}_-\Delta^{}_+\right) -
		e^{\gamma^{}_+\Delta^{}_+}{\rm Ei}\left(-\gamma^{}_+\Delta^{}_+\right) \nonumber\\
& & \hspace{.7cm}+\left. i\pi \left[ \theta(-\Delta^{}_-)\left(e^{\gamma^{}_-\Delta^{}_-} - e^{\gamma^{}_+\Delta^{}_-}\right) -
    \theta(-\Delta^{}_+)\left(e^{\gamma^{}_-\Delta^{}_+} -e^{\gamma^{}_+\Delta^{}_+}\right) \right]\right\} \label{FFrenEi}
\end{eqnarray}
with $\Delta^{}_\pm \equiv (|x|+|x'|)\pm (t-t')$, $\Gamma$ defined below (\ref{Qsol}), and $\gamma^{}_\pm\equiv \gamma + 
\tilde{\gamma}\pm \Gamma >0$. 

\begin{figure}
\includegraphics[width=5.5cm]{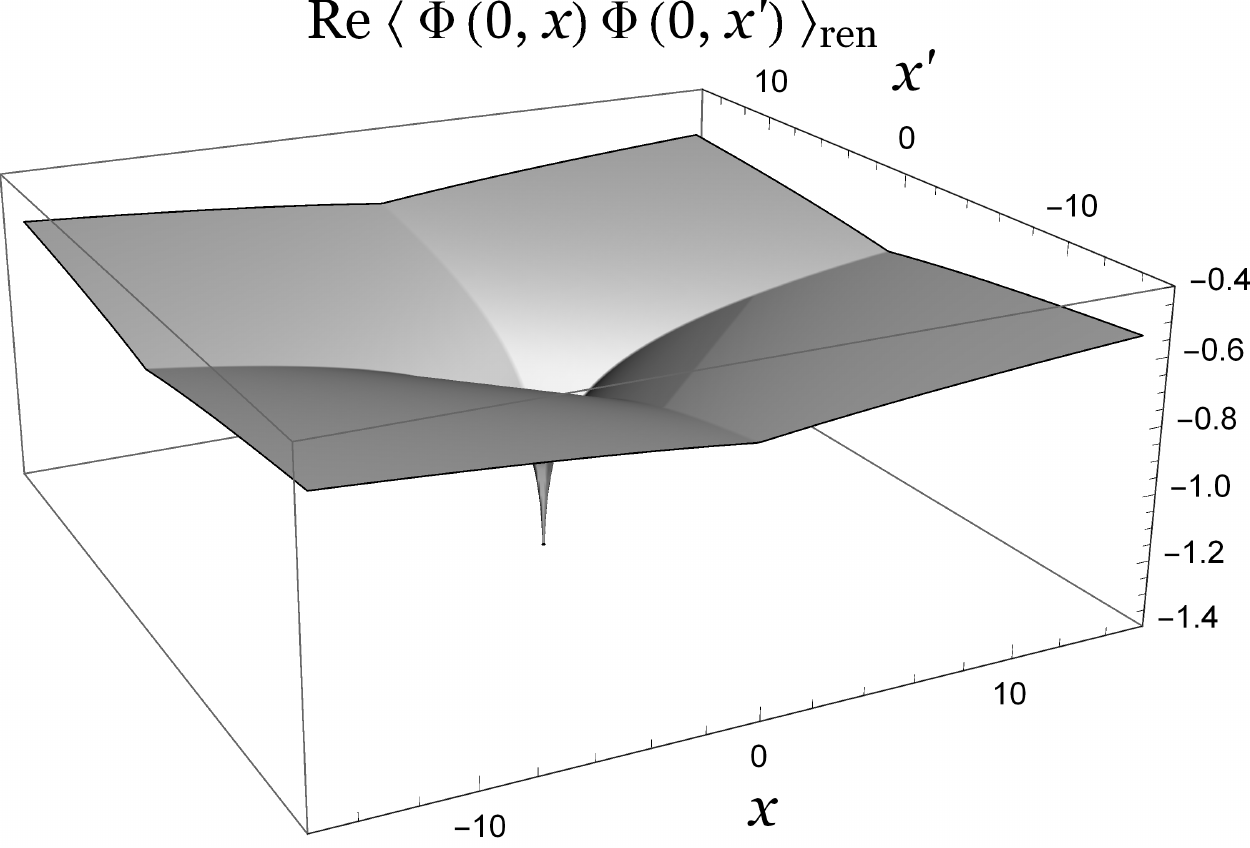}
\includegraphics[width=5.5cm]{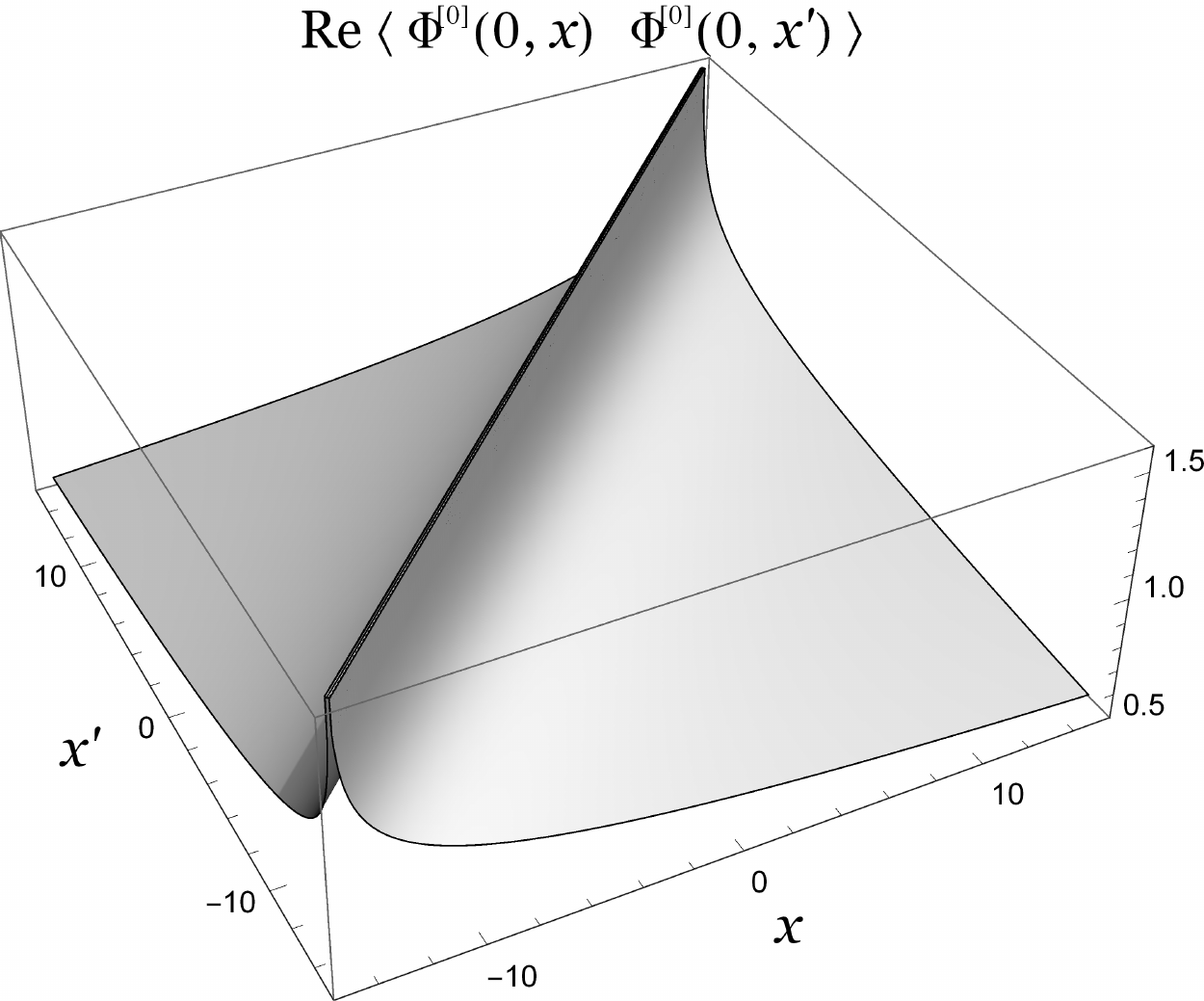}
\includegraphics[width=5.5cm]{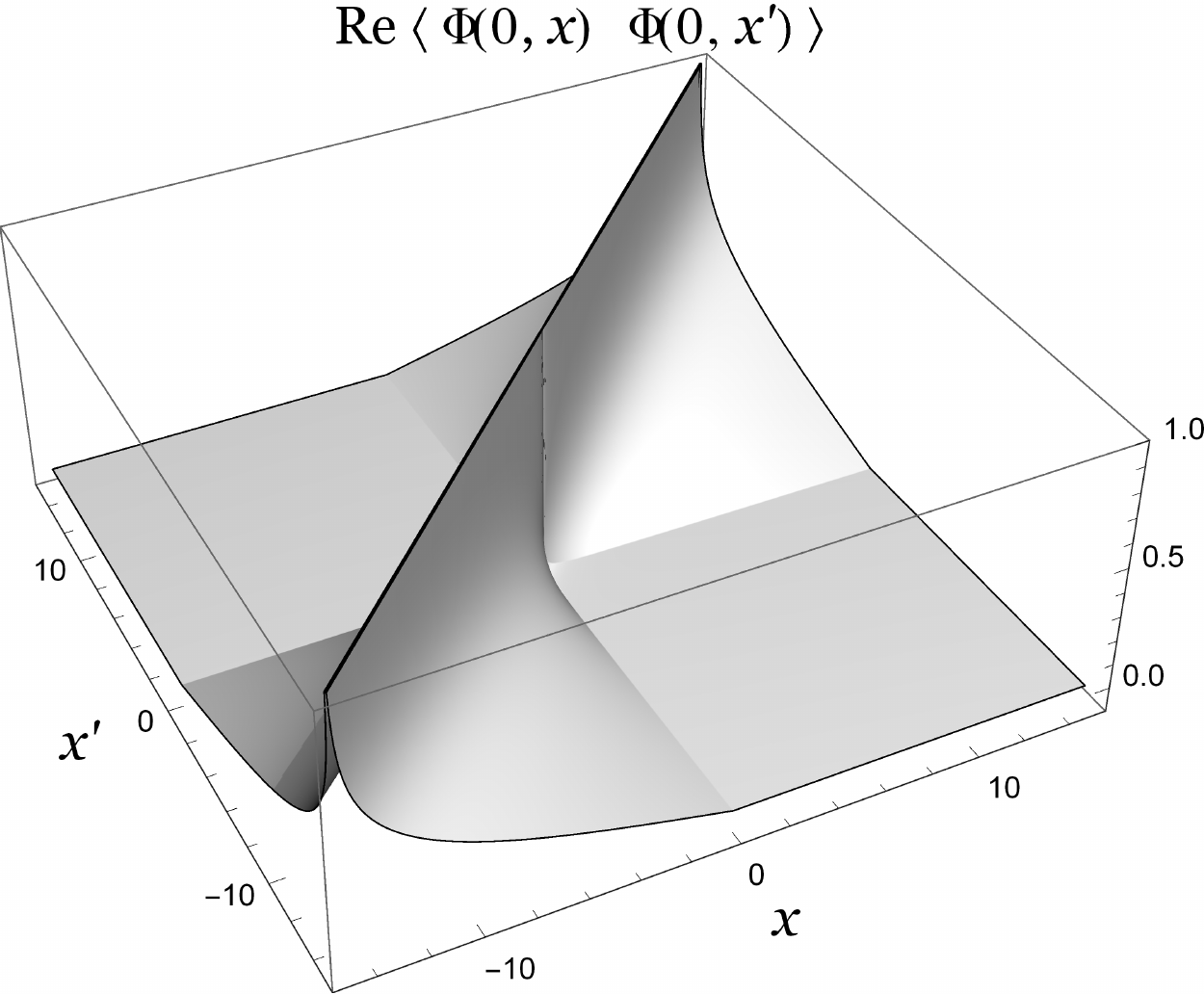}\\
\includegraphics[width=5.5cm]{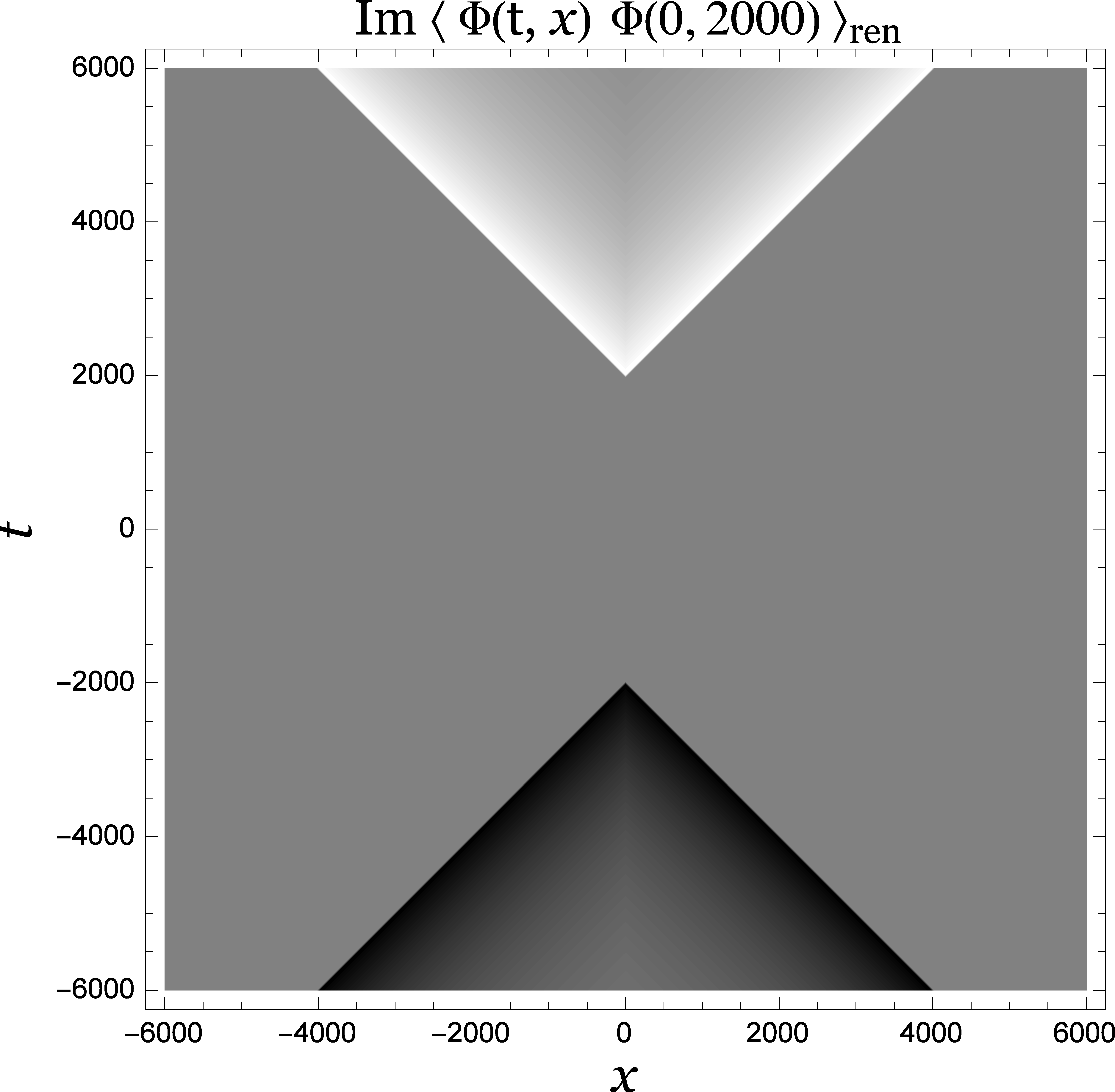}
\includegraphics[width=5.5cm]{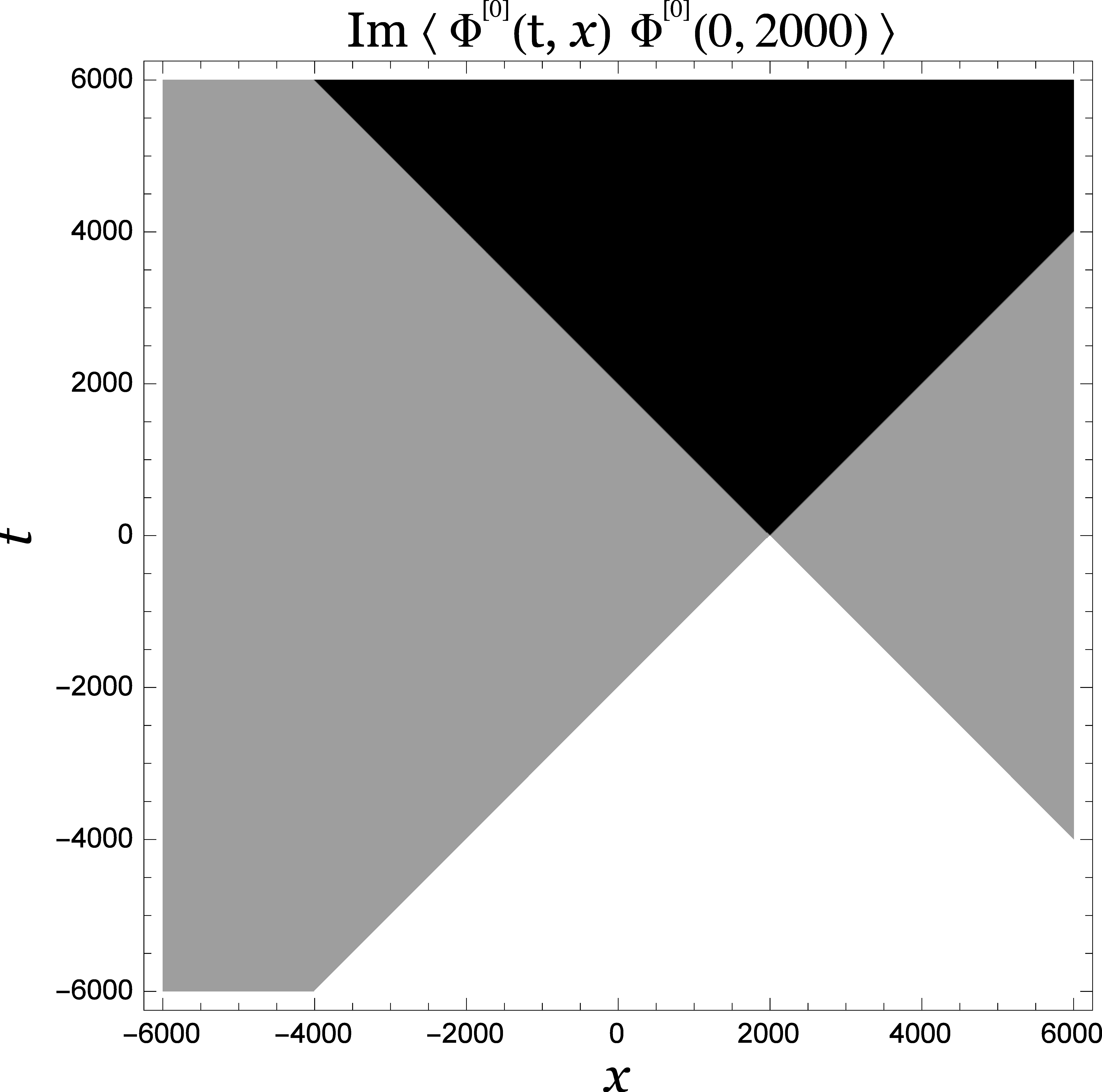}
\includegraphics[width=5.5cm]{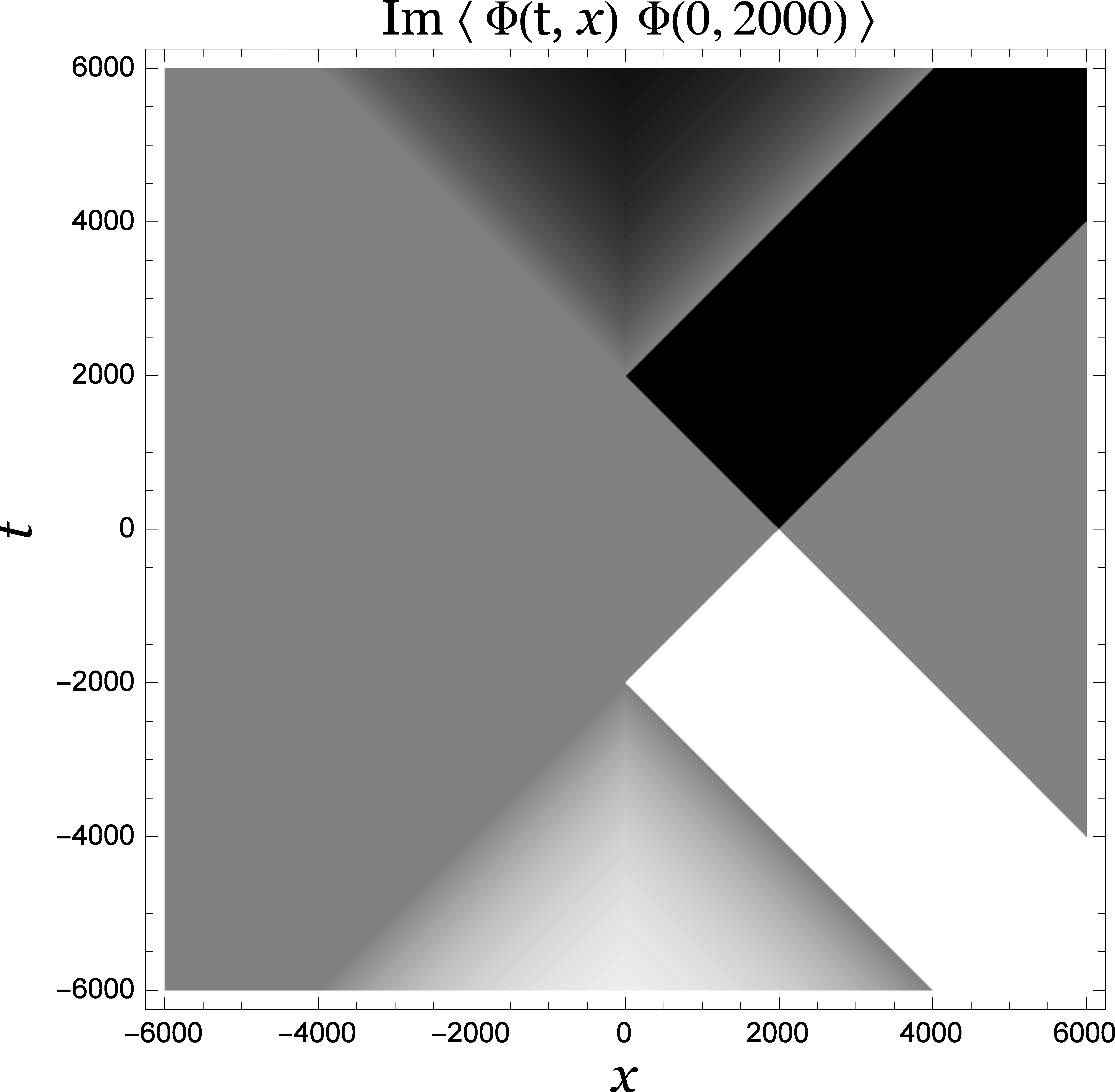}
\caption{The real parts (upper row) and imaginary parts (lower row) of the late-time renormalized correlator of the field in the presence of 
the detector mirror (left plots, Eq. (\ref{FFrenEi})), the correlator of the free field (middle, (\ref{F0F0})), and the full correlator 
(right, the sum of (\ref{FFrenEi}) and (\ref{F0F0})). Since (\ref{FFrenEi}) and (\ref{F0F0}) are stationary, we have shifted $t$ and $t'$ 
from large ($t,t'\gg t_{\rm rlx}$ at late times) to small values for presentation. We choose $t=t'=0$ (equal time) in the upper row, and 
$(t',x')=(0,2000)$ in the lower row, where the gray scale from black to white represents the values from $-1/4$ to $1/4$, and the values 
outside the past and future light cones of $(t',x')$ are exactly zero. Here $\gamma=10$, $\tilde{\gamma}=1$, $\Omega_0=0.1$, and $c=\hbar=1$.}
\label{FFrenRI}
\end{figure}

In the strong OF couplings, over-damping regime, $\gamma \gg \Omega_0$, $\tilde{\gamma}$, one has $\Gamma\approx\gamma$, and 
$\gamma^{}_+ \approx 2\gamma \gg 1\gg \gamma^{}_- \approx \Omega_0^2/(2\gamma)$. For $0 < \frac{\Omega_0^2}{2\gamma}
|\Delta^{}_\pm| \ll 1 \ll 2\gamma |\Delta^{}_\pm|$, the above late-time renormalized field correlator approximately reads
\begin{eqnarray}
 \langle\hat{\Phi}_x(t) \hat{\Phi}_{x'}(t')\rangle_{\rm ren} 
 &\to&  \frac{\hbar}{4\pi} \left( \ln \left|(|x|+|x'|)^2-(t-t')^2\right| + 2\ln \frac{\Omega_0^2}{2\gamma}+2\gamma^{}_e \right)
    \nonumber\\ & & + \frac{i\hbar}{4} \left[ \theta(-\Delta^{}_-) -\theta(-\Delta^{}_+) 
		\right] + O\left( s\ln s, 1/s'\right)\label{FFrenEiStrong}
\end{eqnarray}
with $s \sim \Omega_0^2|\Delta_\pm|/\gamma$ and $s' \sim \gamma|\Delta_\pm|$ (given $e^s {\rm Ei}(-s) \to \ln s + \gamma^{}_e + O(s \ln s)$ 
as $s\to 0$ and $e^{s'}{\rm Ei}(-s') \to -1/s' + O(s'^{-2})$ as $s'\to \infty$). 
On the other hand, the two-point correlator of the free massless scalar field in (1+1)D Minkowski space is given by 
\begin{eqnarray}
    \langle \hat{\Phi}^{[0]}_x(t) \hat{\Phi}^{[0]}_{x'}(t')\rangle &=& - \frac{\hbar}{4\pi} \ln |\sigma| + \hbar C \nonumber\\
		& & - \frac{i\hbar}{4} \left[ \theta(t-t'-(x-x')) + \theta(t-t'+(x-x'))\right], \label{F0F0}
\end{eqnarray}
up to a complex constant $C$. Here $\sigma = -(x_\mu - x'_\mu)(x^\mu - x'^\mu)/2$ is Synge's world function.
Comparing (\ref{F0F0}) with (\ref{FFrenEiStrong}), one can see that the constant $C$ should be chosen as $(2\ln [\Omega_0^2/(2\gamma)] + 
2 \gamma^{}_e )/(4\pi) + (i/4)$ to cancel similar constants in (\ref{FFrenEiStrong}) when $x x'<0$ in the strong OF coupling limit.
With this choice, adding (\ref{FFrenEi}) to (\ref{F0F0}), one finds that the real part of the full equal-time correlation of the 
field amplitudes on different sides of the mirror, Re $\langle \hat{\Phi}_{x}(t)\hat{\Phi}_{x'}(t) \rangle$ with $x x'<0$, will indeed 
be suppressed for small $|x|$ and $|x'|$ at late times (Figure \ref{FFrenRI} (upper-right)). However, when $|x|$ or $|x'|$ gets greater,
the correlation would not be largely corrected since $\langle\hat{\Phi}_x(t)\hat{\Phi}_{x'}(t)\rangle_{\rm ren}$ goes to zero as $|x|+|x'| 
\to \infty$ while $\langle \hat{\Phi}^{[0]}_x(t) \hat{\Phi}^{[0]}_{x'}(t')\rangle$ does not (Figure \ref{FFrenRI} (upper-left) and 
(upper-middle)). 

Actually the real part of the equal-time correlator of the field amplitudes on the same side of the mirror ($x x' >0$) is also reduced 
since the real part of (\ref{FFrenEi}) for $t=t'$ is a negative function of $|x|+|x'|$ only. This may be interpreted as a consequence of the 
image ``point charge" in the Green's function of the field in the presence of the detector mirror.

Regarding to the imaginary part of the field correlator, the renormalized correlator simply adds the effect of the mirror to the retarded 
and advanced Green's functions of the field in free space. In Figure \ref{FFrenRI} (lower right) one can see the reflected and transmitted 
fields generated by the detector mirror at $x=0$. In the presence of the detector mirror, the translational symmetry of the system is broken. 

Anyway, comparing (\ref{FFrenEiStrong}) and (\ref{F0F0}), one can see that for $x$ and $x'$ fixed at finite values with $x x' <0$, which 
implies $(|x|+|x'|)^2= (x-x')^2$, one has the full correlator $\langle \hat{\Phi}^{}_{x}(t) \hat{\Phi}^{}_{x'}(t)\rangle\to 0$ as 
$\gamma\to\infty$ (such that $s\to 0$ and $s'\to \infty$ in (\ref{FFrenEiStrong})). This is exactly the property we mentioned: A perfect 
mirror will suppress the correlations of the field amplitudes on different sides of the mirror.

\subsection{Field spectrum}

\begin{figure}
\includegraphics[width=5.8cm]{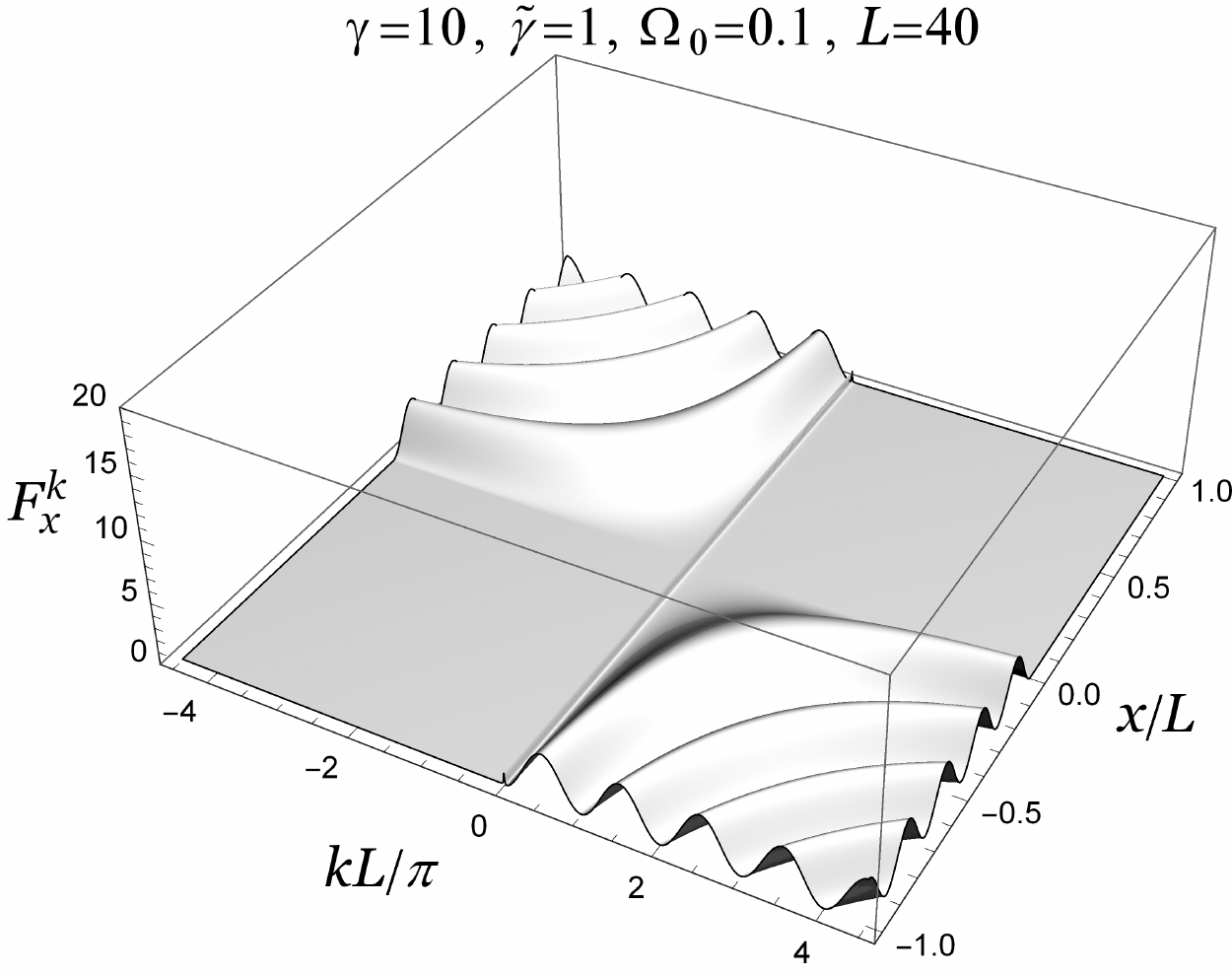}
\includegraphics[width=5.5cm]{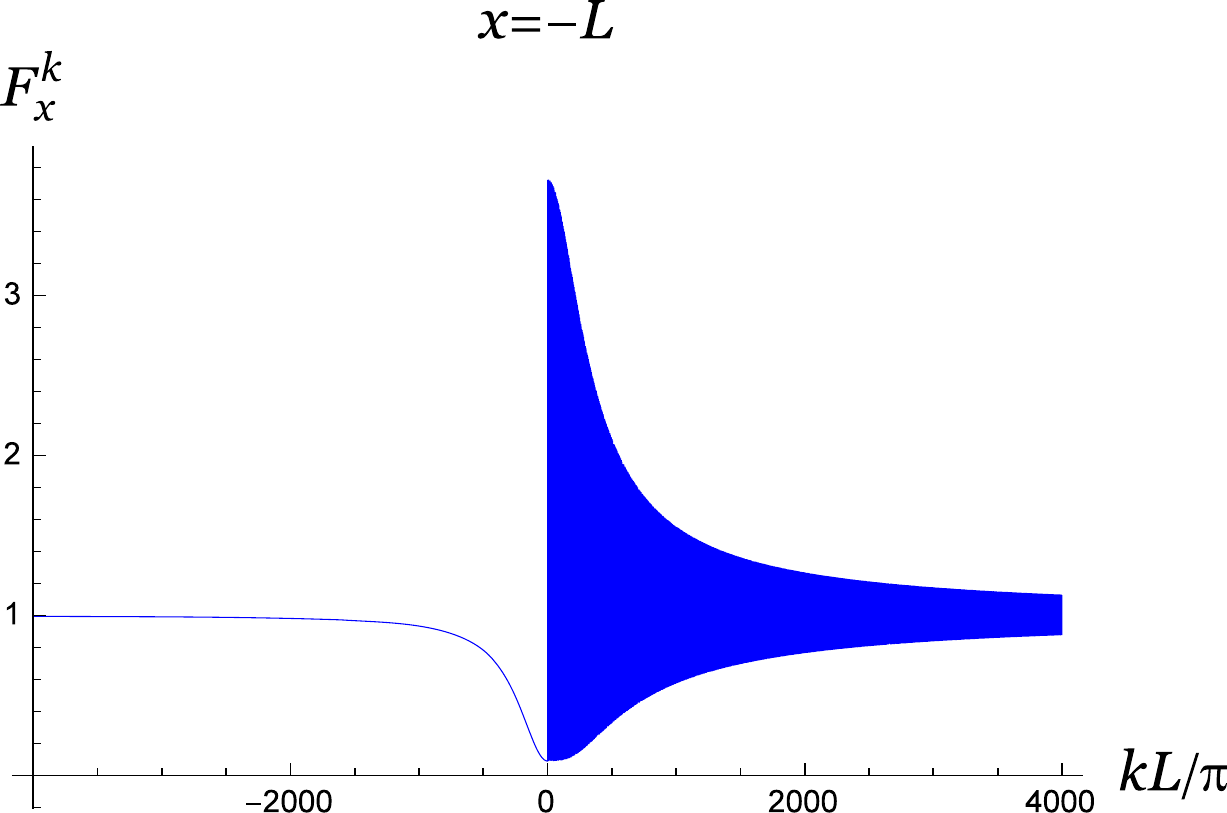}
\includegraphics[width=5.5cm]{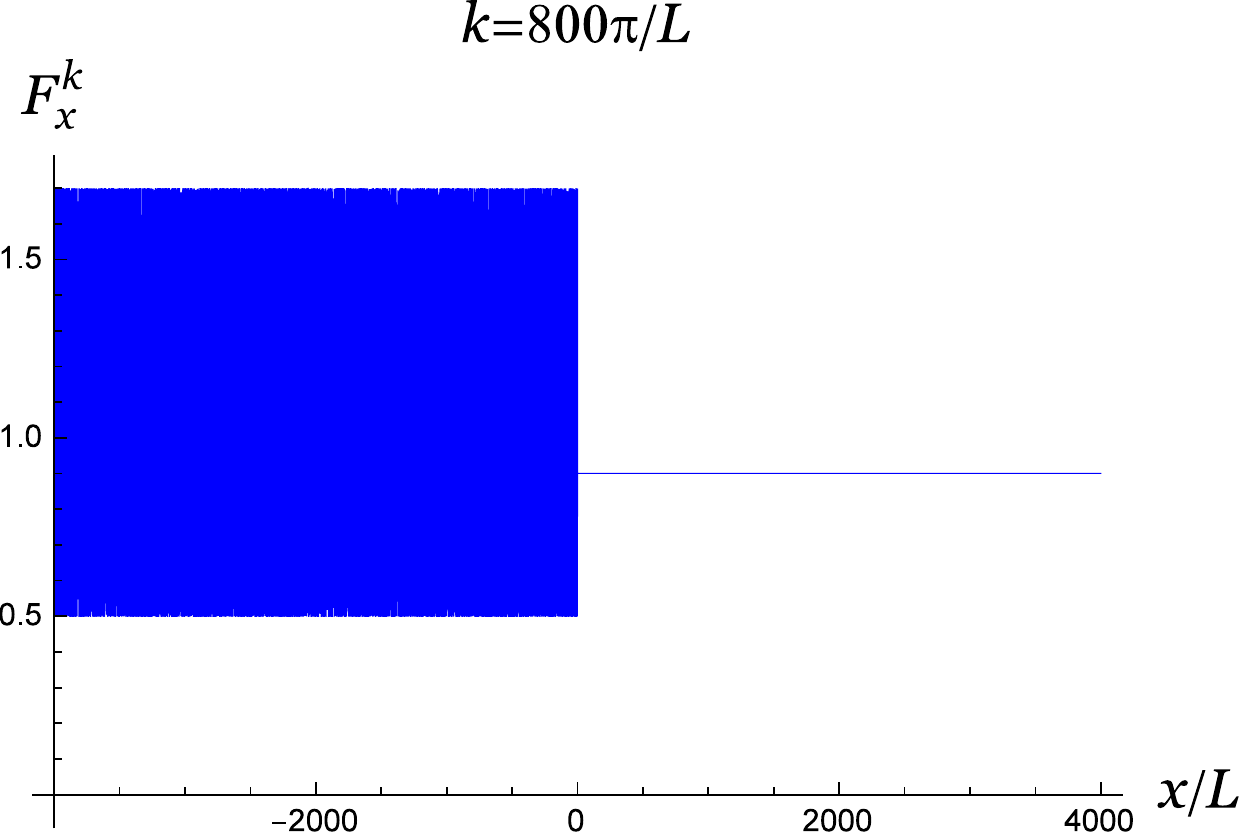}
\caption{The late-time field spectrum $F_x^k$ of a single mirror in Eq. (\ref{FSpec1}) against $k$ and $x$, where $\gamma=10$, 
$\tilde{\gamma}=1$, $\Omega_0 =0.1$ (over-damping), and $c=\hbar=1$. $L=40$ is simply a scaling parameter here for convenience 
of comparison with Figure \ref{Phi2xkOvr}.}
\label{Phi2xk1}
\end{figure}

From (\ref{FFformal}) we define the field spectrum $F_x^k$ by looking at the full correlators of the field in the coincidence limit, 
\begin{equation}
  \langle \hat{\Phi}_{x}(t)^2\rangle = \lim_{t'\to t, x'\to x}\langle \hat{\Phi}_{x}(t),\hat{\Phi}_{x'}(t')\rangle
	\equiv \int_{-\infty}^{\infty} \frac{dk}{2\pi} \frac{\hbar}{2\omega} F_x^k, \label{Flatecoin}
\end{equation}
with $\omega\equiv |k|$ such that
\begin{equation}
	 F_x^k(t) = \left|\varphi_x^{\rm k}(t)\right|^2_{{\rm k}=k} + \left|\varphi_x^{\tilde{\rm k}}(t)\right|_{\tilde{\rm k}=k}^2 + 
	 \int_{-\infty}^{\infty} d\tilde{x} e^{-ik(\tilde{x}-x)} \frac{\omega}{\Omega_0} \varphi_{\tilde{x}}^A(t) \varphi_{x}^{A*}(t)\label{FxKdef}
\end{equation} 
in the presence of our single detector mirror. Note that $k$ is simply a dummy variable in the integral of (\ref{Flatecoin}) and $F_x^k$ is 
not only contributed by the vacuum fluctuations of the field $\Phi$. 
At late times, the last term in (\ref{FxKdef}) decays out and the field spectrum becomes
\begin{equation}
	F_x^k \to 1 -\gamma \left[ \left(2 e^{i(\omega|x|-kx)}-1\right)\chi^{}_{\omega} + 
	    \left(2 e^{-i(\omega|x|-kx)}-1\right)\chi^*_{\omega} \right], \label{FSpec1}
\end{equation}
which is independent of $t$, from (\ref{FFkint}). An example in the over-damping regime is shown in Figure \ref{Phi2xk1}. For $kx<0$, the 
factor $e^{i(\omega|x|-kx)} =e^{-2ikx}$ produces the ripple structure. For $kx>0$, $F_x^k=1-\gamma(\chi_\omega +\chi^*_\omega)$ is 
independent of $x$ (Figure \ref{Phi2xk1} (right), in particular). 
In this case, for $\tilde{\gamma}\ll\gamma$, one has $F_x^k\approx |{\cal T}(k)|^2$, which is the transmittivity defined in (\ref{TT}) with 
${\cal Z}^{^{[0]}}_y(t)=0$. Thus one may interpret that $F_x^k$ for $kx>0$ is small in our example because the low-$|k|$ modes are almost 
totally reflected in the over-damping regime, while the ripple structure of $F_x^k$ for $kx<0$ is due to the interference of the incident 
and the reflected waves. The minimum values in the valleys of the ripple in the low-$|k|$ regime can be very close to zero, which is 
significantly deviated from the value $1$ for the field vacuum in free space. In contrast, the field spectrum at fixed $x$ goes to $1$ as 
$|k|\to\infty$, so the detector mirror is almost transparent to the short-wavelength fluctuations (Figure \ref{Phi2xk1} (middle)).

\subsection{Renormalized energy density of the field}

The expectation value of the energy density of the field is given by
\begin{eqnarray}
  \langle \hat{T}_{00}(t,x)\rangle &=& \frac{1}{2} \left\{ \langle [\partial_t\hat{\Phi}_{x}(t)]^2\rangle +  
	\langle [\partial_x \hat{\Phi}_{x}(t)]^2\rangle \right\} \nonumber\\
	&=& \lim_{(t',x')\to(t,x)} \frac{1}{2} \left( \partial_t \partial_{t'} + \partial_x \partial_{x'}\right)
	    \langle \hat{\Phi}_{x}(t),\hat{\Phi}_{x'}(t')\rangle . \label{T00def}
\end{eqnarray}
While the above expression formally diverges in the coincident limit $(t',x')\to (t,x)$, we are only interested in the renormalized energy 
density of the field with the contribution by the free field subtracted,
\begin{equation}
  \langle \hat{T}_{00}(t,x)\rangle_{\rm ren} = \langle \hat{T}_{00}(t,x)\rangle -\langle \hat{T}_{00}^{^{[0]}}(t,x)\rangle, \label{T00renDef}
\end{equation}
which can be obtained from (\ref{T00def}) with the full correlator of the field $\langle \hat{\Phi}_{x}(t),\hat{\Phi}_{x'}(t')\rangle$ 
replaced by the renormalized one, $\langle\hat{\Phi}_{x}(t),\hat{\Phi}_{x'}(t')\rangle_{\rm ren}$.

When substituted into (\ref{T00def}) and (\ref{T00renDef}), the late-time correlator $\langle\hat{\Phi}_{x}(t),\hat{\Phi}_{x'}
(t')\rangle_{\rm ren}$ in (\ref{FFrenInt}) is always a function of $t-t'$ and $x+x'$ since $x$ and $x'$ must have the same sign in the 
coincidence limit for $x\not= 0$. This implies $\partial_t \partial_{t'}\langle\hat{\Phi}_{x}(t), \hat{\Phi}_{x'}(t')\rangle_{\rm ren} =
 -\partial_x \partial_{x'} \langle\hat{\Phi}_{x}(t),\hat{\Phi}_{x'}(t')\rangle_{\rm ren}$ at late times, and thus $\langle \hat{T}_{00}(t,x)
\rangle_{\rm ren} \to 0$ for $x\not=0$, namely, the late-time energy density of the field outside the detector is the same as the vacuum energy density, though the field spectra are quite different. This is not surprising: It is well known that the late-time stress energy tensor of the field for a uniformly accelerated UD$'$ detector (without coupling to ${\cal Z}$) is exactly zero \cite{HR00}.

Right at the position of the detector $x=0$, if we choose the regularization $|x|= \sqrt{x^2+\epsilon^2}$, $\epsilon= 0+$, then 
$\partial_x |x|$ will vanish at $x=0$ for any finite regulator $\epsilon$ and we will end up with $\langle \hat{T}_{00}(t,x=0)
\rangle_{\rm ren}\to\frac{1}{2}\lim_{t'\to t}\partial_t\partial_{t'}\langle \hat{\Phi}_{0}(t),\hat{\Phi}_{0}(t')\rangle_{ren} = -(\gamma/2) 
\langle \hat{P}_A^2\rangle$ at late times, with the late-time result of $\langle \hat{P}_A^2\rangle$ given in (\ref{P2LT}).

\section{Cavity of detector mirrors}
\label{detCav}

With the knowledge about a detector mirror, we are ready to model a cavity with two detector mirrors coupled to a common scalar 
field in (1+1)D Minkowski space while each detector mirror couples to its own mechanical environment. Our model is described by the action
\begin{eqnarray}
  S &=& -\int dt dx \frac{1}{2}\partial_\mu\Phi_x(t) \partial^\mu\Phi_x(t) \nonumber\\ & &
	    + \sum_{{\bf d}=A,B}\left\{ \frac{1}{2}\int d\tau^{}_{\bf d} \left[ \dot{Q}_{\bf d}^2(\tau^{}_{\bf d})-
			    \Omega_{\bf d}^2 Q_{\bf d}^2(\tau_{\bf d})\right]
	    -\int d\tau^{}_{\bf d} dy^{}_{\bf d} \frac{1}{2}\partial^{}_{\nu^{}_{\bf d}}
			  {\cal Z}_{y^{}_{\bf d}}(\tau^{}_{\bf d}) \partial_{}^{\nu^{}_{\bf d}} {\cal Z}_{y^{}_{\bf d}}(\tau^{}_{\bf d}) \right.\nonumber\\
		&	& \hspace{1cm}-\int d\tau^{}_{\bf d} \int dt dx \lambda^{}_{\bf d}(\tau^{}_{\bf d}) Q^{}_{\bf d}(\tau^{}_{\bf d}) 
		    \frac{d}{d\tau^{}_{\bf d}}\Phi_x(t) \delta(t-z_{\bf d}^0(\tau^{}_{\bf d}))\delta(x-z_{\bf d}^1(\tau^{}_{\bf d}))\nonumber\\ & &
		\hspace{1cm}\left. -\int d\tau^{}_{\bf d} dy^{}_{\bf d} \tilde{\lambda}^{}_{\bf d}(\tau^{}_{\bf d})Q^{}_{\bf d}(\tau^{}_{\bf d}) 
		\frac{d}{d\tau^{}_{\bf d}} {\cal Z}_{y^{}_{\bf d}}(\tau^{}_{\bf d}) \delta(y^{}_{\bf d}-\vartheta^{}_{\bf d}) \right\}.
	\label{Stot2}
\end{eqnarray}
Suppose the two detector mirrors with internal oscillators $Q^{}_A$ and $Q^{}_B$ are at rest in space, and located at $x=0$ and 
$x=L >0$, respectively. In other words, $\tau^{}_A=\tau^{}_B=t$, $z_A^\mu(\tau^{}_A) = (t, 0)$ and $z_B^\mu(\tau^{}_B) = (t, L)$. Let the 
two detector mirrors be identical, $\Omega_A = \Omega_B = \Omega_0$, $\lambda^{}_A(t) = \lambda^{}_B(t) = \lambda(t)$, and 
$\tilde{\lambda}^{}_A(t)=\tilde{\lambda}^{}_B(t)=\tilde{\lambda}(t)$. Generalizing the operator expansions (\ref{QAexpan})-(\ref{Xexpan}) to
$\kappa = A, B, \{{\rm k}\}, \{\tilde{\rm k}_A\}, \{\tilde{\rm k}_B\}$, 
one can write down the equations of motion for the mode functions
\begin{eqnarray}
  \left( \partial_t^2 -\partial_x^2\right) \varphi_x^\kappa(t) &=& \partial_t\left[ \lambda(t) q_A^\kappa(t) \delta(x)+ 
	    \lambda(t) q_B^\kappa(t) \delta(x-L) \right], \label{EOMfxK}\\
  \left( \partial_t^2 -\partial_{y^{}_{\bf d}}^2\right) \zeta_{{\bf d},y^{}_{\bf d}}^\kappa(t) &=&\partial_t\left[\tilde{\lambda}_{\bf d}(t) 
    q_{\bf d}^\kappa(t) \delta(y^{}_{\bf d} - \vartheta^{}_{\bf d}) \right], \label{EOMXyK}\\ 
	\left( \partial_t^2 +\Omega_0^2\right) q_{\bf d}^\kappa(t) &=& -\lambda(t)\partial_t \varphi_{z^1_{\bf d}}^\kappa(t) 
	    - \tilde{\lambda}(t)\partial_t \zeta_{{\bf d},\vartheta^{}_{\bf d}}^\kappa(t) . \label{EOMqdK}
\end{eqnarray}
Similar to the cases of single detectors, inserting the solutions for (\ref{EOMfxK}) and (\ref{EOMXyK}),
\begin{eqnarray}
  \varphi_x^\kappa(t) &=& \varphi_x^{\kappa^{[0]}}(t) +\frac{1}{2} \lambda(t-|x|)q_A^\kappa (t-|x|) +
	  \frac{1}{2} \lambda(t-|x-L|)q_B^\kappa (t-|x-L|), \label{fxKsol}\\
  \zeta_{{\bf d},y_{\bf d}}^\kappa(t) &=& \zeta_{{\bf d},y_{\bf d}}^{\kappa^{[0]}}(t) +\frac{1}{2} 
	\tilde{\lambda}(t-|y^{}_{\bf d}-\vartheta^{}_{\bf d}|)q_{\bf d}^\kappa (t-|y^{}_{\bf d}-\vartheta^{}_{\bf d}|), \label{XyKsol}
\end{eqnarray}
into (\ref{EOMqdK}), one obtains
\begin{eqnarray}
  & & \ddot{q}_{\bf d}^\kappa(t) + 2\left[\gamma(t)+\tilde{\gamma}(t)\right]\dot{q}_{\bf d}^\kappa(t) + 
	\left[\Omega_0^2 +2\dot{\gamma}(t)+2\dot{\tilde{\gamma}}(t)\right]q_{\bf d}^\kappa(t) \nonumber\\ &=& -\frac{\lambda(t)}{2}\partial_t 
	\left[ \lambda(t-L) q_{\bf \bar{d}}^\kappa(t-L) \right] - \lambda(t)\dot{\varphi}_{z^1_{\bf d}}^{[0]\kappa}(t)
	 - \tilde{\lambda}(t)\dot{\zeta}_{{\bf d},\vartheta^{}_{\bf d}}^{[0]\kappa}(t), \label{EOMqdKBR}
\end{eqnarray}
where $\bar{A}\equiv B$ and $\bar{B}\equiv A$.

\subsection{Relaxation and resonance}
\label{relax2}

Suppose the combined system is going through a process similar to the one in Sec. \ref{QTUD}: It is started with the product of the 
ground states of the free internal HOs and the vacuum states of the free field and of the free mechanical environments, and the OE couplings of both detector mirrors have been switched on for a long time ($\tilde{t}_0 \to -\infty$) when their OF couplings are switched on at 
$t=t_0=0$. In (\ref{fxKsol}) and (\ref{EOMqdKBR}), one can see that only half of the retarded field emitted by one detector mirror of the 
cavity in (1+1)D Minkowski space will reach the other detector mirror of the cavity. The other half will go all the way to the null infinity 
and never return. Carried by the retarded field, it seems that all the initial information in the internal HO and the switching function of 
the OF coupling would eventually dissipate into the deep Minkowski space, so that there would be no initial information around $t=0$ kept 
in our cavity at late times. Nevertheless, as we will see below, in the absence of the OE coupling ($\tilde{\gamma}=0$), there 
can exist late-time non-steady states of the combined system which may depend on the initial conditions around $t=0$, if the internal HOs 
of the detector mirrors are resonant with their mutual influences via the field.

Let $q_\pm^\kappa = (q_A^\kappa \pm q_B^\kappa)/\sqrt{2}$. Then (\ref{EOMqdKBR}) can be rewritten as
\begin{eqnarray}
 \ddot{q}_{\pm}^\kappa(t) + 2\left[\gamma(t)+\tilde{\gamma}(t)\right]\dot{q}_{\pm}^\kappa(t) + 
	\left[\Omega_0^2 +2\dot{\gamma}(t)+2\dot{\tilde{\gamma}}(t)\right]q_{\pm}^\kappa(t) & & \nonumber\\
  =\, \mp \frac{\lambda(t)}{2}\partial_t \left[ \lambda(t-L) q_{\pm}^\kappa(t-L) \right] +f_\pm^\kappa(t), & &
\label{EOMqpm}
\end{eqnarray}
where the driving force is defined as $f_\pm^\kappa(t)\equiv-\lambda(t)\dot{\varphi}_{\pm}^{[0]\kappa}(t)-\tilde{\lambda}(t)
\dot{\zeta}_{\pm}^{[0]\kappa}(t)$ with $\varphi_\pm^{[0]\kappa} \equiv ( \varphi_{0}^{[0]\kappa} \pm \varphi_{L}^{[0]\kappa} )/\sqrt{2}$ and
$\zeta_\pm^{[0]\kappa}\equiv ( \zeta_{A, \vartheta^{}_A}^{[0]\kappa} \pm \zeta_{B, \vartheta^{}_B}^{[0]\kappa} )/\sqrt{2}$.
Now $q_+^\kappa$ and $q_-^\kappa$ decouple and each is driven by a nonlocal force.

Suppose $q_\pm^\kappa (t) = \sum_\Omega \alpha_\pm^\kappa (\Omega) e^{-i\Omega t}$ for $t \gg L > 0 \gg \tilde{t}_0$ and $T\ll L$ in 
(\ref{thetaT}), so that $\gamma$ and $\tilde{\gamma}$ have become constants of time. Since $f_\pm^\kappa(t)$ are zero for $\kappa =A,B$ 
and simple harmonic for $\kappa= \{\rm k\}, \{\tilde{\rm k}_A\}, \{\tilde{\rm k}_B\}$ (cf. the expressions below Eq. (\ref{chisol})), for 
those $\Omega\not={\rm w}(\equiv |{\rm k}|)$ for $\kappa={\rm k}$, or $\Omega\not= \tilde{\rm w}_{\bf d}(\equiv|\tilde{\rm k}_{\bf d}|)$ 
for $\kappa = \tilde{\rm k}_{\bf d}$, Eq. (\ref{EOMqpm}) requires
\begin{equation}
   \Omega^2 + 2 i\Omega \left[\tilde{\gamma} + \gamma \left(1\pm e^{i\Omega L}\right) \right]- \Omega_0^2 = 0 \label{EqOm}
\end{equation}
for nonzero $\alpha_\pm^\kappa(\Omega)$. 
Let $\Omega = R + i I$ with $R, I \in {\bf R}$. Then the real and imaginary parts of (\ref{EqOm}) read
\begin{eqnarray}
  R^2-I^2-\Omega_0^2-2 I\left[\tilde{\gamma} +\gamma\left( 1\pm e^{-I L}\cos RL\right)\right]\mp 2\gamma R\, e^{-I L}\sin R L &=& 
	0\label{ReW}\\
	2RI+2R\left[\tilde{\gamma} +\gamma\left( 1\pm e^{-I L}\cos R L\right)\right]\mp 2\gamma I\, e^{-I L}\sin R L &=& 0. \label{ImW}
\end{eqnarray}

The real solutions for $\Omega$, if they exist, will have $I=0$ and so (\ref{ImW}) implies
\begin{equation}
   \mp\cos RL = 1 + \frac{\tilde{\gamma}}{\gamma},
\end{equation}
which will not be true unless $\tilde{\gamma}=0$ since $|\cos RL| \le 1$ and $\gamma, \tilde{\gamma} \ge 0$. For $\tilde{\gamma}=0$,
the real solution for (\ref{EqOm}) is $\Omega = \Omega_0$ for $q_-^\kappa$ when $\Omega_0 = 2n\pi/L$ for some positive integer $n$, or 
for $q_+^\kappa$ when $\Omega_0=(2n-1)\pi/L$. When one of these happens, the internal HOs in the detector mirrors are resonant with their mutual influences, while $q_\pm^\kappa (t)$ will never both settle down to steady states of constant amplitudes. 
This makes the late-time field spectrum ($\sim |\varphi_x^k(t)|^2$; see Sec. \ref{cavmodLT}) inside the cavity restless forever in a 
range of frequency $|k|$ of the driving force $f_\pm^k(t)$ ($k={\rm k}, \tilde{\rm k}_A, \tilde{\rm k}_B$), due to the mixing of the driving 
and the resonant frequencies. Outside the cavity, the late-time field spectrum at the same frequencies will never settle down, either, 
though the changes in time are less significant in magnitude than those inside the cavity. 
These time-varying patterns of the field spectrum at late times may depend on the initial conditions such as the time-scale 
and the functional form of the switching function $\gamma(t)$ for the OF coupling.

If there exist purely imaginary solutions, which have $R=0$, then (\ref{ImW}) will be trivial ($0=0$) and (\ref{ReW}) will become 
\begin{equation} 
  I^2 + \Omega_0^2 = -2 I [\tilde{\gamma} + \gamma(1\pm e^{-I L})], \label{condR0}
\end{equation}
which implies that $I\not= 0$ and $ I [\tilde{\gamma} + \gamma(1\pm e^{-I L})]$ must be negative.
If $I>0$, then $1\pm e^{-I L} >0$ and so $I [\tilde{\gamma} + \gamma(1\pm e^{-I L})] > 0$, which contradicts (\ref{condR0}). 
Thus $I$ must be negative here. Similarly, when both $R$ and $I$ are nonzero, (\ref{ReW}) and (\ref{ImW}) yield
\begin{equation}
  (R^2+I^2)\{ 1+ 2 I^{-1}[\tilde{\gamma} +\gamma(1\pm e^{-I L}\cos RL)]\} = -\Omega_0^2, \label{condRn0}
\end{equation}
which implies that the expression in the curly brackets must be negative. If $I>0$, then $1+ (2/I)[\tilde{\gamma} + 
\gamma(1\pm e^{-I L}\cos RL)]> 0$ and (\ref{condRn0}) cannot hold. So $I$ must be negative here, too. 
Therefore, the imaginary parts of the complex solutions for $\Omega\not={\rm w}$, $\tilde{\rm w}_{\bf d}$, or $\Omega_0$ if $\Omega_0=
n\pi/L$ for some positive integer $n$, must be negative, and the corresponding modes $e^{-i\Omega t} = e^{-|I| t}e^{-i R t}$ will decay out 
as $t\to \infty$. At late times, only the oscillations of $\Omega={\rm w}$ and $\tilde{\rm w}_{\bf d}$ for all values of $\Omega_0$, and 
additionally $\Omega=\Omega_0$ when $\Omega_0$ happens to be $n\pi/L$ for some positive integer $n$, will survive.

Longer relaxation times would occur in the cases with $\Omega\approx \Omega_0 \approx n\pi/L$ for some positive integer $n$. 
In these near-resonance cases, one may write
\begin{equation}
  \Omega = \frac{n\pi}{L} + \epsilon_{n} + i I_{n} ,
\end{equation}
where $|\epsilon_n|, |I_n| \ll n\pi/L$ and $I_{n}<0$.
Assuming $|\epsilon_n|$ and $|I_n|$ are roughly the same order and $|\epsilon_n L|, |I_n L| \ll 1$, 
then they can be approximated by 
\begin{eqnarray} 
  \epsilon^{}_{n} \approx \frac{\Omega_0^2 - (n\pi/L)^2}{2(n\pi/L)(1+\gamma L)}, \hspace{.5cm} & &
	I_n \approx \frac{1}{\gamma L^2}\left\{ J_n-\sqrt{ J_n^2+ 2\gamma L^2\left(\tilde{\gamma}+\frac{\gamma\epsilon_n^2 L^2}{2}\right)}\right\},
	\nonumber\\ & & J_n\equiv 1+\gamma L+(-1)^n\frac{\gamma\epsilon_n L^2}{n\pi},
\end{eqnarray}
from (\ref{ReW}) and (\ref{ImW}). To keep the above approximate expression of $\epsilon^{}_n$ small, one should take a large value of 
$\gamma L$, and/or $\Omega_0$ should be very close to $n\pi/L$ with some positive integer $n$. This can be achieved more easily when the 
separation of the mirrors $L$ is large, since $|\Omega_0 - n\pi/L| \le \pi/(2L)$ will be small for a general $\Omega_0$ and the integer $n$ 
closest to the value of $\Omega_0 L/\pi$. For a very large $L$ the approximation can be good even for $|\Omega_0 -n' \pi/L|$ being a few 
times of $\pi/L$ for some $n'\not= n$. Note that $I_n|_{\epsilon_n=0}$ vanishes for $\tilde{\gamma}=0$, when we return to the resonant cases. 

As we have known in (\ref{condR0}), besides $\Omega\approx n\pi/L$ with positive integer $n$, there may exist purely imaginary solutions 
$\Omega = i I_0$ for $q_+^\kappa$ (in general) and for $q_-^\kappa$ (in some particular parameter ranges). 
According to (\ref{condR0}), indeed, when $|I_0 L| \ll 1$, one has 
\begin{equation}
   I_0 \approx {\rm Re}\left[ \frac{-(\tilde{\gamma}+2\gamma)+\sqrt{ (\tilde{\gamma}+2\gamma)^2-\Omega_0^2(1-2\gamma L)}}{1-2\gamma L} 
	\right] \label{Image0}
\end{equation}
for $q_+$, which is always closer to zero than the counterpart for $q_-$ (if any) is. 
These would be clear by arranging (\ref{condR0}) 
into $I^2 +2I(\tilde{\gamma} +\gamma) +\Omega_0^2 = \mp 2 I e^{-I L}$ for $q_\pm$, and then observing that the left-hand side is a concave-up 
parabola with the minimum at some negative $I$ while the right-hand side is zero at $I=0$ and monotonically decreasing (increasing) for 
$q_+^\kappa$ ($q_-^\kappa$) as $I$ approaches to $0$ from a negative value.

The relaxation time for our cavity with a not-too-small separation of the mirrors 
could be estimated by the inverse of the minimal $|I_{n'}|$ ($n'=0,1,2,3,\cdots$) among the above solutions.
In the cases with the minimal $|I_n|\not= |I_0|$ (namely, $n>0$), when the separation is sufficiently large 
so that $\gamma L \gg \tilde{\gamma}$, and $\Omega_0$ is close enough to $n\pi/L$, one has
\begin{equation}
  t_{\rm rlx}\approx (1+\gamma L)/\tilde{\gamma}
\end{equation}
for the HO pair in the weak OE coupling and strong OF coupling, over-damping regime. Compared with (\ref{rlxtime}) for the HO in a 
single detector mirror in the same regime, we see that a stronger OF coupling still makes the relaxation time longer and 
$t_{\rm rlx}\sim \gamma$ for very large $\gamma$ in both cases, but a stronger HO environment here plays the opposite role to those in the 
single-mirror cases and shorten the relaxation time of the cavity near resonance. 

Note that, unlike the (3+1)D case in Ref. \cite{LH09}, there is no instability in the small $L$ limit here since the retarded field is 
independent of the distance $L$ from the source in (1+1)D, while it is proportional to $1/L$ in \cite{LH09}. As $L\to 0$, the equations
of motion in (\ref{EOMqpm}) simply become regular, ordinary differential equations without delay.

\subsection{Cavity modes at late times}
\label{cavmodLT}

With a non-vanishing coupling to the environment $\tilde{\gamma}$, one can get rid of the late-time non-steady states
described in Sec. \ref{relax2}. After the OF coupling is switched on, if we look at the field amplitudes only in the cavity, 
the field spectrum will appear to evolve from continuous to nearly discrete in the neighborhood of the resonant frequency. 

\begin{figure}
\includegraphics[width=4.2cm]{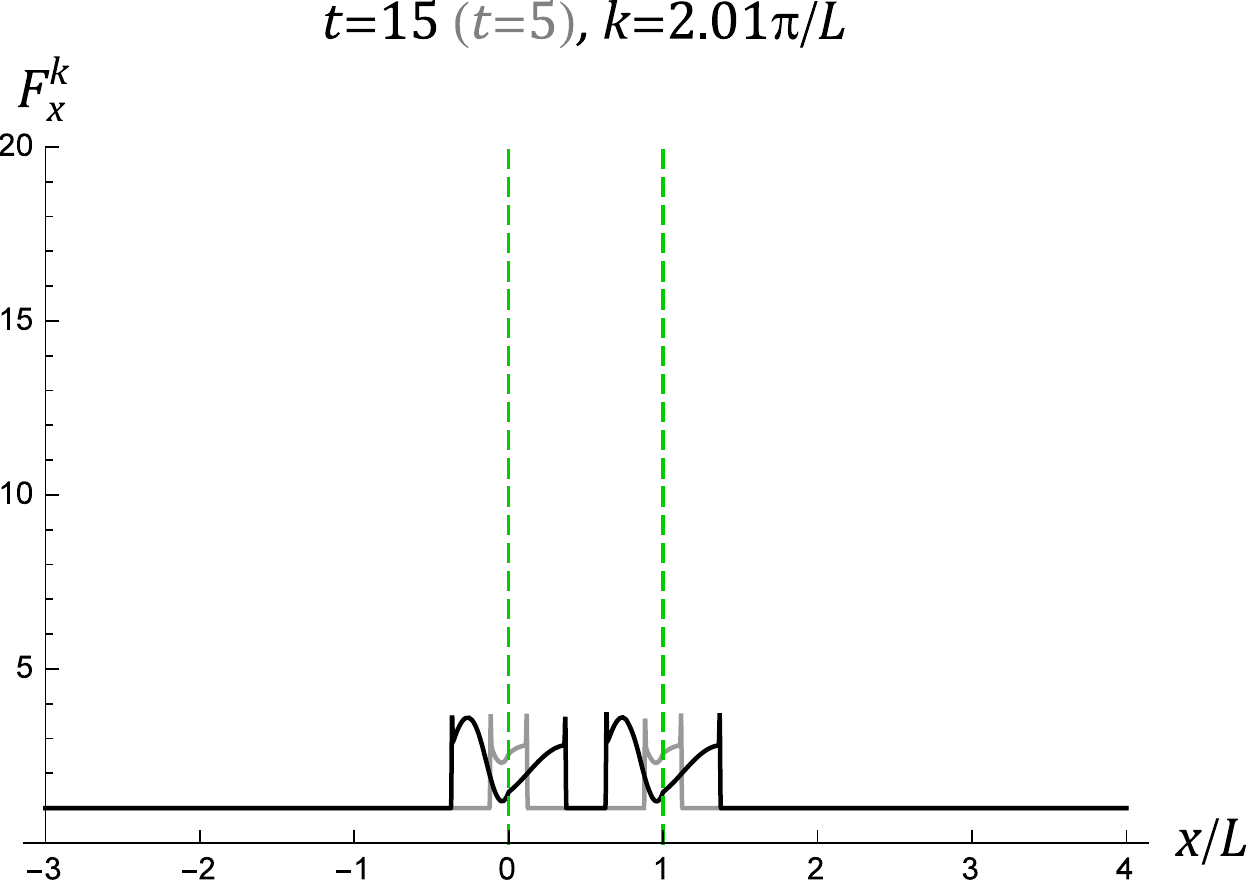}
\includegraphics[width=4.2cm]{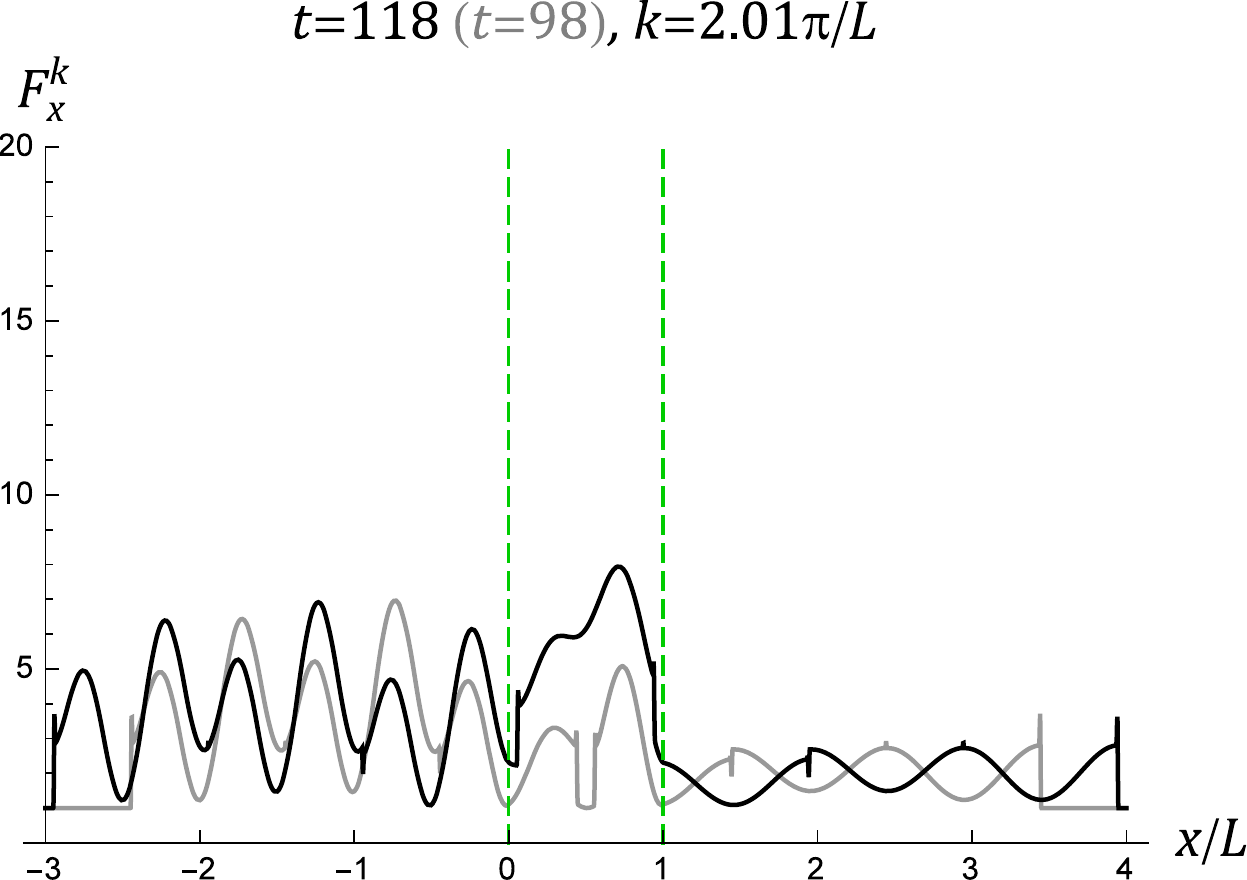}
\includegraphics[width=4.2cm]{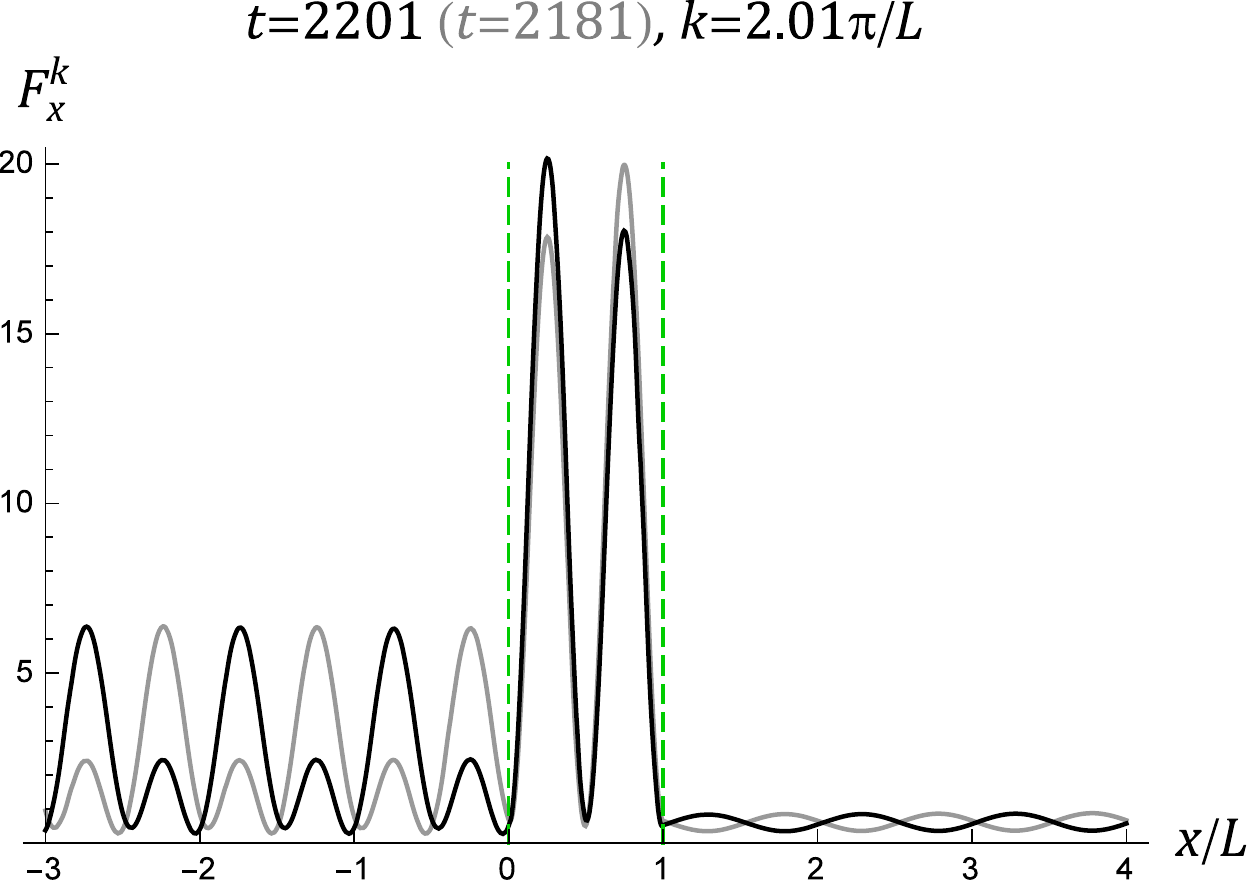}
\includegraphics[width=4.2cm]{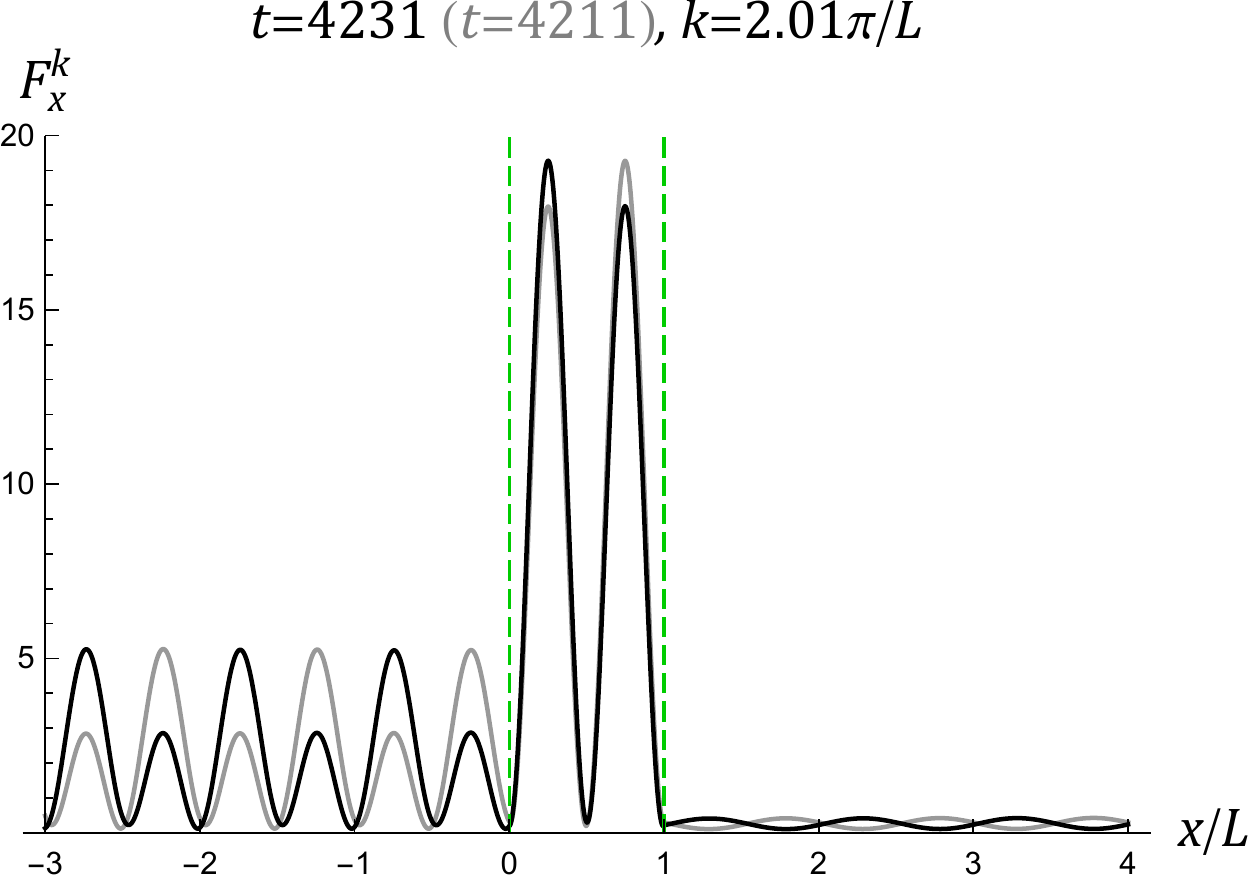}\\
\includegraphics[width=4.2cm]{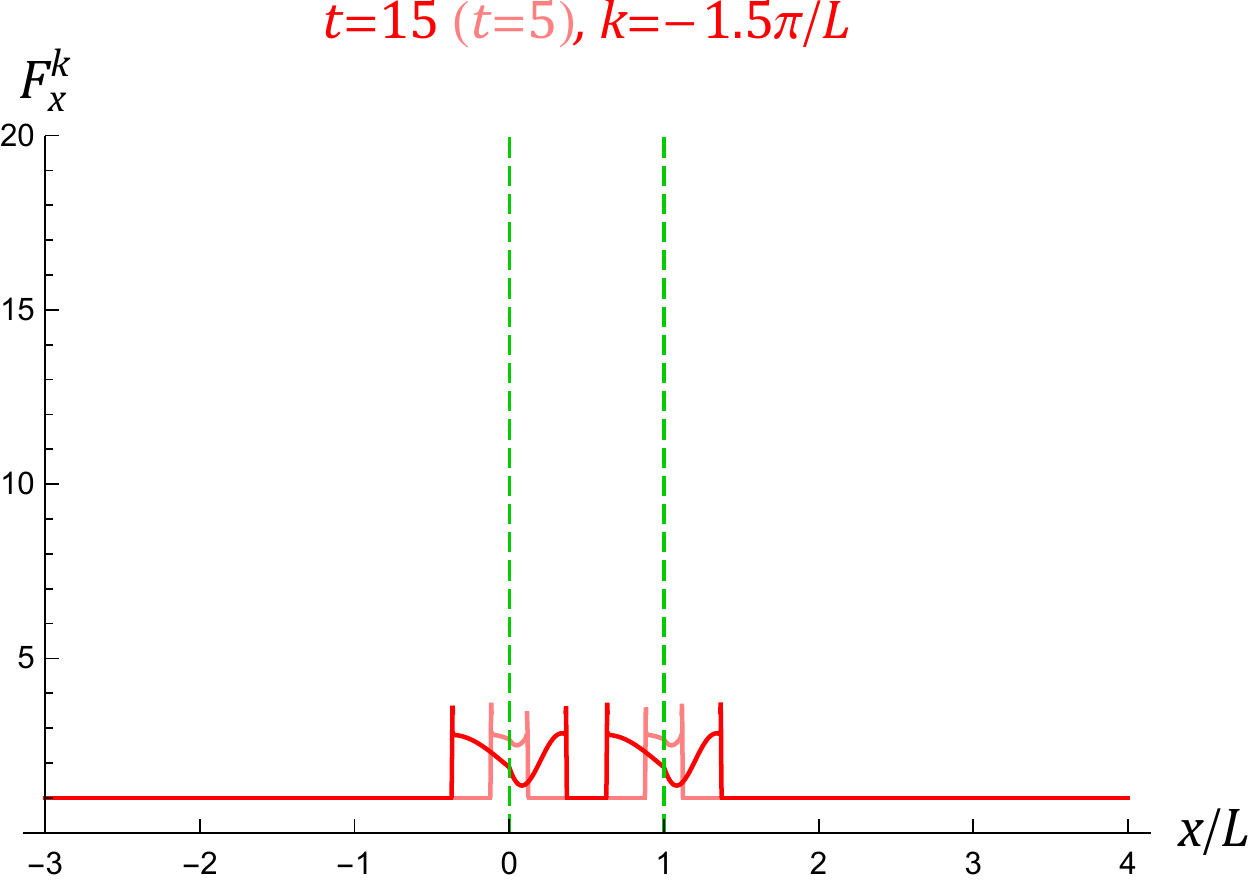}
\includegraphics[width=4.2cm]{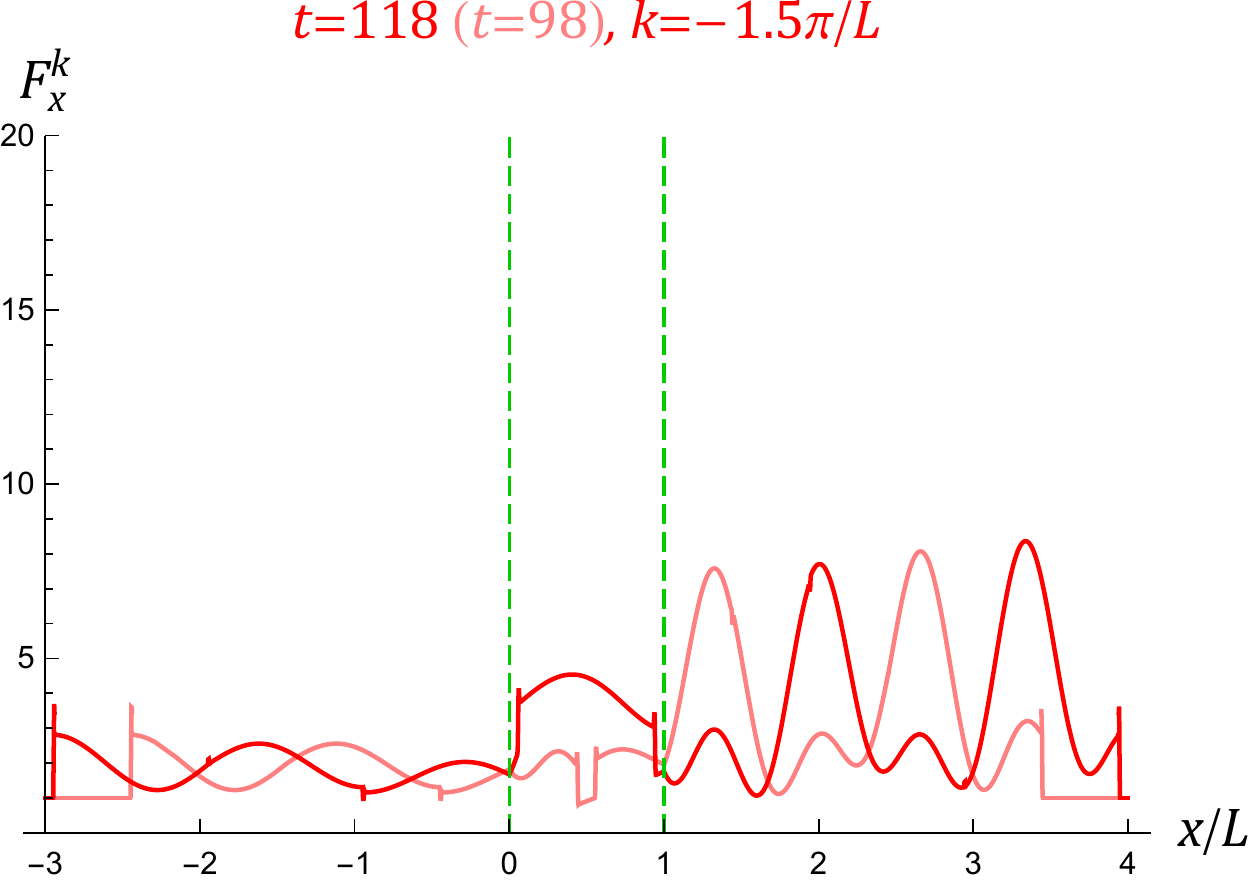}
\includegraphics[width=4.2cm]{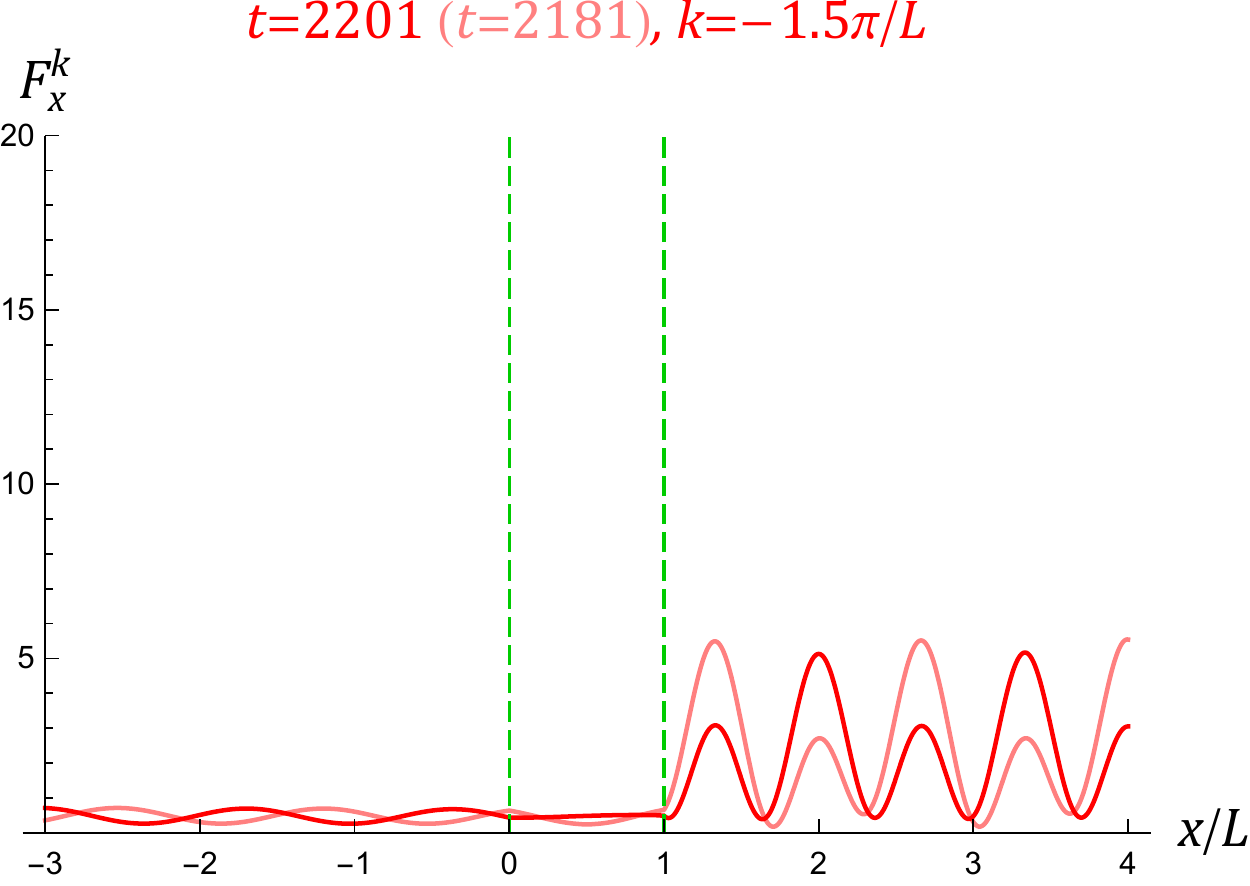}
\includegraphics[width=4.2cm]{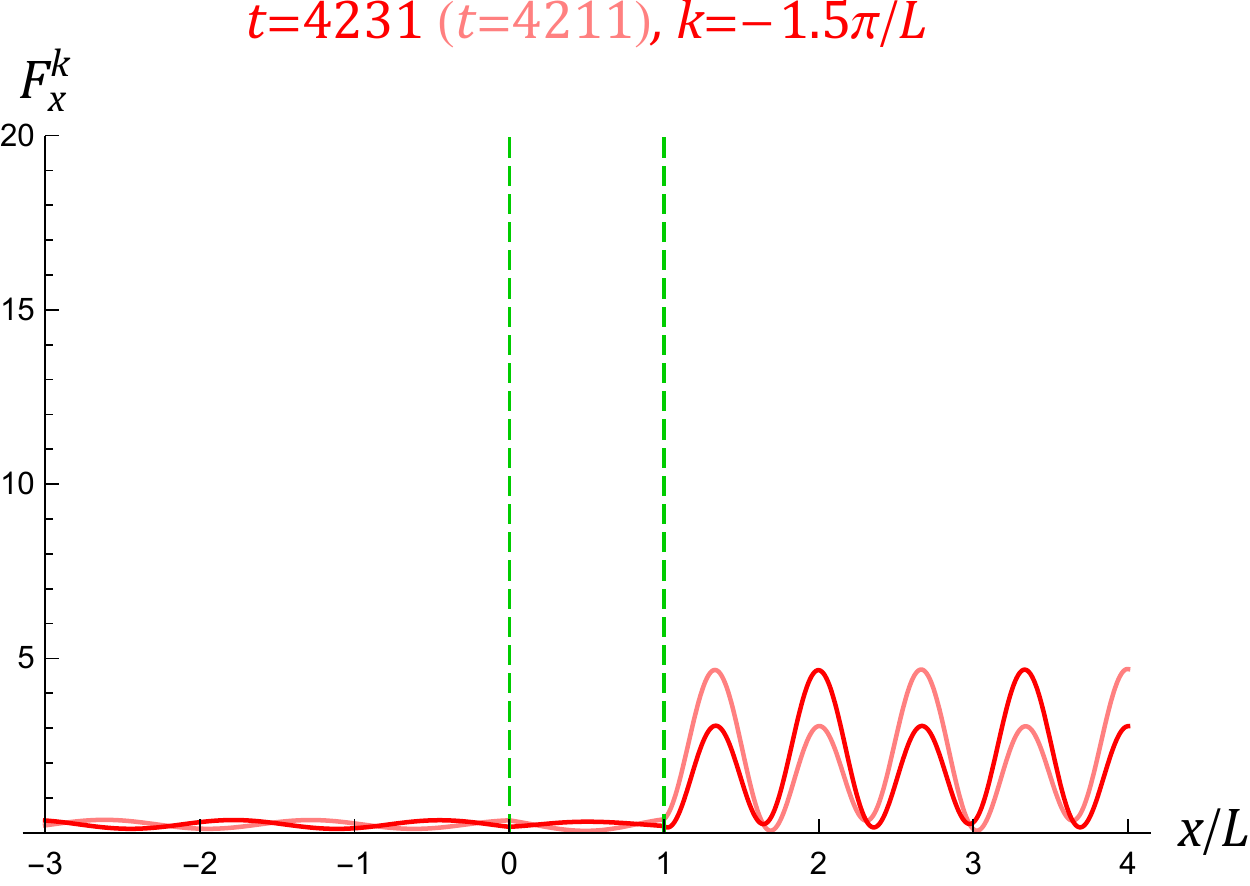}
\caption{Time evolution of the field spectrum $F_x^k$ defined in (\ref{Flatecoin}) and read off from (\ref{FxtFyt}) for $k=2.01\pi/L$ 
(right mover, upper row) and $k=-1.5\pi/L$ (left mover, lower row) against $x$. Here $\gamma=10$, $\tilde{\gamma}=1$, $\Omega_0=1/10$, 
$L=40$, and $c=\hbar=1$. The green dashed lines mark the locations of the detector mirrors at $x=0$ and $x=L=40$. 
Here the relaxation time for each single mirror is $t^{(1)}_{\rm rlx} \approx 2200$ according to (\ref{rlxtime}),
while the relaxation time for the cavity is $t^{(2)}_{\rm rlx} = 1/|I_0| \approx 4219 \approx 2t^{(1)}_{\rm rlx}$ from (\ref{Image0}). 
The third and the fourth plots from the left in each row are $F_x^k$ at $t\approx t^{(1)}_{\rm rlx}$ and 
$t^{(2)}_{\rm rlx}$, respectively.}
\label{FxkEvo}
\end{figure}

For $t>0$, the field spectrum defined in  (\ref{Flatecoin}) can be read off from the coincidence limit of the symmetrized two-point 
correlator of the field,
\begin{eqnarray}
  \langle \hat{\Phi}_{x}(t),\hat{\Phi}_{x'}(t')\rangle &=& {\rm Re}\left\{
	\int \frac{d\rm k}{2\pi} \frac{\hbar}{2\rm w} \varphi_x^{\rm k}(t)\varphi_{x'}^{\rm k*}(t') \right.\nonumber\\ & & + \left. 
	\int \frac{d\tilde{\rm k}^{}_A}{2\pi} \frac{\hbar}{2\tilde{\rm w}^{}_A}\varphi_x^{\tilde{\rm k}^{}_A}(t)	
	    \varphi_{x'}^{\tilde{\rm k}^{}_A*}(t')+ 
  \int \frac{d\tilde{\rm k}^{}_B}{2\pi} \frac{\hbar}{2\tilde{\rm w}^{}_B}\varphi_x^{\tilde{\rm k}^{}_B}(t)
	    \varphi_{x'}^{\tilde{\rm k}^{}_B*}(t')
  \right\} \label{FxtFyt}
\end{eqnarray}
in the presence of the cavity.
An example on the time evolution of the field modes is given in Figure \ref{FxkEvo}, where we consider a case with a larger value of 
$\tilde{\gamma}$, namely, $\Omega_0, L^{-1} <\tilde{\gamma} \ll \gamma$ to reach the late-time steady states sooner while a wide range of 
the cavity modes can still be generated. In this example, the evolution of each single field mode from the initial moment to late times can 
roughly be divided into four stages:
(i) At very early times, the shock waves produced by the switching-on of the OF coupling propagate freely in space;
(ii) after the waves produced by two different mirrors collide, violent changes of the field amplitude squared occur;
(iii) after a timescale comparable with the relaxation time of the cavity, the interference pattern of the cavity mode is basically built up, 
but the field amplitude squared keeps ringing down with small oscillations in time;
(iv) after a longer timescale the shape of the field spectrum against $x$ gets into the late-time steady state.
The resonant modes ($\omega \approx n \pi/L$, $n=1,2,3,\cdots$) will survive, while the off-resonant modes will be suppressed in the cavity. 

\begin{figure}
\includegraphics[width=6.3cm]{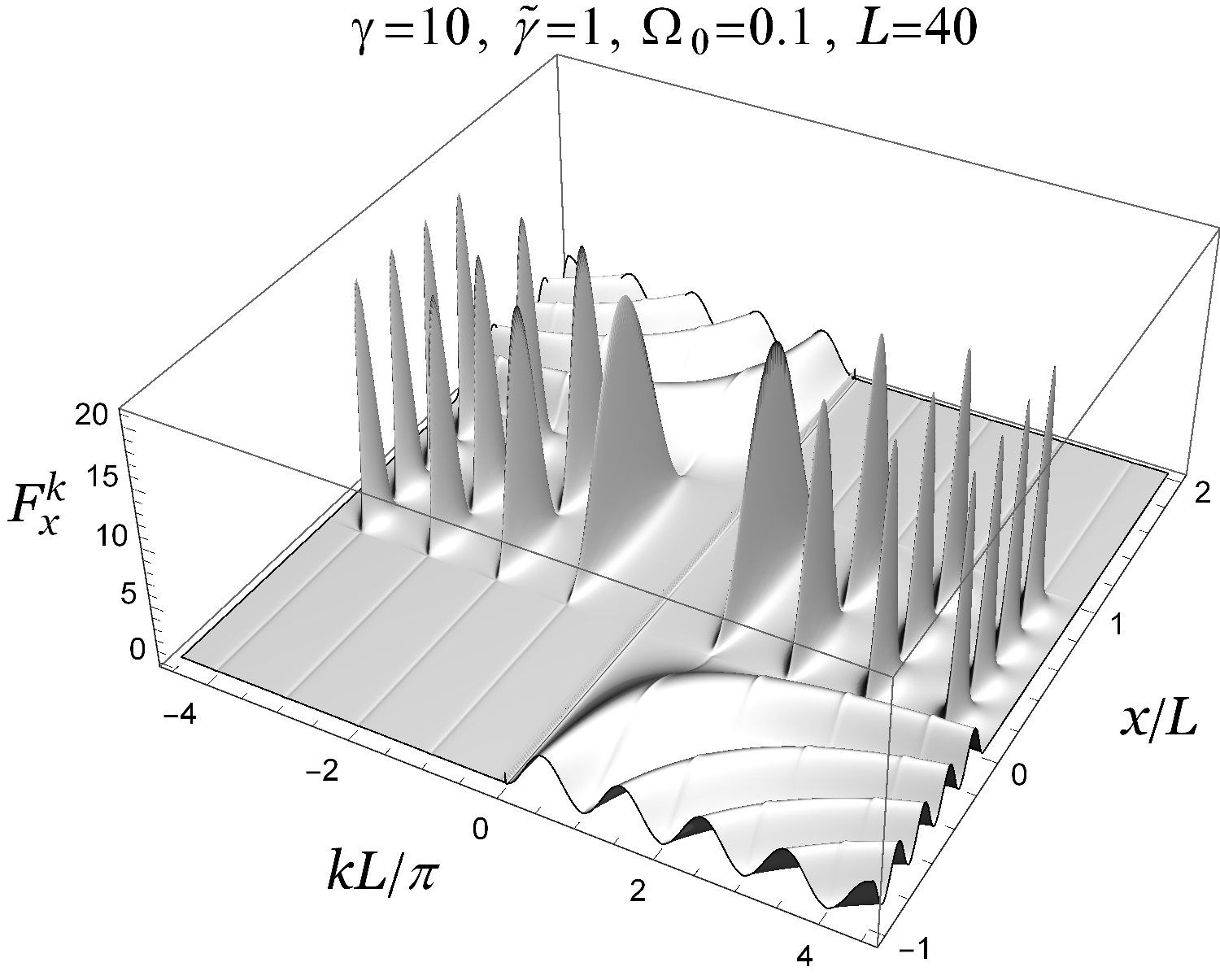}
\includegraphics[width=5.5cm]{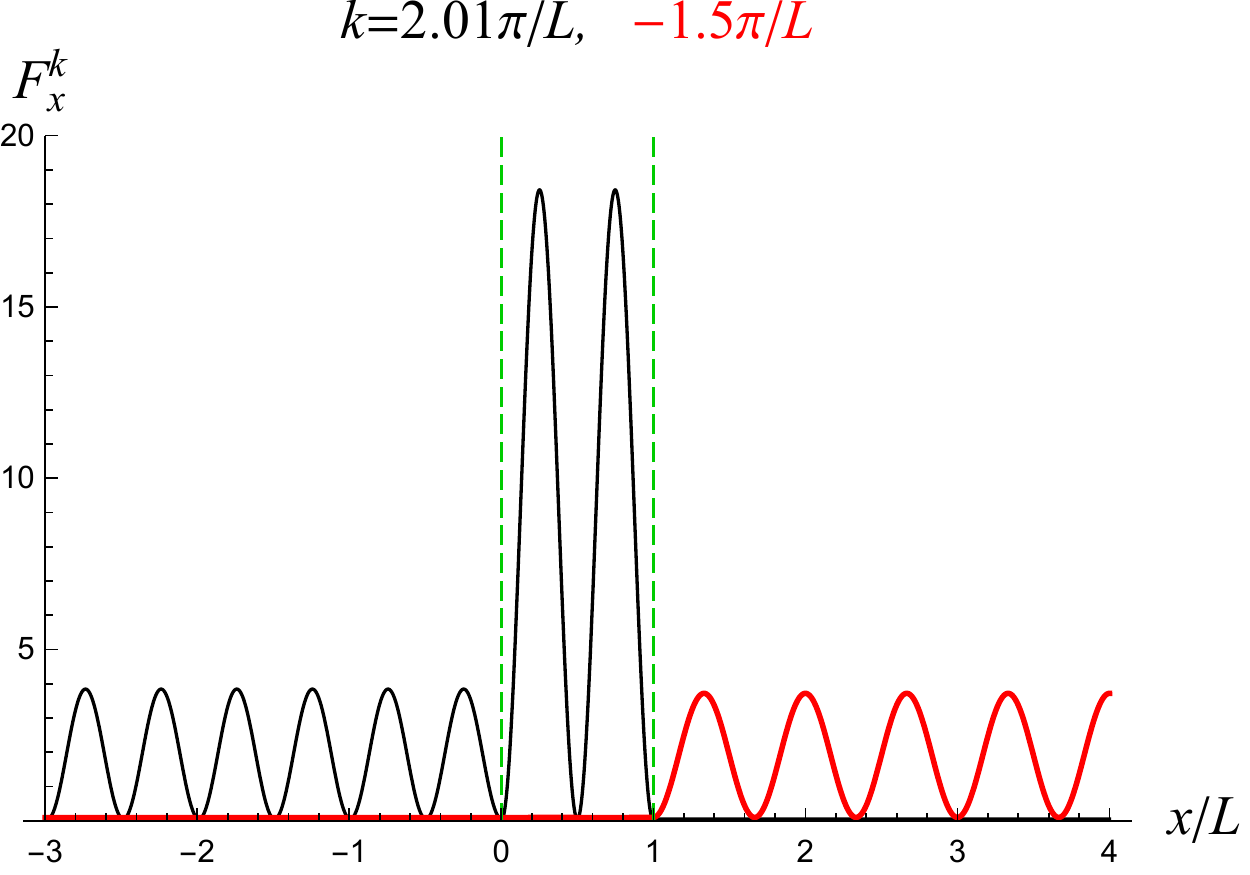}
\includegraphics[width=5.5cm]{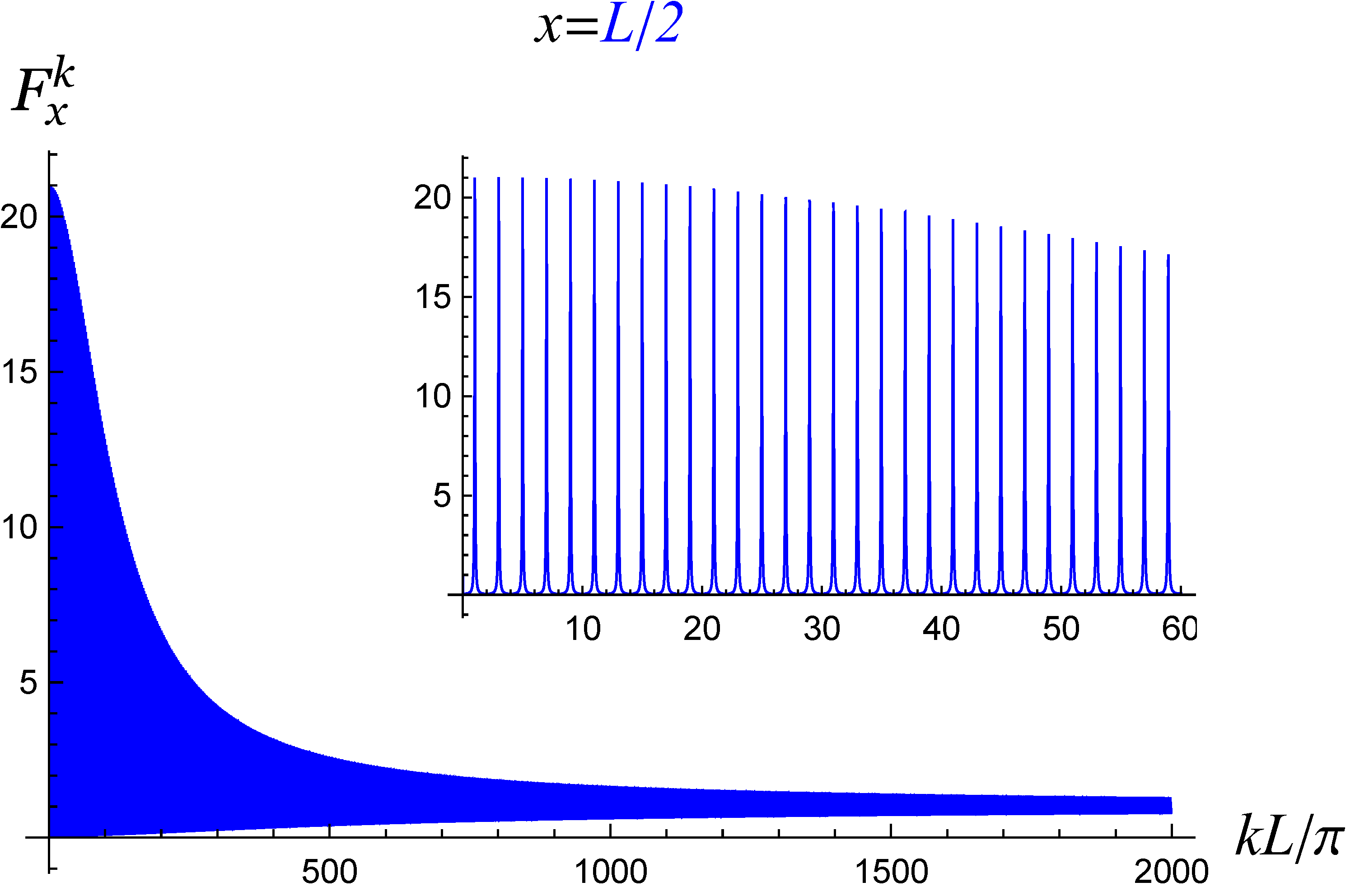}
\caption{(Left)　The late-time field spectrum $F_x^k(t)$ against $k$ and $x$ in the over-damping regime, with the same parameter values as 
those in Figure \ref{FxkEvo}. (Middle) The late-time results of $F_x^k$ in Figure \ref{FxkEvo} for $k=2.01\pi/L$ (black line) and $k=-1.5\pi/L$ (red line). 
(Right) $F_x^k$ at the cavity center $x=L/2=20$ shows that the field spectrum in the cavity is nearly discrete in the low-$|k|$ regime 
(inset), while the sharpness and the contrast of the comb teeth around $|k|=(2n-1)\pi/L$, $n=1,2,3,\cdots$, decrease as $|k|$ increases.}
\label{Phi2xkOvr}
\end{figure}

\begin{figure}
\includegraphics[width=6.3cm]{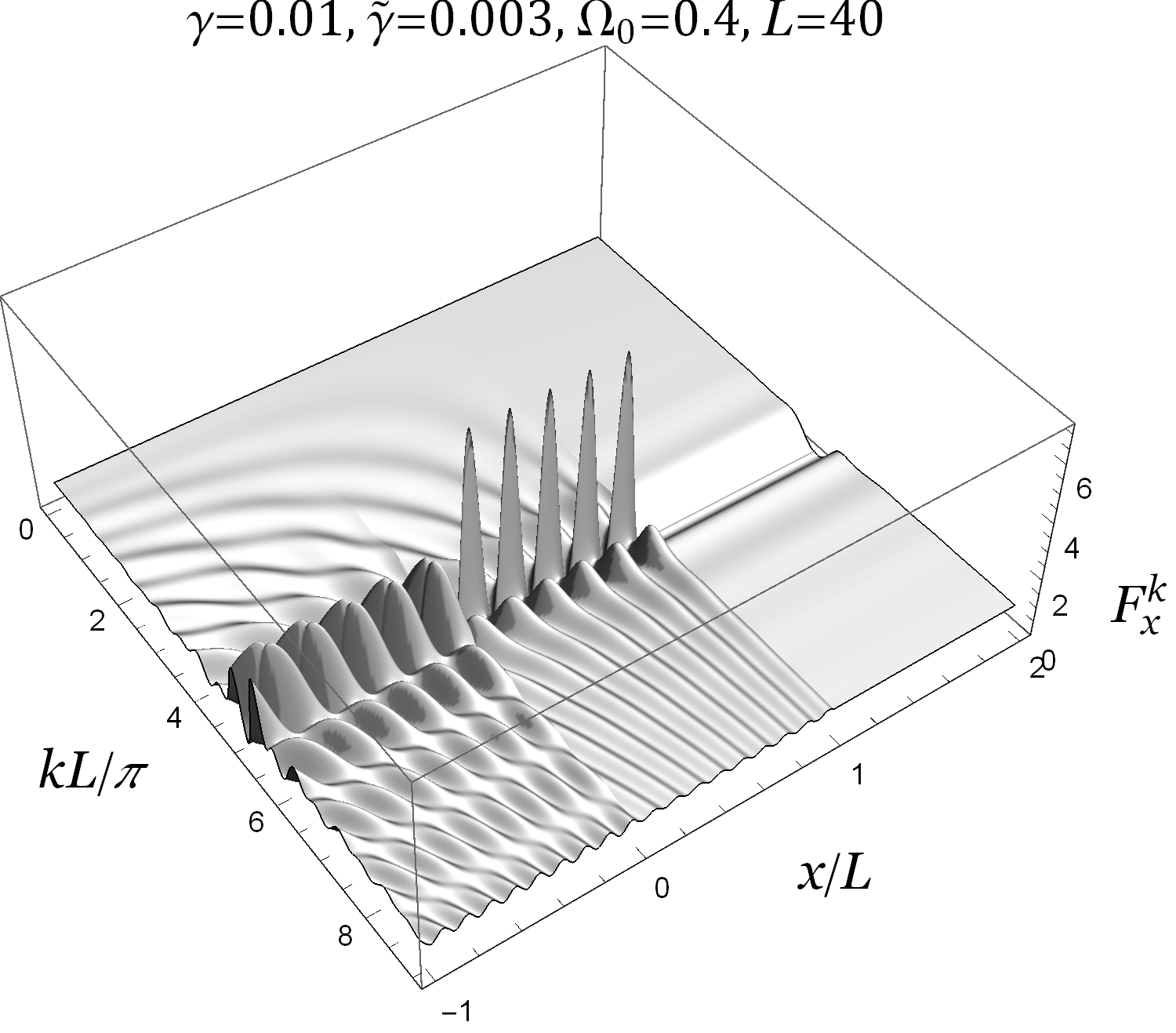}
\includegraphics[width=5.5cm]{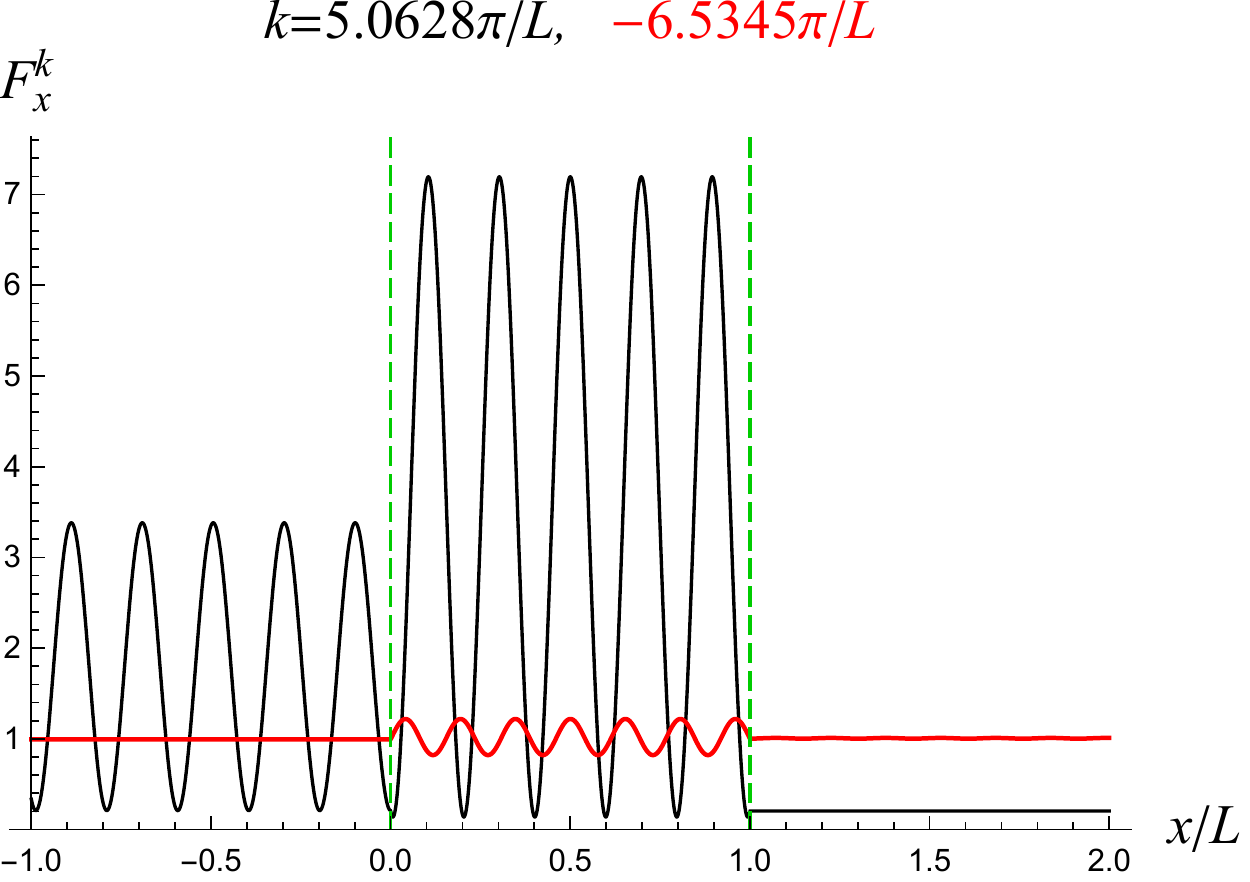}
\includegraphics[width=5.5cm]{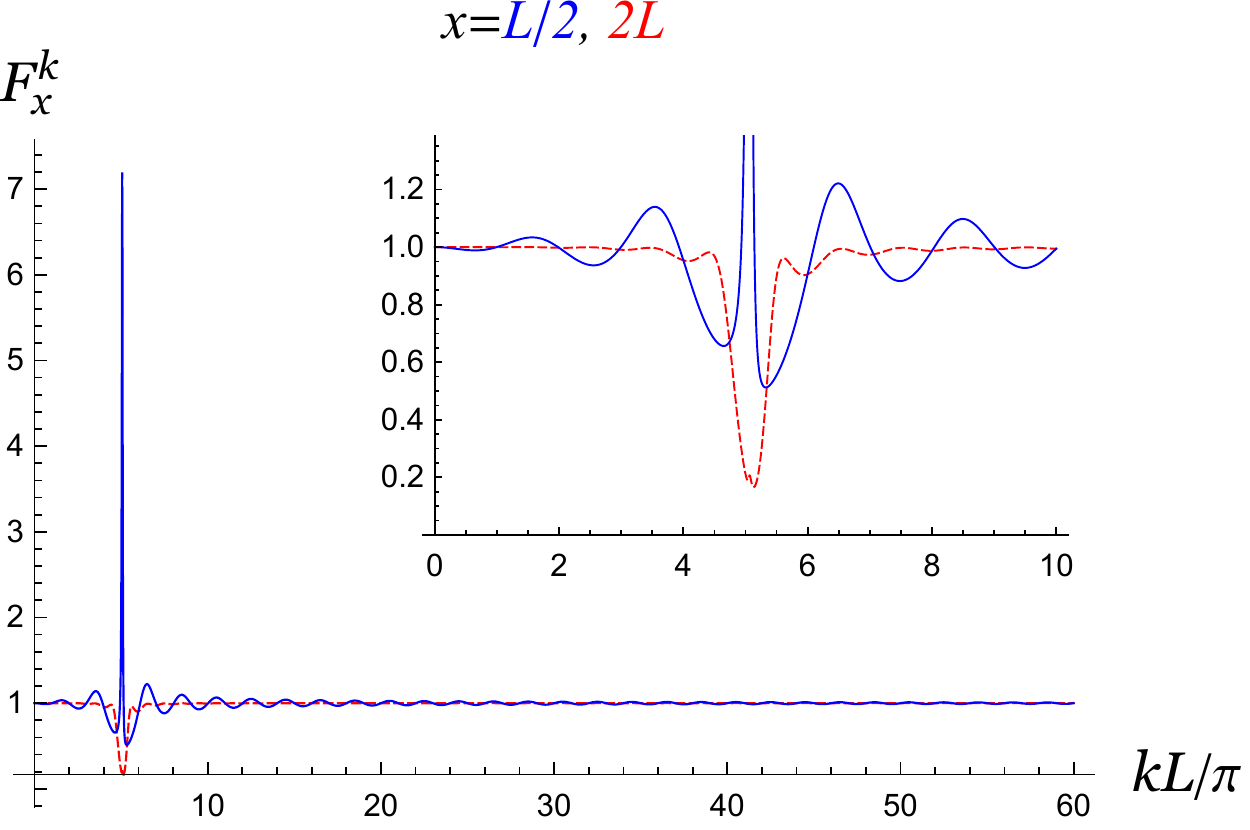}
\caption{(Left) The late-time field spectrum $F_x^k(t)$ against $k$ and $x$ in the under-damping regime, where $\gamma=0.01$, 
$\tilde{\gamma}=0.003$, $\Omega_0=0.4$, $L=40$, and $c=\hbar=1$. Here we only show the domain of $k>0$ (right movers). 
(Middle) The field spectrum against $x$ for the cavity mode of $k =5.0628\pi/L \approx 0.3976 \approx \Omega_0-\tilde{\gamma}$ (black) 
and the field mode of $k=-6.5345\pi/L \approx -(7-(1/2))\pi/L$ (red line, resonant transmission from right to left). 
(Right) $F_x^k$ at the cavity center $x=L/2=20$ (blue line) and $x=2L=80$ (red dashed line) for $k>0$. The blue curve shows that the only significant cavity mode for $k>0$ is peaked at $k\approx 5.0628L/\pi$.
The red dashed curve indicates that the transmittivity through the cavity is suppressed around the cavity mode, and close to $1$ around the resonant transmissions at $k\approx (n-1/2)\pi/L$, $n=1,2,3,\cdots$.}
\label{Phi2xkUnd}
\end{figure}

At late times, the mode functions in (\ref{FxtFyt}) become
\begin{eqnarray}	
	\varphi_x^{\rm k}(t) &\to& e^{-i{\rm w} t}\left\{ e^{i{\rm k}x} - \gamma\left[ (1+e^{i{\rm k}L}){\cal E}^+_{\rm w}(x)\chi^{+}_{\rm w} 
	    +(1-e^{i{\rm k}L}){\cal E}^-_{\rm w}(x) \chi^{-}_{\rm w} \right] \right\}, \\
	\varphi_x^{\tilde{\rm k}^{}_A}(t)	&\to&  -\sqrt{\gamma\tilde{\gamma}} e^{i\tilde{\rm k}^{}_A\vartheta^{}_A-i\tilde{\rm w}^{}_A t}
	    \left[{\cal E}^+_{\tilde{\rm w}^{}_A}(x) \chi^{+}_{\tilde{\rm w}^{}_A} + 
			{\cal E}^-_{\tilde{\rm w}^{}_A}(x)\chi^{-}_{\tilde{\rm w}^{}_A}  \right], \\
	\varphi_x^{\tilde{\rm k}^{}_B}(t)	&\to&  -\sqrt{\gamma\tilde{\gamma}} e^{i\tilde{\rm k}^{}_B\vartheta^{}_B-i\tilde{\rm w}^{}_B t}
	    \left[{\cal E}^+_{\tilde{\rm w}^{}_B}(x) \chi^{+}_{\tilde{\rm w}^{}_B} - 
			{\cal E}^-_{\tilde{\rm w}^{}_B}(x)\chi^{-}_{\tilde{\rm w}^{}_B}  \right],
\end{eqnarray}
with 
\begin{eqnarray}
 {\cal E}^\pm_\omega (x) &\equiv& e^{i\omega |x|}\pm e^{i\omega |x-L|}, \\
 \chi^{\pm}_\omega &\equiv& \frac{-i\omega}{\Omega_0^2 - \omega^2 -2i\omega\left[ \tilde{\gamma}+\gamma(1\pm e^{i\omega L})\right]},
    \label{chipm}
\end{eqnarray}
such that $q_\pm^k =\chi_{\omega}^\pm (-\lambda\varphi_\pm^{[0]k}-\tilde{\lambda}\zeta_\pm^{[0]k})$, $k=\{{\rm k}\}, \{\tilde{\rm k}_A\}, 
\{\tilde{\rm k}_B\}$, $\omega=|k|$, from (\ref{EOMqpm}). Then the coincidence limit $(t',x')\to (t,x)$ gives the late-time field spectrum:
\begin{eqnarray}	
	F^k_x = 1 +\gamma &{\rm Re}& \left\{ 
	\chi^+_\omega {\cal E}_\omega^+(x)\left[{\cal E}_\omega^{+*}(x) -2\left(e^{-ikx}+e^{-ik(x-L)}\right)\right] \right. \nonumber\\
	& & + \left.\chi^-_\omega {\cal E}_\omega^-(x)\left[{\cal E}_\omega^{-*}(x) -2\left(e^{-ikx}-e^{-ik(x-L)}\right)\right] \right\} ,
	\label{Flate}
\end{eqnarray}
as defined in (\ref{Flatecoin}). Here we have used the identity
\begin{equation}
 \chi^\pm_\omega +\chi^{\pm*}_\omega = 4\left[\tilde{\gamma}+\gamma\left( 1\pm\cos\omega L\right)\right]|\chi^\pm_\omega|^2 \label{FDR2}
\end{equation}
similar to (\ref{FDR}).
Note that the odd functions of $k$ in the integrand for the late-time $\langle \hat{\Phi}_{x}(t),\hat{\Phi}_{x'}(t')\rangle$ 
do not contribute to the $k$ integral and so they are not included in the above $F^k_x$. 

Examples of the late-time field spectra in the over- and under-damping regimes are shown in Figures \ref{Phi2xkOvr} and \ref{Phi2xkUnd}, 
respectively. Figure \ref{Phi2xkOvr} is the late-time result of the case considered in Figure \ref{FxkEvo}. One can see that there are 
indeed many cavity modes inside the cavity ($0<x<L$) in the strong OF coupling, over-damping regime. The standing waves due to the
interference of the incident and reflected waves outside the cavity, similar to those in the single mirror case in Figure 
\ref{Phi2xk1}, can also be seen. Sampling at the center of the cavity $x=L/2$, the field spectrum $F_{L/2}^k$ looks discrete in the low-$|k|$ 
regime. In this example, $\Omega_0^2 \ll 2 \tilde{\gamma} \pi/L$ and so the peak values of the comb teeth of $F_{L/2}^k$ with small $n$ are 
about $2\gamma/\tilde{\gamma}$, 
while in the high-$|k|$ regime $F_{L/2}^k \approx 1+4 (\gamma/\omega)\sin\omega L$ looks continuous and goes to 
the free-space value $1$ as $\omega=|k|\to \infty$. 
The working range of this detector mirror is about $0< k < 150\pi/L$ from Figure \ref{Phi2xkOvr} (right).

When our attention is restricted in the cavity, it appears that all the two-point correlators of an off-resonant mode in the 
cavity, $\langle \Phi_k, \Phi_{-k}\rangle$, $\langle \Pi_k, \Pi_{-k}\rangle$, and $\langle \Phi_k, \Pi_{-k}\rangle$, are suppressed in the 
strong OF coupling regime, and the uncertainty relation of that mode would be violated. This is not true since in looking at those 
correlators in the $k$ space we have to consider the field spectrum outside the cavity as well.

As we discussed in Sec. \ref{secReflect} and illustrate in Figure \ref{Phi2xkUnd}, there are only one or a few pairs of significant cavity 
modes at late times in the weak OF coupling, under-damping regime. In Figure \ref{Phi2xkUnd} the only significant cavity modes are peaked 
around $|k| \approx 5 \pi/L$, which is nearly resonant with the natural frequency $\Omega_0$ of the internal HO in this example. 
The reflectivity in the vicinity of the resonant frequency is high enough to suppress the transmitted wave on the other side of the cavity,
while the detector mirrors become almost transparent for the field modes away from this narrow resonance.
Outside the cavity, one can see the interference pattern of the incident wave and the reflected waves by the two detector mirrors if the 
reflectivity of the mirror for that field mode is not too small or too large. The interferences of the waves reflected by the two detector 
mirrors are destructive for $k\approx \pm (n-(1/2))\pi/L$, $n=1,2,3,\cdots$, where the resonant transmission occurs, and constructive for 
$k \approx \pm n\pi/L$, which is the basis of Bragg reflection \cite{CGS11, CJGK12, CG16, SB16}. 
The result in the over-damping regime in Figure \ref{Phi2xkOvr} (left) does not show this feature because the reflectivity of the 
detector mirrors in the plot is so close to $1$ that the waves (say, from $x<0$) transmitted through the first mirror (at $x=0$) and 
reflected by the second mirror (at $x=L$), and then transmitted through the first mirror again to the incident region ($x<0$), are negligible. In the same conditions as those in Figure \ref{Phi2xkOvr} but now going to the high-$|k|$ regime where the reflectivity 
is lower, similar destructive and constructive interferences of the incidence and reflected waves outside the cavity can also be observed.  

\subsection{Casimir effect}
\label{CasimirEff}

Inserting the results (\ref{FxtFyt})-(\ref{chipm}) into (\ref{T00def}) and (\ref{T00renDef}), one obtains the late-time renormalized field 
energy density in the presence of the cavity mirrors:
\begin{equation}
	\langle \hat{T}_{00}(t,x)\rangle_{\rm ren}
	\to \lim_{(t',x')\to (t,x)}\frac{1}{2}\left( \partial_t\partial_{t'}+\partial_x\partial_{x'} \right)
	\int_{0}^{\omega^{}_M} \frac{d\omega}{2\pi}\frac{\hbar}{2\omega}F^\omega(t,x;t',x'), 	\label{T00int}
\end{equation}
where
\begin{eqnarray}	
	& & F^\omega(t,x;t',x') = - 2\gamma\cos\omega(t-t'){\rm Re}\left[ \chi^+_\omega {\cal E}_+(x){\cal E}_+(x') + 
	\chi^-_{\omega} {\cal E}_-(x){\cal E}_-(x') \right] \label{Fomegalate} 
\end{eqnarray}
and $\omega^{}_M$ is the UV cutoff, which should be identical to the ones for the internal HOs of our detector mirrors 
(will be introduced in Sec. \ref{EntMirOsc}) since (\ref{T00int}) has included the back-reaction of the detector mirrors to the field.  
A straightforward calculation shows that at late times $\langle \hat{T}_{00}(t,x)\rangle_{\rm ren} =0$ outside the cavity ($x<0$ or $x>L$), 
and inside the cavity
\begin{eqnarray}
    & & \left. \langle \hat{T}_{00}(t,x)\rangle_{\rm ren}\right|_{0<x<L} \to \nonumber\\
		& & -\hbar {\rm Re} \int_0^{\omega^{}_M}
		\frac{d\omega}{2\pi}\frac{8 \gamma^2 \omega^3 e^{2i\omega L}}{\left[ \omega^2 + 2i\omega(\gamma+\tilde{\gamma})-\Omega_0^2\right]^2+
		4\gamma^2\omega^2 e^{2i\omega L}} \stackrel{\omega^{}_M\to\infty}{\longrightarrow} \rho^{}_\Phi, \label{E0inCav}
\end{eqnarray}
which is a finite constant independent of $x$.
For $\Omega_0=0.1$, $\gamma=10$, $\tilde{\gamma}=1$, and $L=40$ in Figures \ref{FxkEvo} and \ref{Phi2xkOvr}, we have $\rho^{}_\Phi\approx 
-0.0000483163 <0$ ($c=\hbar=1$). This is the Casimir effect in our cavity of imperfect mirrors.

\begin{figure}
\includegraphics[width=8cm]{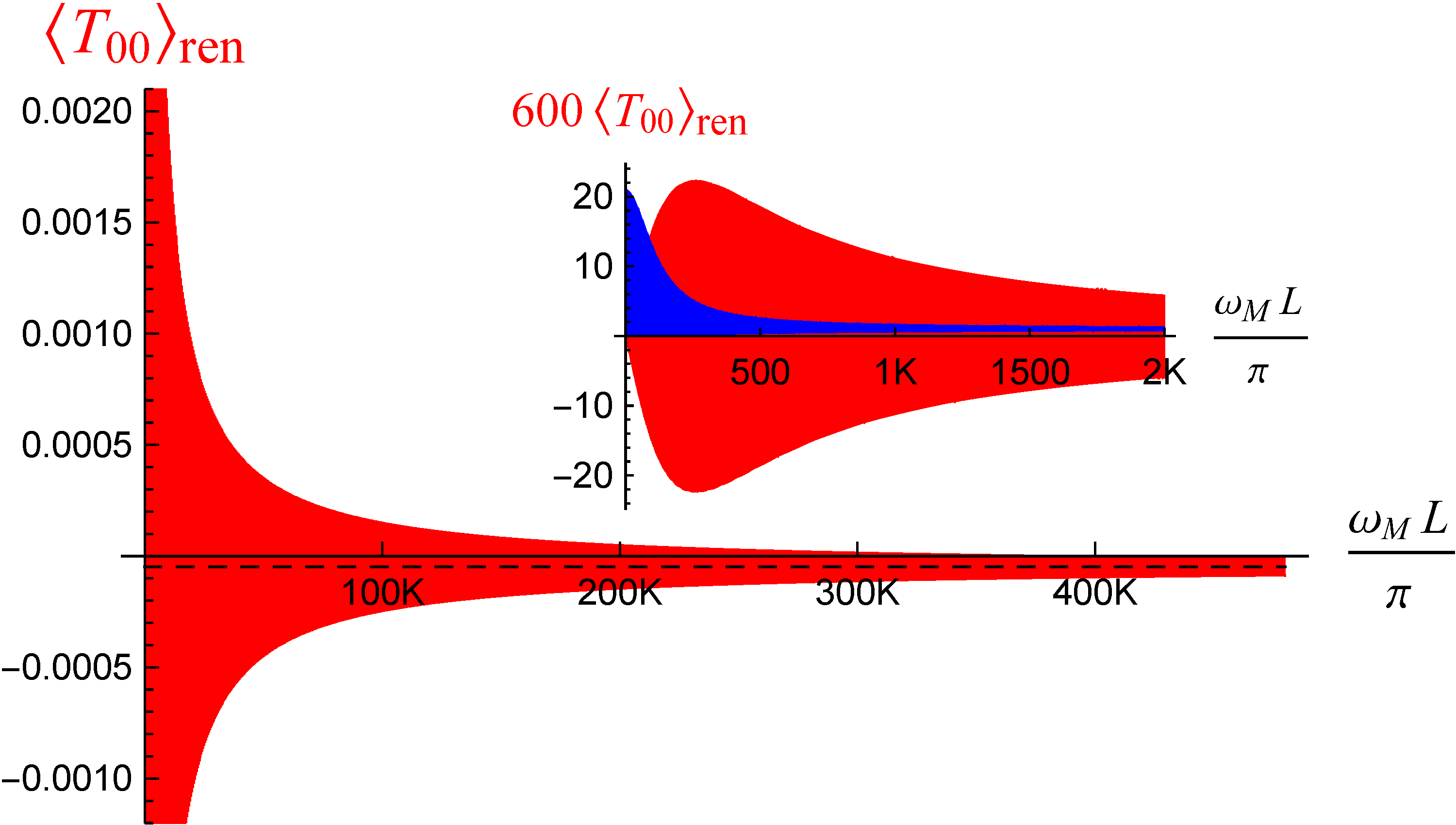} \hspace{.5cm}
\includegraphics[width=5cm]{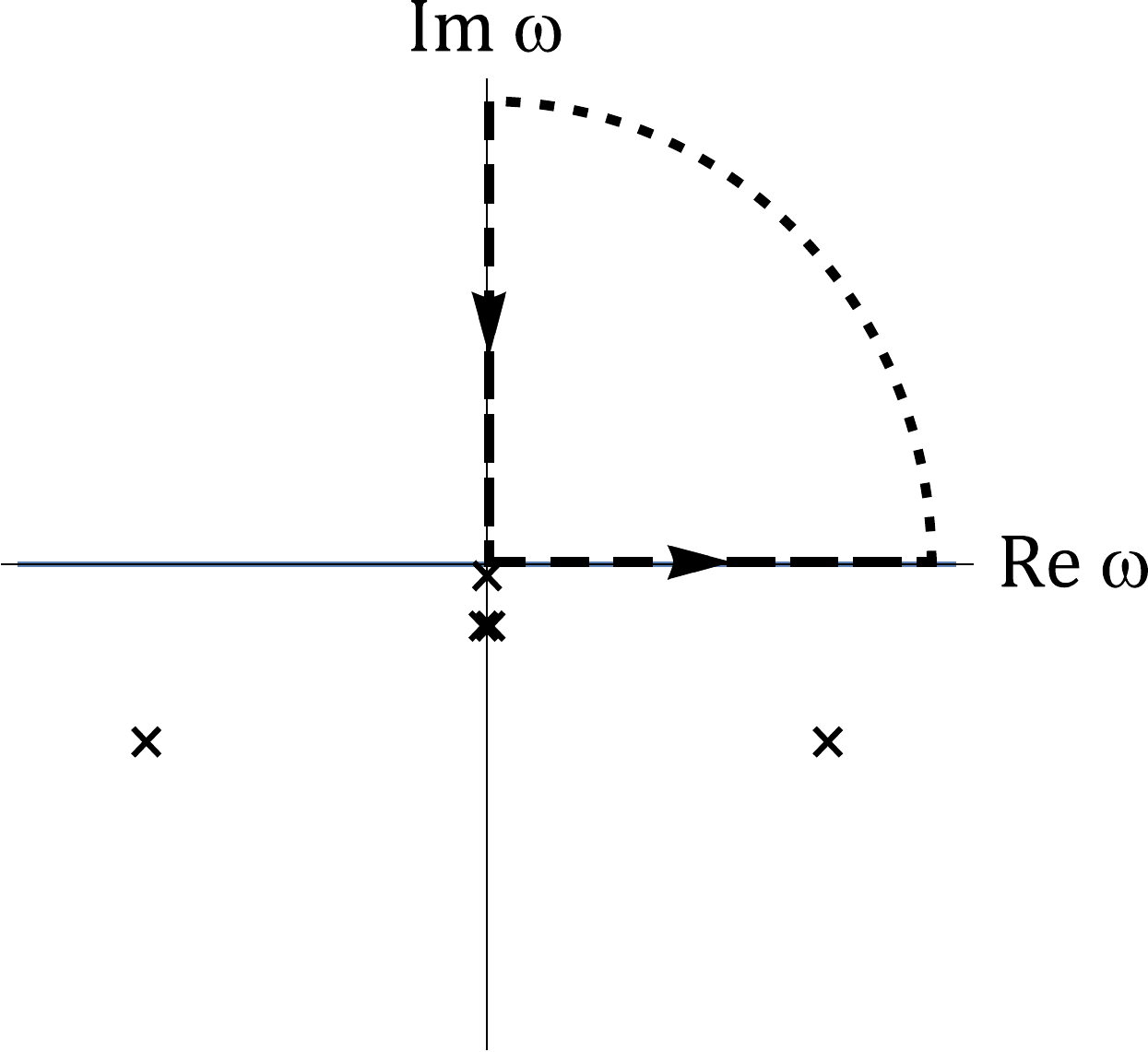}
\caption{(Left) Late-time energy density of the field $\langle \hat{T}_{00}\rangle_{\rm ren}$ in (\ref{E0inCav}) inside the cavity against 
the UV cutoff $\omega^{}_M$ scaled by $\pi/L$ (red). Here $\gamma=10$, $\tilde{\gamma}=1$, $\Omega_0=0.1$, $L=40$, and $c=\hbar=1$.
The value of $\langle \hat{T}_{00}\rangle_{\rm ren}$ oscillates between negative and positive values for $\omega^{}_M$ less than about 
$4.2\times 10^5 \pi/L$, and then converges to $-0.0000483163$ (black dashed line) as $\omega^{}_M$ increases further. 
The blue curve in the inset is the field spectrum in Figure \ref{Phi2xkOvr} (right). The largest amplitude of the oscillating $\langle 
\hat{T}_{00}\rangle_{\rm ren}$ occurs around $(\omega^{}_M L/\pi) \approx 250$, namely, $\omega^{}_M \approx 2\gamma = 20$, where the peak 
values of the field spectrum have dropped significantly from the maximum at low $\omega^{}_M$. 
(Right) The poles (represented in ``$\times$") in the integrand of (\ref{E0inCav}) are all in the lower half of the complex $\omega$ plane.
Thus the integral along the closed contour (dashed and dotted lines) must vanish.}
\label{E0Lam}
\end{figure}

The integral in (\ref{E0inCav}) for small UV cutoff $\omega^{}_M$ oscillates between negative 
and positive values as $\omega^{}_M$ increases (Figure \ref{E0Lam} (left)). 
The amplitude of this oscillation remains large until $\omega^{}_M$ gets much greater than $\gamma$, $\tilde{\gamma}$, and $\Omega_0$, when 
the $\omega^4$ term dominates the denominator of the integrand in (\ref{E0inCav}) for $\omega$ close to $\omega^{}_M$ and makes the integral 
evolving like $-\hbar{\rm Re}[\int^{\omega^{}_M} d\omega 8 \gamma^2 e^{2i\omega L}/(2\pi\omega)] = -4\hbar\gamma^2 {\rm Ci}(2L\omega^{}_M)/
\pi\approx -2\hbar\gamma^2 (\pi L\omega^{}_M)^{-1}\sin(2L\omega^{}_M)$ on top of the lower-UV-cutoff result, so that $\langle \hat{T}_{00}
(t,x)\rangle_{\rm ren}$ in the cavity oscillates roughly about the constant $\rho_\Phi$ with the amplitude decreasing as $\omega_M^{-1}$.
One cannot see whether the value of the renormalized energy density is negative or positive if the UV cutoff is not large enough.
If $\rho^{}_\Phi <0$, one should take the value of $\omega^{}_M$ much greater than $2\hbar\gamma^2/(\pi L|\rho^{}_\Phi|)$ to resolve the 
negativity of $\rho^{}_\Phi$. This reminds us about the fact that the Casimir effect is a finite-size effect of constraints on quantum fluctuations \cite{HO87}, which is not a purely IR or UV phenomenon. It depends not only on the field modes of long wavelengths comparable with the scale of the background geometry. One has to sum over all the cavity modes in a perfect cavity to obtain the conventional result of the Casimir energy density \cite{Ca48}. 

If one introduces a normalizable, smooth switching function such as a Gaussian or Lorentzian function of time for the coupling of an 
apparatus to the cavity field, it will suppress the contribution from the short-wavelength modes \cite{LCH16} and makes the ``observed" 
energy density not so negative \cite{Fo91, FR95, Fl97}. 
In our model the spectrum of the short-wavelength modes is closer to the ones in free space than those in a perfect cavity. 
One may wonder if there exists some choice of the parameter values which leads to a non-negative late-time energy density in our cavity
for $\omega^{}_M$ sufficiently large. To answer this question, one needs to know the exact sign of $\rho^{}_\Phi$,
which looks very hard in calculating (\ref{E0inCav}) numerically when $\rho^{}_\Phi$ is extremely close to zero.

Fortunately, the poles in the integrand of (\ref{E0inCav}) are all located in the lower half of the complex plane. Thus the integral along a 
closed contour from $\omega = 0\to \infty\to i\infty\to 0$ in the upper complex plane (Figure \ref{E0Lam} (right)) gives zero. 
Since $L > 0$ in the factor $e^{2iL\omega}$ in the numerator of the integrand in (\ref{E0inCav}), which suppressed the contribution 
around $\omega \sim i\infty$ (the dotted part of the contour in Figure \ref{E0Lam} (right)), we have
\begin{equation}
  \rho^{}_\Phi = -\hbar \int_0^\infty \frac{d\beta}{2\pi}\frac{8 \gamma^2 \beta^3 e^{-2L\beta}}
	{\left[ \beta^2 + 2\beta(\gamma+\tilde{\gamma})+\Omega_0^2\right]^2 -4\gamma^2 \beta^2 e^{-2L\beta}}, \label{E0inCavWR}
\end{equation} 
which is Wick-rotated from (\ref{E0inCav}) by letting $\omega =i\beta$ \cite{GJ02, GJ03, GJ04, Ja05}.
Eq. (\ref{E0inCavWR}) converges much faster than (\ref{E0inCav}) in numerical calculations. 
Further, the integrand in (\ref{E0inCavWR}) is positive definite for $\beta\ge 0$, so $\rho^{}_\Phi$ must be negative for all regular, non-resonant choices of the parameter values in our model (in the resonant case with $\tilde{\gamma}=0$ and $\Omega_0=n\pi/L$ for some positive integer $n$, the system will never settle down to the late-time steady state with (\ref{E0inCav}); see Sec. \ref{cavmodLT}). 
Note that we did not take the strong OF coupling limit in obtaining (\ref{E0inCav}) and (\ref{E0inCavWR}). 
Even in the weak OF coupling regime where the working range of our detector mirrors is narrow (recall Figures \ref{reflec} and 
\ref{Phi2xkUnd}), the Casimir energy density in our cavity with sufficiently large $\omega^{}_M$ is still negative, though it may be very 
close to zero. In the example in Figure \ref{Phi2xkUnd}, indeed, one has $\rho^{}_\Phi \approx -6.9096\times 10^{-10} <0$ in the cavity,
while only one pair of the cavity modes are significant in the under-damping regime there.

It is obvious in (\ref{E0inCav}) and (\ref{E0inCavWR}) that the Casimir energy density goes to zero as the OF coupling $\gamma\to 0$. 
Going to the other extreme, if one takes the limit $\gamma\to \infty$ before doing integration \cite{GJ02,GJ03,GJ04,Ja05}, then
\begin{eqnarray}
    \rho^{}_\Phi &\to& -\hbar {\rm Re} \int_0^\infty
		\frac{d\omega}{2\pi}\frac{8 \gamma^2 \omega^3 e^{2i\omega L}}{- 4\omega^2\gamma^2+
		4\gamma^2\omega^2 e^{2i\omega L}} =-\hbar {\rm Re} \int_0^\infty
		\frac{d\omega}{\pi}\frac{\omega e^{2i\omega L}}{-1 + e^{2i\omega L}} \nonumber\\
		&=& \hbar {\rm Re} \int_0^\infty\frac{d\omega}{\pi} \omega \sum_{n=1}^\infty e^{2i\omega Ln} 
		=\frac{\hbar}{\pi} {\rm Re} \sum_{n=1}^\infty \int_0^\infty d\omega \omega e^{2i\omega Ln} 
		\nonumber\\ &=& \frac{\hbar}{\pi}\sum_{n=1}^\infty \frac{-1}{4 L^2 n^2}
		= - \frac{\hbar \pi}{24L^2}, \label{ECconven}
\end{eqnarray}
and one recovers the conventional result for a perfect cavity in (1+1)D \cite{BD82}. In the above calculation a regularization 
$L\to L+i\epsilon$ with $\epsilon\to 0+$ is understood. For $L=40$, $\rho^{}_\Phi \approx -0.0000818123$ in (\ref{ECconven}), which is
the same order of magnitude as the Casimir energy density in Figure \ref{E0Lam}.

Right at the position of a detector mirror ($z^1_A=0$ or $z^1_B=L$), one has the late-time renormalized energy density of the field
\begin{equation}
    \langle \hat{T}_{00}(z^\mu_{\bf d})\rangle_{\rm ren} \to  -\frac{\gamma}{2} \langle \hat{P}^2_{\bf d}(t) \rangle + 
		\left.\langle \hat{T}_{00}(t,x)\rangle_{\rm ren}\right|_{0<x<L} 
\end{equation}
which appears to have a logarithmic divergence in the first term if we did not introduce a UV cutoff $\omega^{}_M$
for $\langle \hat{P}^2_{\bf d}(t) \rangle$ (see (\ref{PA2CavLT}) and below). With a finite $\omega^{}_M$, while the above energy density of 
the field has a large negative value, its contribution to the field energy is about $\{-(\gamma/2) \langle \hat{P}^2_{\bf d}(t)\rangle 
+\langle \hat{T}_{00}\rangle_{\rm ren}|_{0<x<L}\} dx$, which is small compared with the detector energy 
$E_{\bf d} =(\langle \hat{P}^2_{\bf d}(t) \rangle + \Omega_0^2\langle \hat{Q}^2_{\bf d}(t) \rangle)/2$. 
Thus the total energy around is still positive. Also the total Casimir energy of the field is still
\begin{equation}
  E^{}_\Phi =\int_{-\infty}^\infty dx\langle\hat{T}_{00}(t,x)\rangle_{\rm ren}=L \left.\langle \hat{T}_{00}\rangle_{\rm ren}\right|_{0<x<L}
\label{CasEn}
\end{equation}
since the contribution by the finite $\langle \hat{T}_{00}(z^\mu_{\bf d})\rangle_{\rm ren}$ at $x=0$ and $x=L$ are infinitesimal in the 
integral.

When $L\to 0$, the conventional result for the Casimir energy diverges like $L\times (-L^{-2}) = -L^{-1}$ from (\ref{ECconven}) and 
(\ref{CasEn}). In contrast, $\rho^{}_\Phi$ in (\ref{E0inCavWR}) behaves like $\ln L$ when $L$ is small, so the total Casimir energy $E_\Phi 
\sim L\times \ln L$ goes to zero as the separation $L\to 0$ in our model. 
The total energy of our HO-field system (with the field energy radiated in transient ignored) is thus finite and cutoff dependent, and would be positive when the UV cutoff is sufficiently large.

\subsection{Late-time entanglement between mirror oscillators}
\label{EntMirOsc}

For our cavity of two detector mirrors, the symmetric two-point correlators of the internal HOs of the detectors can be formally represented 
as
\begin{eqnarray}
  \langle \hat{Q}^{}_{\bf d}(t), \hat{Q}^{}_{\bf d'}(t') \rangle &=& \frac{1}{2} {\rm Re} \left[ 
	  \sum_{\tilde{\bf d},\tilde{\bf d}'=A,B}\frac{\hbar}{2\Omega_0} q_{\bf d}^{\tilde{\bf d}}(t)q_{\bf d'}^{\tilde{\bf d}'*}(t') + 
		\int \frac{d\rm k}{2\pi} \frac{\hbar}{2\rm w} q_{\bf d}^{\rm k}(t)q_{\bf d'}^{\rm k*}(t')	\right. \nonumber\\ & & +\left.
		\int \frac{d\tilde{\rm k}_A}{2\pi} \frac{\hbar}{2\tilde{\rm w}^{}_A} q_{\bf d}^{\tilde{\rm k}_A}(t)q_{\bf d'}^{\tilde{\rm k}_A*}(t')+
\int \frac{d\tilde{\rm k}_B}{2\pi} \frac{\hbar}{2\tilde{\rm w}^{}_B} q_{\bf d}^{\tilde{\rm k}_B}(t)q_{\bf d'}^{\tilde{\rm k}_B*}(t')\right],
\end{eqnarray}
and so on. After some algebra, the late-time correlators of the oscillators are found to be
\begin{eqnarray}
  &&\langle \hat{Q}^{2}_A(t)\rangle = \langle \hat{Q}^{2}_B(t)\rangle = 2{\rm Re} \left( {\cal F}^{}_{0+} + {\cal F}^{}_{0-}\right),
	  \label{QAQACavLT}\\
  &&\langle \hat{Q}^{}_A(t), \hat{Q}^{}_B(t)\rangle = 2{\rm Re} \left( {\cal F}^{}_{0+} - {\cal F}^{}_{0-}\right), \\
  &&\langle \hat{P}^{2}_A(t)\rangle = \langle \hat{P}^{2}_B(t)\rangle = 2{\rm Re} \left( {\cal F}^{}_{2+} + {\cal F}^{}_{2-}\right), 
	  \label{PA2CavLT}\\
  &&\langle \hat{P}^{}_A(t), \hat{P}^{}_B(t)\rangle = 2{\rm Re} \left( {\cal F}^{}_{2+} - {\cal F}^{}_{2-}\right), \label{PAPBCavLT}
\end{eqnarray}
and $\langle\hat{Q}_{\bf d}(t), \hat{P}_{\bf d'}(t)\rangle = 0$. Here
\begin{equation}
    {\cal F}^{}_{c\pm} \equiv \frac{\hbar}{4\pi} \int_0^{\omega^{}_{M}} d\omega \, \omega^{c-1} \chi_\omega^\pm . \label{calFdef}
\end{equation}
with the UV cutoff $\omega^{}_M$ and the susceptibility functions $\chi_\omega^\pm$ defined in (\ref{chipm}). 
The above late-time results are actually constants of $t$ and very similar to Eqs. (48)-(52) in Ref. \cite{LH09} except the oscillating term 
($\propto \gamma e^{i\omega L}$ in the denominator of $\chi_\omega^\pm$) due to the differences in the coupling and the number of spatial 
dimensions. Unlike its counterpart in \cite{LH09}, the oscillating term here keeps the denominator of the integrand of 
${\cal F}^{}_{c\pm}$ regular as $L\to 0$ for every finite $\omega$. 

For $L=0$, the integrals of ${\cal F}^{}_{c\pm}$ can be done analytically to get
\begin{eqnarray}
 \left.{\cal F}^{}_{0\pm}\right|_{L=0} &=& \left.\frac{\hbar i}{4\pi\Gamma^{}_\pm}\tan^{-1}\frac{\omega+i\gamma^{}_\pm}{\Gamma^{}_\pm}
    \right|^{\omega^{}_{M}}_{\omega=0} \nonumber\\ &&\stackrel{\omega^{}_M \gg\gamma_\pm,\Omega_0}{\longrightarrow}
     \frac{\hbar i}{4\pi\Gamma^{}_\pm}\left( \frac{\pi}{2} - 
     \tan^{-1}\frac{i\gamma^{}_\pm}{\Gamma^{}_\pm} \right), \label{F0pmL0}\\
 {\rm Re} \left.{\cal F}^{}_{2\pm}\right|_{L=0} &=& (\Omega_0^2 -2\gamma_\pm^2){\rm Re}\left.{\cal F}^{}_{0\pm}\right|_{L=0}
  +\frac{\hbar\gamma_\pm}{8\pi} \ln \frac{(\omega_M^2 - \Omega_0^2)^2 + 4 \gamma_\pm^2 \omega_M^2}{\Omega_0^4} \nonumber\\
&& \stackrel{\omega^{}_M\gg\gamma_\pm,\Omega_0}{\longrightarrow}
    (\Omega_0^2 -2\gamma_\pm^2){\rm Re}\left.{\cal F}^{}_{0\pm}\right|_{L=0} + 
    \frac{\hbar\gamma^{}_\pm}{2\pi} \Lambda_1, \label{F2pmL0} 
\end{eqnarray}
with $\gamma^{}_\pm \equiv \tilde{\gamma} + \gamma (1\pm 1)$ and $\Gamma^{}_\pm \equiv \sqrt{\gamma_\pm^2 - \Omega_0^2}$, which can be 
real (over-damping) or imaginary (under-damping). 
Here we set $\omega^{}_M = \Omega_0 e^{\Lambda_1}$ to recover Eq. (A12) in Ref. \cite{LH07} after the $\Lambda_1$ there is redefined
as $\Lambda_1 = -\gamma_e - \ln \Omega_r|\tau-\tau'|$, as we discussed in Sec. \ref{OFEnt}
\footnote{Below Eq. (A9) in Appendix A of \cite{LCH16}, $\omega^{}_M$ is put as $2\pi\Omega e^{\Lambda_0+\gamma_e}$ or $2\pi\Omega 
e^{\Lambda_1+\gamma_e}$. To exactly recover Eqs. (A9)-(A12) in \cite{LH07}, they should be corrected to $\omega^{}_M =\Omega e^{\Lambda_0}$ or 
$\Omega e^{\Lambda_1}$.}.
While ${\rm Re}\,{\cal F}^{}_{2\pm}|_{L=0}$ is UV divergent as $\omega^{}_{M}\to\infty$, when the UV cutoff $\omega^{}_M$ and so $\Lambda_1$ are set to be finite and not too large, the internal HOs of the two UD$'$ detectors can be entangled. 
For example, when $\Lambda_1=100$, 
$\gamma = 10$, $\tilde{\gamma}=0.01$, $\Omega_0 = 0.1$, and $c=\hbar=1$, we find $c_-^2  -(\hbar^2/4)\approx -0.18$ 
with $c_-^2 \equiv 16{\rm Re} {\cal F}^{}_{0+} {\rm Re} {\cal F}^{}_{2-}$
\footnote{The condition $16{\rm Re}{\cal F}^{}_{0+}{\rm Re}{\cal F}^{}_{2-} -(\hbar^2/4) <0$ in the context above 
Eq.(58) in Ref. \cite{LH09} obviously should be corrected to $16{\rm Re} {\cal F}^{}_{0-} {\rm Re}{\cal F}^{}_{2+} - (\hbar^2/4) < 0$. 
Here in the UD$'$ detector theory with the derivative coupling in (1+1)D, however, we have $(c_-^2, c_+^2) = (16{\rm Re} {\cal F}^{}_{0+} 
{\rm Re} {\cal F}^{}_{2-}, 16{\rm Re} {\cal F}^{}_{0-} {\rm Re} {\cal F}^{}_{2+})$, in contrast to those expressions in 
\cite{LH09} with the minimal coupling in (3+1)D. So $16{\rm Re} {\cal F}^{}_{0+} {\rm Re}{\cal F}^{}_{2-} - (\hbar^2/4) < 0$ implies
entanglement here.}, and
the separability function $\Sigma\equiv\left(16{\rm Re} {\cal F}^{}_{0+} {\rm Re} {\cal F}^{}_{2-} - (\hbar^2/4)\right) \times 
\left(16{\rm Re}{\cal F}^{}_{0-} {\rm Re} {\cal F}^{}_{2+} - (\hbar^2/4)\right) \approx -1076$ 
is negative \cite{LH09}, while the uncertainty function $\Upsilon \equiv \left(16{\rm Re} {\cal F}^{}_{0+} {\rm Re} {\cal F}^{}_{2+} - 
(\hbar^2/4)\right) \left(16{\rm Re} {\cal F}^{}_{0-} {\rm Re} {\cal F}^{}_{2-} -(\hbar^2/4)\right) \approx 370$ 
is positive. This implies that the reduced state of the oscillator pair, which is a Gaussian state, is well behaved and the 
oscillators are entangled (with the logarithmic negativity $E_{\cal N} = \max \{0, -\log_2 (2c_-/\hbar) \} \approx 0.94$) 
\cite{Si00, DG00, VW02, Pl05, LH09}. 
If we increase the value of $\Lambda_1$ while keeping all other parameters unchanged, the oscillators will be entangled until $\Lambda_1$
exceeds about $400$.

\begin{figure}
\includegraphics[width=7cm]{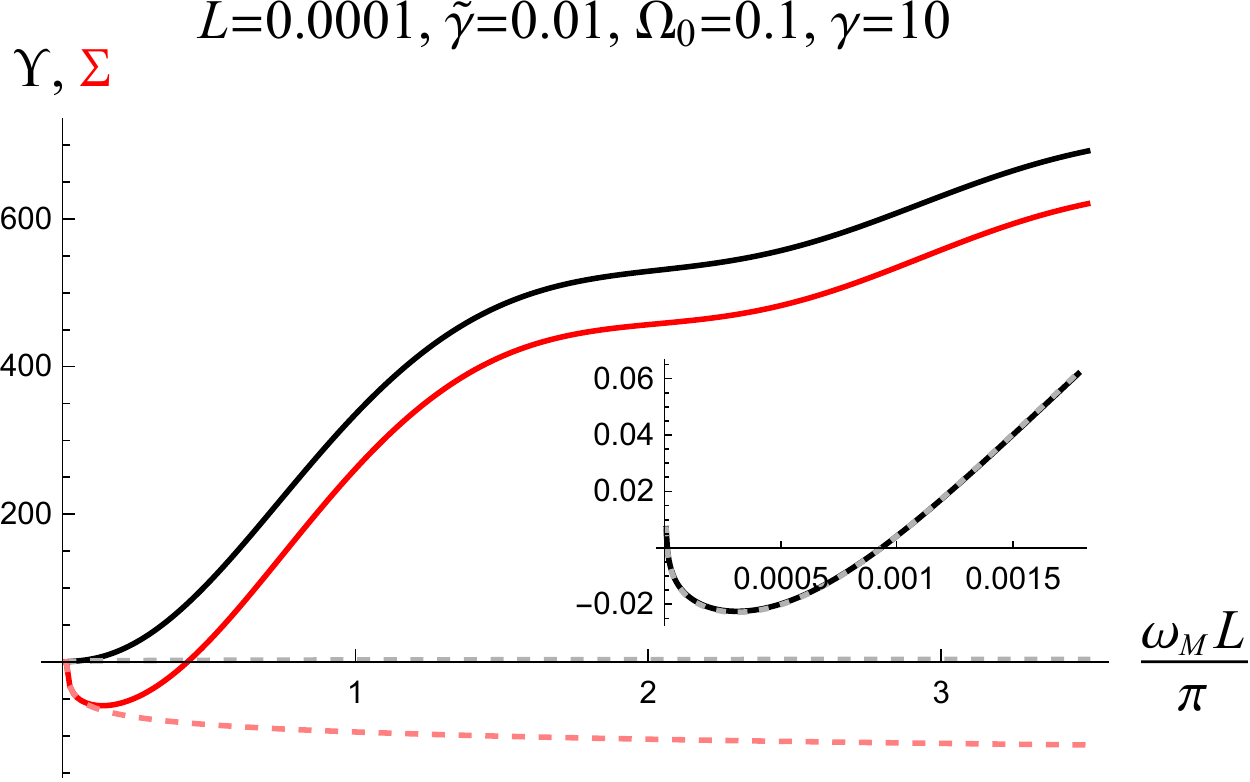}\hspace{.4cm}
\includegraphics[width=7cm]{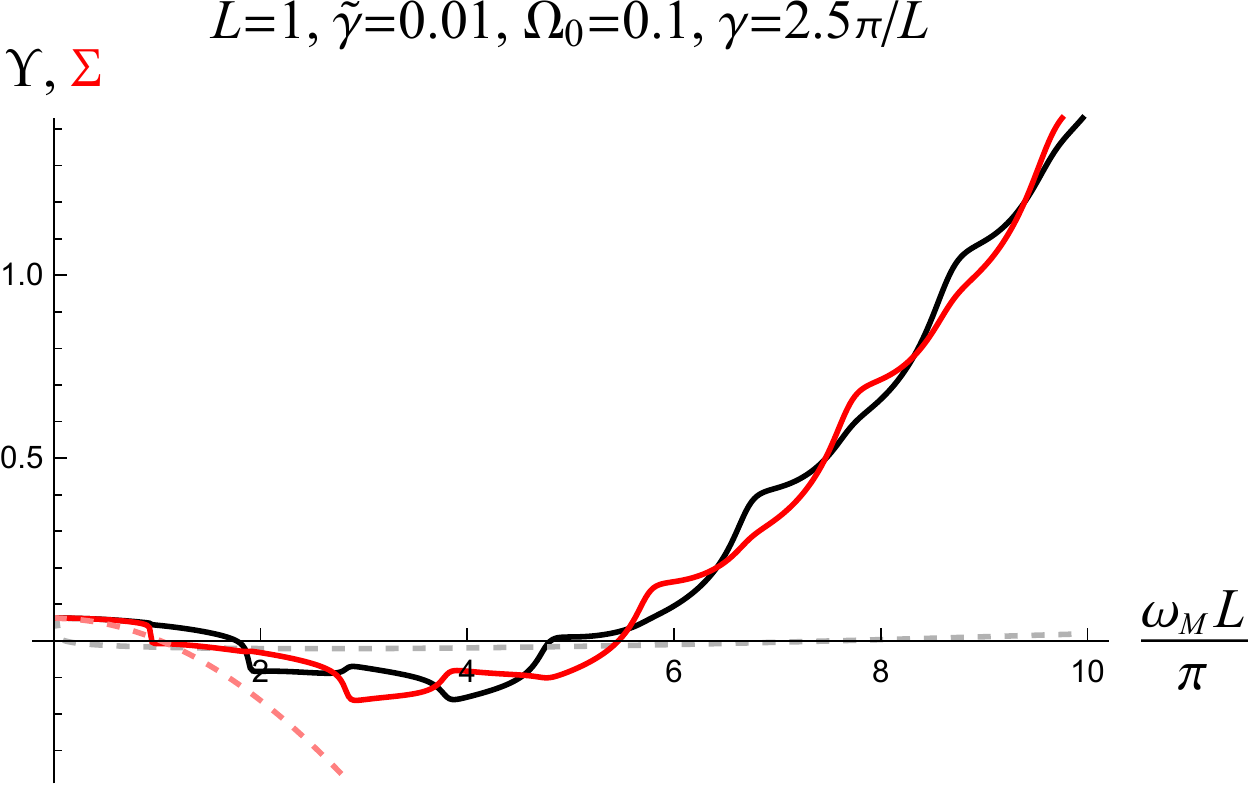}
\caption{The uncertainty function $\Upsilon$ (black lines) and the separability function $\Sigma$ (red lines) with $L=0.0001$ (left) and
$L=1$ (right) against the UV cutoff $\omega^{}_M$ (scaled by $L/\pi$ in the plot) at late times. 
The gray dashed and pink dashed curves are $\Upsilon$ and $\Sigma$, respectively, for $L=0$ with the same $\omega^{}_M$ (obtained from 
Eqs. (\ref{F0pmL0}) and (\ref{F2pmL0})).}
\label{UpSig}
\end{figure}

\begin{figure}
\includegraphics[width=5.5cm]{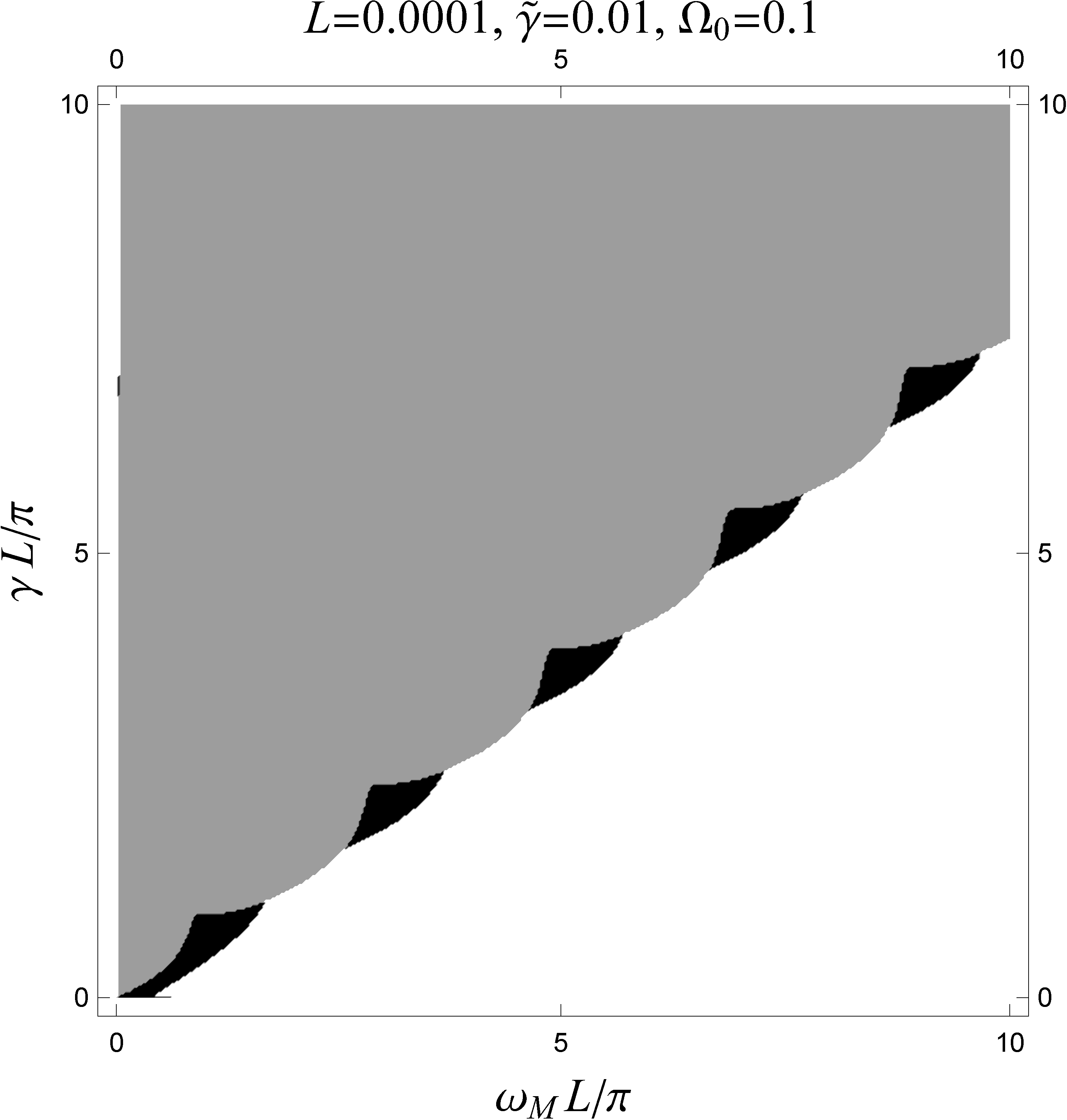}
\includegraphics[width=5.5cm]{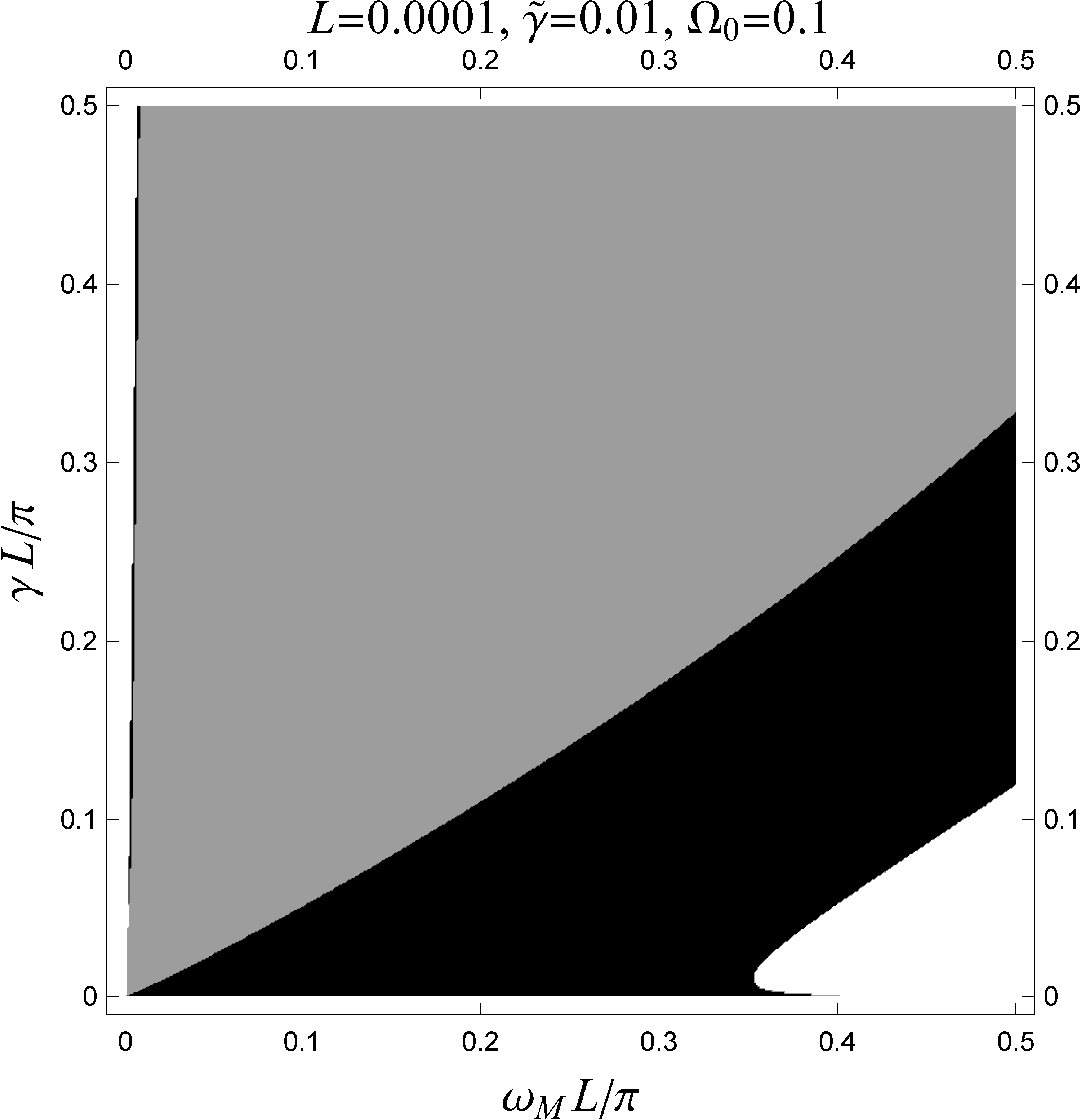}
\includegraphics[width=5.5cm]{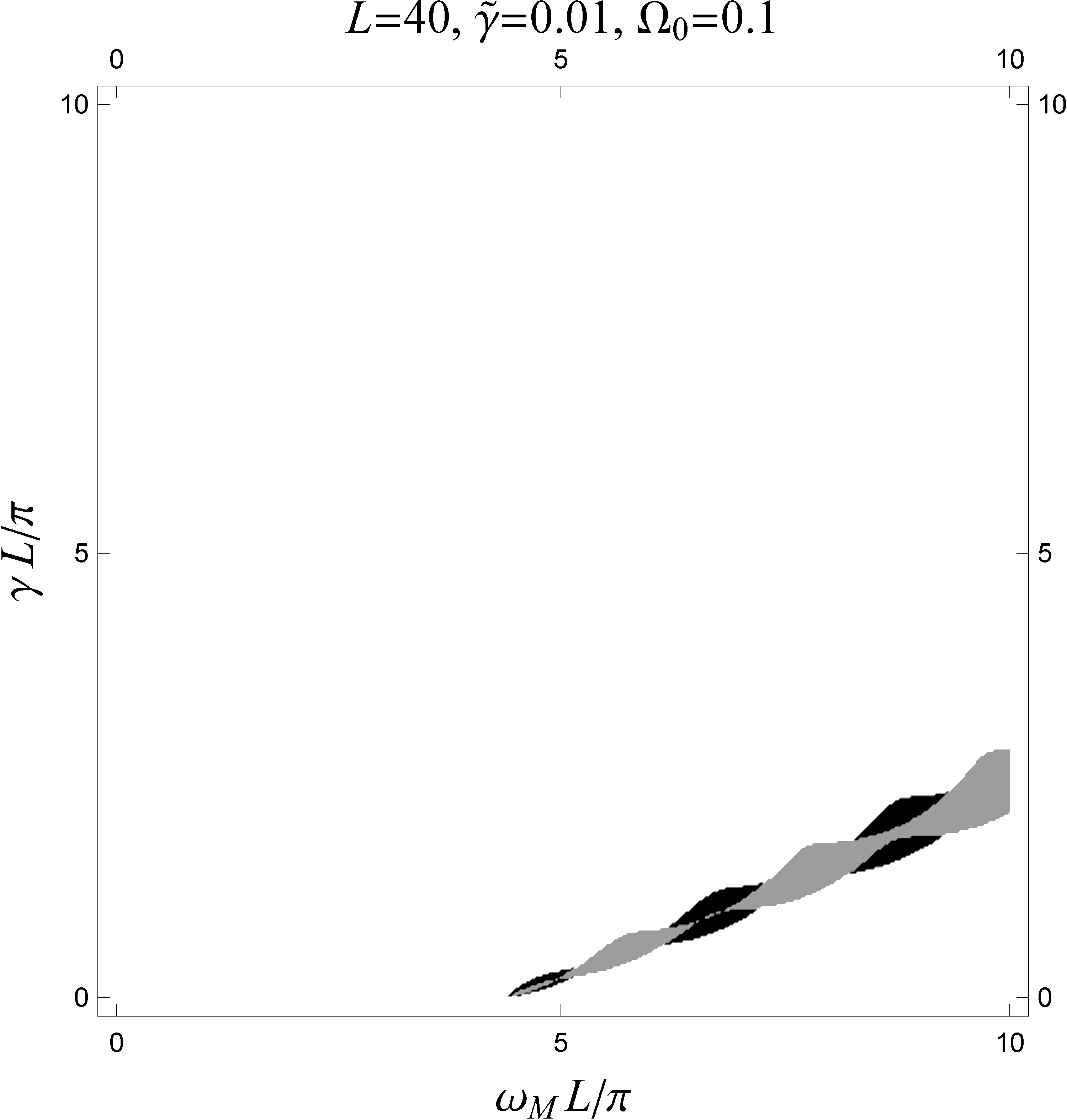}\\ \vspace{.5cm}
\includegraphics[width=5.5cm]{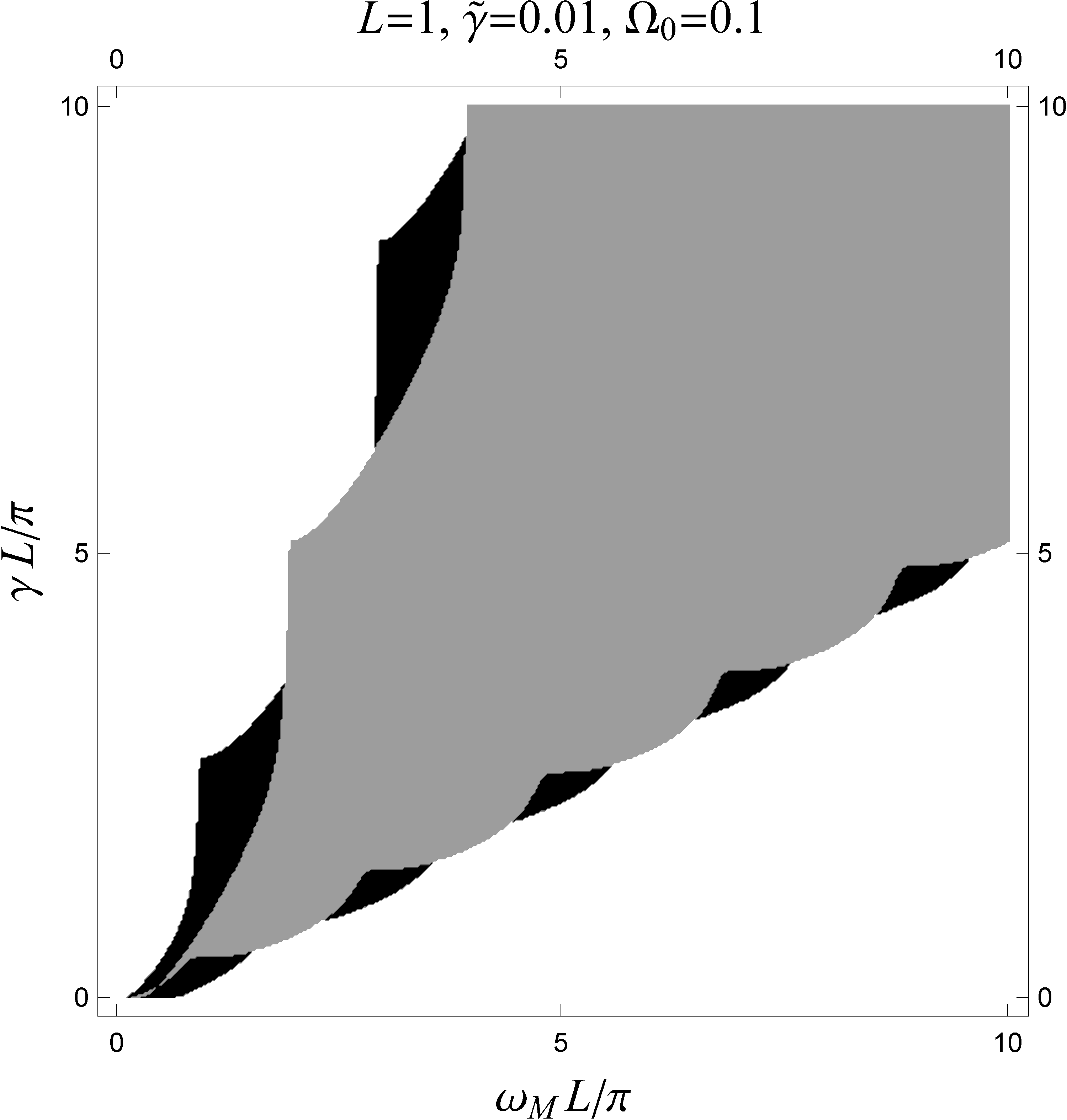}
\includegraphics[width=5.5cm]{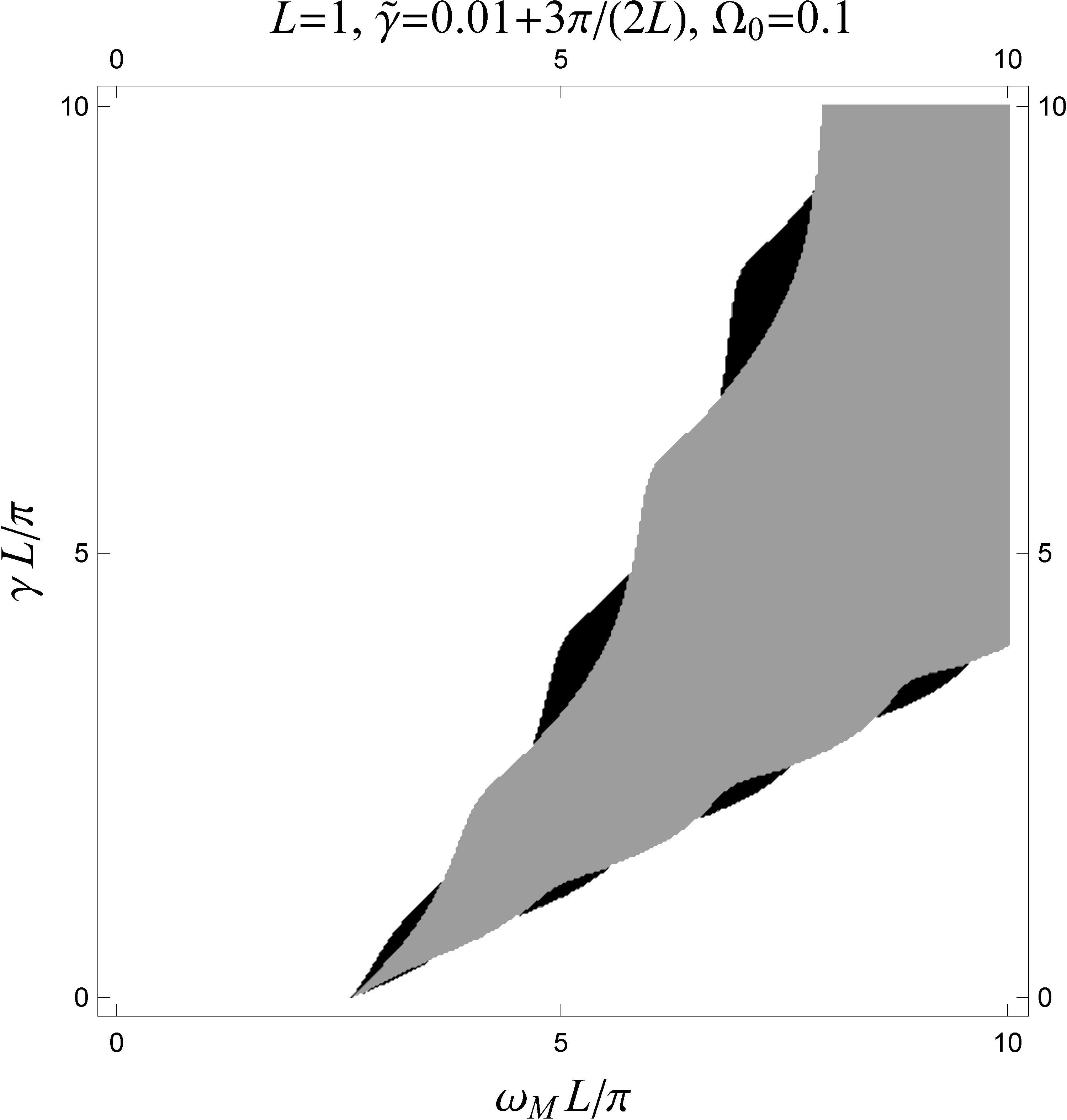}
\includegraphics[width=5.5cm]{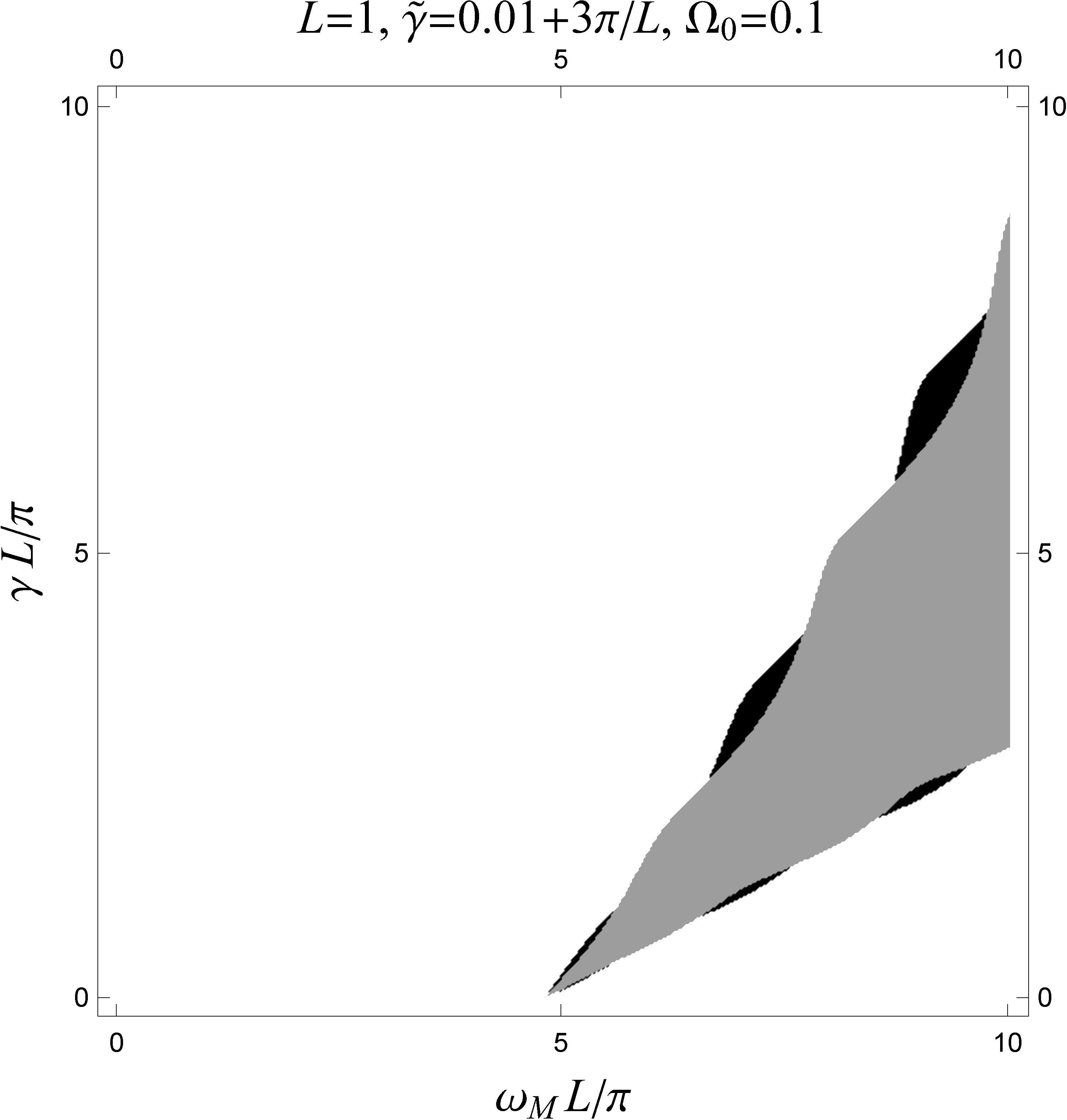}
\caption{The HO pair with ($\omega^{}_M$, $\gamma$) in the dark regions is entangled at late times ($\Sigma<0$ and $\Upsilon\ge 0$),
while in the gray regions the uncertainty relation of the reduced state of the HOs is violated ($\Upsilon <0$) and so unphysical.
The upper-middle plot is an enlargement of the lower-left corner of the upper-left plot.
The result along the horizontal lines $\gamma L/\pi =0.001$ in the upper-left and upper-middle plots and the line $\gamma L/\pi =2.5$ in the 
lower-left plot can be compared with Figure \ref{UpSig} (left) and (right), respectively.} 
\label{UpSig2}
\end{figure}

For $L>0$, the integrals of ${\cal F}^{}_{c\pm}$ deviate significantly from those with $L=0$ for $\omega^{}_M>O(\pi/L)$ (Figure \ref{UpSig}). 
When we fix $\tilde{\gamma}$, $\Omega_0$, and $L$, the unphysical negative-$\Upsilon$ region in which the uncertainty relation 
$\Upsilon\ge 0$ is violated looks like a wedge in the $\omega^{}_M\gamma$-plane in our examples with either $\omega^{}_M$ or $\gamma$ not 
too large (gray regions in Figure \ref{UpSig2}). The angle and the slopes of the two boundaries of the wedge decrease as $L$ increases 
(compare the upper-left, upper-right, and lower-left plots in Figure \ref{UpSig2}). Around the boundary of the negative-$\Upsilon$ region 
there are islands of parameter values in which one has $\Sigma<0$ while the uncertainty relation $\Upsilon\ge 0$ holds
(dark regions). The late-time quantum entanglement between the oscillators of the two mirrors only occurs when the point $(\omega^{}_M, 
\gamma)$ with the fixed values of $\tilde{\gamma}$, $\Omega_0$, and $L$ is located in one of these islands in the parameter space. 
The islands look disconnected in the $\omega^{}_M\gamma$-plane because $\Upsilon(\omega^{}_M)$ and $\Sigma(\omega^{}_M)$ are alternating 
when $\omega^{}_M \sim O(\gamma)$; namely, if $\Upsilon(\omega^{}_M)>\Sigma(\omega^{}_M)$ for some $\omega^{}_M = \Lambda$ then $\Upsilon(\omega^{}_M)<\Sigma(\omega^{}_M)$ for $\omega^{}_M \approx \Lambda\pm \pi/L$, as shown in Figure \ref{UpSig} (right). This is due to the 
alternating nature of the $\gamma(1\pm e^{i\omega L})$ term in the denominators of $\chi^\pm_\omega$ in ${\cal F}_{c\pm}$. 
As $\tilde{\gamma}$ increases, the projections of the islands on the $\omega^{}_M$-axis are roughly invariant, while the whole wedge of the 
$\Upsilon<0$ region shifts along the $+\omega^{}_M$ direction (from left to right in the lower row of Figure \ref{UpSig2}). 
The width of those islands in $\omega^{}_M$ is about $O(\pi/L)$; thus, the larger $L$ would give a smaller scale of the 
islands in the $\omega^{}_M\gamma\tilde{\gamma}$-space. 

For any UV cutoff $\omega^{}_M$, no matter how large it is, the above result suggests that one still has a chance to find an 
OF coupling strength $\gamma \sim O(\omega^{}_M)$ while adjusting the UV cutoff around $\omega^{}_M \pm \pi/L$ 
(with $\tilde{\gamma}$, $\Omega_0$, and $L$ fixed) to make the two internal HOs entangled at late times. 
However, this is extremely fine-tuned and the result cannot be trusted in this regime since the interaction energy could easily exceed the validity range of this model.  
Moreover, when $\gamma$ and $\omega^{}_M$ have the same order of magnitude while $\tilde{\gamma}$, $\Omega_0$, and $1/L$ are relatively small, the denominator of the integrand in (\ref{E0inCav}) is approximately $\omega^4 +4i\gamma\omega^3 +4\gamma^2\omega^2(e^{2i\omega L}-1)$, whose three terms are roughly the same order of magnitude, namely, $O(\gamma^4)$, so the energy density of the field in the cavity around this parameter range oscillates largely between positive and negative values as $\omega^{}_M$ increases. 
Indeed, in Figure \ref{E0Lam} (left) one can see that the maximum amplitude of the oscillating value of the field energy density occurs
around $\omega^{}_M \approx 2\gamma$, and the oscillation will not be suppressed until $\omega^{}_M$ is much larger.
Such a large UV cutoff ($\omega^{}_M \gg O(\gamma)$) is also desirable to get rid of the violation of the uncertainty relation,
by noting that the small dark islands are always neighboring to the gray regions in Figure \ref{UpSig2}. 
Thus the late-time entanglement between the HOs of the cavity mirrors is very unlikely to exist for physically reasonable values of the UV cutoff in our model. 

\section{Summary}
\label{Summa}

We employed the derivative-coupling Unruh-DeWitt(UD$'$) HO detector theory in (1+1) dimensions to model the atom mirror interacting with a massless quantum field (OF coupling) and an environment of mechanical degrees of freedom (OE coupling). The reflectivity of our 
atom or detector mirror is dynamically determined by the interplay of the detector's internal oscillator and the field. In the strong OF coupling regime, the effect of the mechanical environment is negligible and the detector acts like a perfect mirror at late times, when the energy density of the field outside the detector vanishes while the field spectrum is nontrivial. Compared with the field correlators in free space, in the presence of a detector mirror the late-time correlators are reduced for both the field amplitudes on the same side and those on two different sides of the mirror.

A pair of such UD$'$ detector mirrors can form a cavity. If both oscillators are decoupled from the environment, the system will not settle to a steady state at late times if the two internal HOs of the cavity mirrors are on resonance, namely, the natural frequency of the oscillator is integer times of the frequency for the massless scalar field in the cavity traveling from one detector mirror to the other. 

If the OE coupling is nonvanishing, the field in this cavity will evolve into a steady, quasi-discrete spectrum at late times. Then there will be many cavity modes in the strong OF coupling, over-damping regime but only one or a few pairs of significant cavity modes in the weak OF coupling, under-damping regime. With the UV cutoff sufficiently large, the late-time renormalized field energy density in the cavity converges to a negative value for all positive OF coupling strengths. In the infinite OF coupling limit, the negative field energy density goes to the conventional result in the Casimir effect. In contrast to the conventional result with the perfect mirrors, however, the total energy density in our cavity does not diverge as the separation of the detector mirrors goes to zero. Outside the cavity the renormalized field energy density is again vanishing while the field spectrum is nontrivial. 

Our result shows that the internal oscillators of the two mirrors of our cavity can have late-time entanglement when the OF coupling strength is roughly of the same order of the UV cutoff for the two identical HOs. In this regime, however, the model is nearly broken down, and the field energy density in the cavity does not converge but is very sensitive to the choice of the UV cutoff. When the UV cutoff is large enough to obtain a convergent value of the Casimir energy density and far from  inconsistencies, the HOs in the parameter range of our results are always separable.

\begin{acknowledgments}
I thank Bei-Lok Hu, Larry Ford, and Jen-Tsung Hsiang for illuminating discussions. 
This work is supported by the Ministry of Science and Technology of Taiwan under Grant No. MOST 106-2112-M-018-002-MY3 and in part by the 
National Center for Theoretical Sciences, Taiwan.
\end{acknowledgments}


\end{document}